\documentclass[12pt, letterpaper]{article}

\RequirePackage[hyphens]{url}
\usepackage[utf8]{inputenc}

\usepackage[export]{adjustbox}
\usepackage{amsmath}
\usepackage{amssymb}
\usepackage{animate}
\usepackage{appendix}
\usepackage{array}
\usepackage{authblk}
\usepackage{bm}
\usepackage{booktabs}
\usepackage{caption}
\usepackage{color}
\usepackage{changepage}
\usepackage{chemformula}
\usepackage{csvsimple}
\usepackage{etoolbox}
\usepackage{enumitem}
\usepackage{float}
\usepackage{framed}
\usepackage[margin=1in]{geometry}
\usepackage{graphicx}
\definecolor{links}{HTML}{0074CC} 
\usepackage[colorlinks=true, allcolors=links]{hyperref}
\usepackage{lineno}
\usepackage{lscape}
\usepackage{listings}
\usepackage{makecell}
\usepackage{marginnote}
\usepackage{mathrsfs}
\usepackage{mathtools}
\usepackage{mdframed}
\usepackage{multicol}
\usepackage{multirow}
\usepackage[natbibapa]{apacite}
\usepackage{setspace}
\usepackage{soul}
\usepackage{tabulary}
\usepackage{subcaption}
\usepackage{tikz}
\usetikzlibrary{shapes, arrows, positioning, fit}
\usepackage{wrapfig}
\usepackage{xargs}
\usepackage[normalem]{ulem}
\graphicspath{ {figs/} }
\usepackage{epstopdf}
\AppendGraphicsExtensions{.tif}
\PassOptionsToPackage{hyphens}{url}\usepackage{hyperref}

\definecolor{sandstorm}{rgb}{0.93, 0.84, 0.25}

 \bibpunct[, ]{(}{)}{,}{a}{}{,}%
\newcommand{\instructions}[1]{}

\newcolumntype{C}[1]{>{\centering\arraybackslash}p{#1}}

\providecommand{\keywords}[1]
{
  \noindent \small	
  \textbf{Keywords: } #1
}

\makeatletter
\newcommand*{\centerfloat}{%
  \parindent \z@
  \leftskip \z@ \@plus 1fil \@minus \textwidth
  \rightskip\leftskip
  \parfillskip \z@skip}
\makeatother

\AtBeginEnvironment{tabular}{\footnotesize} 
\usepackage[colorinlistoftodos,prependcaption,textsize=tiny]{todonotes}
\newcommandx{\unsure}[2][1=]{\todo[linecolor=red,backgroundcolor=red!25,bordercolor=red,#1]{#2}}
\newcommandx{\change}[2][1=]{\todo[linecolor=blue,backgroundcolor=blue!25,bordercolor=blue,#1]{#2}}
\newcommandx{\info}[2][1=]{\todo[linecolor=orange,backgroundcolor=orange!25,bordercolor=orange,#1]{#2}}
\newcommandx{\improvement}[2][1=]{\todo[linecolor=sandstorm,backgroundcolor=	sandstorm!25,bordercolor=sandstorm,#1]{#2}}
\newcommandx{\thiswillnotshow}[2][1=]{\todo[disable,#1]{#2}}

\definecolor{miamired}{HTML}{C3142D}

\newif\ifblinded
\newif\iffinal

\newcommand{\setfinal}{
    \blindedfalse
    \finaltrue
}

\newcommand{\authorinfo}{
    \ifblinded
        \author{\large (Authors blinded for peer review)}
    \fi
    \iffinal
        \author[1,*]{\"Ozge S\"urer}
        \affil[1]{Farmer School of Business, Miami University, Oxford, OH 45056, USA}
        \affil[*]{Corresponding author can be reached at surero@miamioh.edu}
    \fi
}

\def\hb#1{#1}

\title{\Large \textbf{Active Learning for Data-Efficient Calibration of Stochastic Simulation Models}}

\usepackage{caption}
\usepackage{subcaption}
\usepackage{graphicx}
\usepackage{tikz}
\usetikzlibrary{trees}
\usepackage{times,scalefnt}
\usepackage{amsmath,amssymb,mathtools}
\usetikzlibrary{shapes.multipart}
\usetikzlibrary{matrix,decorations.pathreplacing, calc, positioning, shapes.callouts, shapes.geometric}

\usepackage{amsmath}

\usepackage[boxruled,vlined,linesnumbered,commentsnumbered]{algorithm2e}

\usepackage{subcaption}

\newcommand{\p}{{p}}

\newcommand{\Cb}{{\pmb{C}}}

\newcommand{\kb}{{\mathbf{k}}} 
\newcommand{\Kb}{{\mathbf{K}}}

\newcommand{\yb}{{\mathbf{y}}}

\newcommand{\simout}{{\zeta}}
\newcommand{\latent}{{\pmb{\eta}}}

\newcommand{\thetav}{{\pmb{\theta}}}
\newcommand{\Sigmav}{{\pmb{\Sigma}}}

\newcommand{\Vb}{{\mathbf{V}}}

\newcommand{\Sv}{{\mathbf{S}}}

\newcommand{\Mb}{{\mathbf{M}}}

\newcommand{\xb}{{\mathbf{x}}}
\newcommand{\xf}{\mathbf{x}^o}
\newcommand{\zf}{\mathbf{z}^o}
\newcommand{\zb}{{\mathbf{z}}}

\newcommand{\tb}{{\pmb{\theta}}}

\newcommand{\tg}{{\pmb{\vartheta}}}
\newcommand{\PHI}{{\pmb{\phi}}}
\newcommand{\GAMMA}{{\pmb{\Gamma}}}

\newcommand{\muv}{{\pmb{\mu}}}

\newcommand{\ab}{{\pmb{a}}}

\newcommand{\tcb}{\textcolor{black}}

\newcommand{\DatCan}{\mathcal{D}^c_t}

\newcommand{\E}{\mathbb{E}_{\latent(\tb) | \mathcal{D}_{t}}}
\newcommand{\V}{\mathbb{V}_{\latent(\tb) | \mathcal{D}_{t}}}

\newcommand{\outputk}{\zeta_k}
\newcommand{\outputnew}{\breve{\zeta}}
\newcommand{\siminput}{\zb}

\newcommand{\VA}[1]{\mathbb{V}_{\latent\left(\tb^{#1}\right) | \mathcal{D}_{t}}}

\usepackage[english]{babel}
\usepackage{amsthm}
\newtheorem{theorem}{Theorem}[section]

\newtheorem{lemma}[theorem]{Lemma}

\setfinal

\date{\small \today}

\begin{document}

\authorinfo

\maketitle

\begin{abstract}
\noindent Simulation-based calibration aims to infer unknown parameters of complex simulation models by aligning model outputs with real-world observations. 
When simulation runs are computationally expensive, statistical emulators trained on simulation data are used to efficiently approximate the model. 
An intelligent, adaptive selection of simulation inputs for building the emulator can substantially improve the efficiency of the calibration process.
This task is particularly challenging for stochastic simulations with noisy outputs, since both selecting new input locations (exploration) and allocating repeated runs at existing inputs (replication) are essential for efficiently learning the input-output relationship.
In this paper, we introduce an active learning framework that adaptively balances exploration and replication for data-efficient calibration.
Our uncertainty-aware acquisition criterion targets learning the posterior density of the unknown simulation parameters, and we derive two corresponding forms of the acquisition function for exploration and replication.
Building on these, we propose a strategy that, at each stage of the sequential design, chooses between exploration and replication to most effectively reduce the uncertainty in the estimate of the posterior density of the simulation parameters.
Experiments on synthetic benchmarks and a real epidemiological model demonstrate that our approach significantly improves learning of the posterior distribution of the simulation parameters while reducing the number of required simulations, making it well-suited for expensive stochastic simulation settings.
\end{abstract}


\thispagestyle{empty}

\newpage 

\doublespacing

\section{Introduction}
\label{sec:intro}

Simulation models are widely used to analyze complex systems, particularly when real-world experimentation is infeasible or costly. 
In many domains, such as epidemiology, climate science, and manufacturing, stochastic simulation models are essential to capture inherent randomness in system behavior. 
These models typically take controllable design inputs and user-defined model parameters (also referred to as calibration parameters) to generate outputs that represent the system. 
Although the simulation model parameters are not directly observable physical quantities, they govern how the simulation model reproduces real-world behavior and therefore support reliable prediction, scenario analysis, and decision-making in regimes where real-world data are limited or sparse.
Calibration \citep{Sung2024} aims to infer such unknown parameters that align simulation outputs with observed data from real-world or field experiments. 
Unlike deterministic simulation models, stochastic models yield different outputs when evaluated repeatedly at the same input due to inherent randomness, known as intrinsic (aleatoric) uncertainty. 
This makes calibration particularly challenging when simulations are both computationally expensive and noisy, as many evaluations may be required to estimate the underlying response accurately.
In the Bayesian paradigm \citep{Ohagan2001}, calibration yields a posterior distribution over the unknown simulation model parameters, combining prior knowledge with information from observed data. Under the assumption of a well-specified simulation model, learning this posterior distribution allows the simulation model to reproduce real-world behavior in distribution, while explicitly quantifying uncertainty in both parameter estimates and downstream predictions.
This uncertainty-aware inference is essential in practice, as it enables principled risk assessment and decision support in the presence of variability and limited observations.

When simulation runs are expensive, surrogate models are frequently employed to replace direct evaluations of the simulation model \citep{gramacy2020surrogates}. 
Gaussian process (GP) models are widely used as surrogates for deterministic models due to their ability to provide predictions with associated uncertainty \citep{Rasmussen2005}.
While GPs can be extended to emulate stochastic simulation models, these extensions face additional complexity due to non-constant intrinsic variance across inputs \citep{Baker2022}. 
\tcb{In this work, we consider the widely used setting of Gaussian simulation noise. Within this framework,}
stochastic kriging (SK) is \tcb{a commonly adopted approach} for emulating the relationship between inputs and noisy outputs \citep{Ankenman2009}. A \tcb{practical} limitation of SK is that it requires a separate GP model and \tcb{replicated observations} at each input to estimate the input-dependent noise (i.e., intrinsic variance). \tcb{To alleviate this issue}, \citet{Binois2018} introduced heteroskedastic GP (hetGP) models that jointly model the mean and noise variance without requiring a fixed replication level. While \tcb{several} alternative methods for modeling input-dependent noise \tcb{have also been proposed} \citep[e.g.,][]{Kersting2007, Gredilla2011}, we adopt the hetGP model in this work \tcb{primarily because it provides} integrated inference for both the mean and noise variance together with an accessible software implementation.

Constructing a surrogate model requires a design, which refers to the selection of input locations at which the simulation model is evaluated. 
For deterministic simulation models, designs are often constructed using space-filling strategies \citep{santner2018design} that aim to cover the input space with distinct input points. 
A common strategy for handling input-dependent noise in stochastic models is replication, in which the simulation model is evaluated multiple times at the same input location. 
Replication enables estimation of intrinsic uncertainty and can also improve emulator accuracy and reduce emulator construction costs by averaging outputs. 
Accordingly, designs for emulating stochastic models typically combine the selection of distinct input locations with replication to account for input-dependent intrinsic noise. These design strategies often build on those used for deterministic simulations---for example, employing space-filling methods to choose distinct input locations and assigning a fixed number of replicates to each.
However, such designs may be suboptimal for calibration tasks, especially in high-dimensional settings, because they are constructed without reference to how well simulation outputs align with observed data.
As a result, such designs may fail to sufficiently explore regions of the input space that are most informative for calibration. 
Moreover, determining the number of replicates at each input requires careful consideration, as an effective replication strategy should account for variability within the calibration region of interest.

To address these challenges, we propose a design strategy that targets input locations near the (unknown) calibration region of interest and allocates an appropriate number of replicates per location using active learning to facilitate efficient posterior inference for the simulation model parameters.
Active learning, also referred to as sequential design in simulation and statistics, offers key advantages over traditional one-shot approaches where all input locations are selected in advance (see, e.g., \cite{Lam2008}). At the core of active learning is an acquisition function, which quantifies the expected benefit of evaluating the simulation model at a specific input and guides the adaptive selection of new input points. Starting from an initial set of simulation data, additional inputs are chosen sequentially by leveraging information gained from prior runs. Active learning is frequently applied to optimization tasks where the objective is to identify an optimizer \citep{Jalali2017, Frazier2018}. In the context of calibration, this often corresponds to finding the maximum a posteriori estimate, which may result in sampling primarily near the mode of the posterior. However, in this work, our goal is to learn the entire posterior distribution of the unknown simulation model parameters that align simulation outputs with real-world observations.

Active learning is also widely used to build globally accurate surrogate models, particularly when the goal is to approximate the full input–output relationship of a complex simulation model rather than identify a single optimum. To support the construction of globally accurate surrogate models, the integrated mean squared prediction error (IMSE) is a commonly used acquisition criterion. IMSE quantifies the predictive uncertainty of an emulator across the entire input space, making it well-suited for general-purpose emulation. For instance, \citet{Ankenman2009} propose a two-stage design strategy for building accurate SK models: the first stage allocates a fixed number of replicates to a Latin hypercube design, while the second stage determines the optimal replication levels to minimize IMSE. Extending this idea, \cite{ChenZhou2015} and \cite{ChenZhou2017} develop sequential IMSE-based approaches that dynamically balance exploration of new inputs and exploitation through additional replication at existing points. In their framework, each active learning iteration is constrained by a fixed simulation budget, and the key decision is whether to assign replications to existing design points or introduce a new one. More recently, \citet{Binois2019} use the hetGP emulator in a fully sequential setting, acquiring one simulation at a time and choosing between replicating an existing input or sampling a new one based on IMSE minimization.

Although active learning with stochastic simulation models is known to enhance global prediction accuracy of emulators, relatively little research has addressed criteria specifically tailored for calibration tasks, where global prediction is not the primary goal. In recent work, \cite{Surer2025+} propose a batch sequential approach that uses an acquisition function based on the aggregated posterior variance over the parameter space to efficiently learn the posterior of the unknown simulation model parameters. Their setting focused on stochastic simulation models that, when run at a given parameter, return high-dimensional outputs evaluated over a fixed set of design inputs. As a result, only the parameters serve as inputs to the simulation model, and acquisition functions are formulated to select a batch of parameters for either replication or exploration. However, many stochastic simulation models are functions of both unknown parameters and design inputs, with the latter also being part of the real-world (field) data collection process. In such cases, field experiments are performed at a fixed set of design inputs---referred to here as field data design inputs---and the corresponding field observations are used for parameter calibration. In this work, we propose a novel acquisition function for stochastic simulation models that enables the sequential selection of input pairs, each comprising a parameter and a design input, to reduce uncertainty in the estimate of the posterior density of the simulation parameters.

A related approach is presented in \cite{Surer2024}, which considers an active learning strategy for selecting parameter-design input pairs to minimize the total uncertainty in the posterior estimate of the model parameters. Their framework is developed for deterministic simulation models, where repeated evaluations at the same input yield identical outputs. Consequently, replication decisions are not considered, and the approach relies on GP emulators tailored to deterministic simulations, with each acquisition stage focused only on exploring new regions of the input space. In contrast, the proposed work addresses stochastic simulation models with intrinsic, input-dependent variability in the outputs. We model this heteroskedastic uncertainty using a hetGP emulator and explicitly account for the trade-off between exploration of new input locations and replication at existing locations. This leads to two distinct forms of the acquisition criterion, corresponding to exploration and replication, which are essential for efficient posterior learning in the presence of simulation noise. To choose between exploration and replication, we further propose a strategy that encourages replication by leveraging its advantages from both design and computational perspectives. The proposed design balances exploration and exploitation by selecting simulation inputs that align with field data inputs near the parameter region of interest (exploration), while increasing replication in noisier regions to better capture the signal–noise relationship (exploitation).

The remainder of the paper is organized as follows. Section~\ref{sec:review} reviews the calibration and emulation framework and introduces the sequential design setting. Section~\ref{sec:acquisition} details the proposed acquisition function and its two forms for exploration and replication, and discusses strategies to guide the sequential selection of simulation inputs. Experimental results are reported in Section~\ref{sec:experiment}, and concluding remarks are provided in Section~\ref{sec:conc}.

\section{Background}
\label{sec:review}
\subsection{Problem setting}

Let $\xb = (x_1, \ldots, x_q)^\top \in \mathcal{X} \subset \mathbb{R}^{q}$ denote the vector of design inputs, and let $\tg = (\vartheta_1, \ldots, \vartheta_p)^\top \in \Theta \subset \mathbb{R}^{p}$ denote the vector of input parameters. The simulation model $\simout(\cdot)$ takes both the design input $\xb$ and the parameter $\tg$ as input and returns an output. For notational simplicity, define $\zb = \left(\xb^\top, \tg^\top\right)^\top$ as the combined simulation input vector of dimension $q+p$. We denote the expected value of the stochastic simulation output at input $\siminput$ by $\eta(\siminput) = \mathbb{E}[\simout(\siminput)]$. The expected output cannot be observed directly; instead, we only have access to noisy realizations of the simulation output.
The simulation output is modeled as the sum of this expected output and a zero-mean noise term:
\begin{equation} \notag
\simout(\siminput) = \eta(\siminput) + \nu, \quad \nu \sim \mathcal{N}(0, r(\siminput)), 
\end{equation}
where the noise $\nu$ has input-dependent variance $r(\siminput)$.
In this work, we assume heteroskedastic simulation noise, where the variance of the noise, $r(\siminput)$, depends on the input $\siminput$. However, our results are equally applicable in the homoskedastic setting, where the noise level is constant across inputs and the problem is generally simpler. In such cases, the surrogate model can learn the constant variance, and the design may assign a uniform number of replicates at all input points.

A design input $\xb$ is common to both the simulation model and the field experiment. To study the physical system, field experiments are conducted at $d$ specific design inputs, where $d$ denotes the total number of field observations. These field data design inputs are denoted $\xf_1, \ldots, \xf_d \in \mathcal{X}$. The observed field data $\yb = \left(y(\xf_1), \ldots, y(\xf_d)\right)^\top$ are used to estimate the unknown calibration parameter $\tb = (\theta_1, \ldots, \theta_p)^\top \in \Theta$, which aligns the simulation model with the observed data $\yb$. Define the input vector $\zb_j^o = \left({\xf_j}^\top, \tb^\top\right)^\top$ as the combination of the field data design input and the calibration parameter. The relationship between the field data and the simulation model is formulated using the following statistical model
\begin{equation}
    y(\xf_j) = \eta(\zb_j^o) + \epsilon, \quad  \epsilon \sim \mathcal{N}\left(0, \sigma^2\right), \label{eq:statmodel}
\end{equation}
where $\epsilon$ represents the residual error. Our design approach operates under the assumption of a well-specified simulation model that, together with $\tb$, governs the expected field data. However, in many real-world applications, even a best-tuned simulation model may not fully represent the physical system. Incorporating a discrepancy term, as formalized in the Kennedy and O’Hagan (KOH, \cite{Ohagan2001}) calibration framework, allows for more flexible modeling of the gap between simulation outputs and field data. While the modular approach \citep{Bayarri2007, Bayarri2009} can be adopted as in the case of deterministic simulation models \citep{Surer2024}, the combined variability arising from noisy simulation outputs, model discrepancy, and observational noise can lead to more severe identifiability challenges for stochastic models \citep{Jenny2014, Tuo2015, Plumlee2017}. Extending the proposed acquisition function to jointly account for noisy simulation outputs and model discrepancy is an important direction for future work. 

In this work, we focus on Bayesian calibration, a form of calibration that quantifies uncertainty in both model parameters and predictions. This is achieved by combining prior knowledge---expressed through a known closed-form prior distribution $p\left(\tb\right)$ on the parameters---with field data information captured via a likelihood function $p\left(\yb\mid\tb\right)$ based on the simulation model. The resulting posterior $p(\tb\mid\yb)$ represents the updated probability of the parameters, reflecting how well they align with both the prior information and the observed data. According to Bayes' rule, the posterior density is expressed as follows
\begin{equation} 
\label{eq:posterior}
    p\left(\tb\mid\yb\right) = \frac{ p\left(\yb\mid\tb\right) p\left(\tb\right)}{\int\limits_{\Theta}  p\left(\yb\mid\tb^\prime\right) p\left(\tb^\prime\right) {\rm d \tb^\prime}} \propto \tilde{p}\left(\tb\mid\yb\right) =  p\left(\yb\mid\tb\right) p\left(\tb\right).
\end{equation}
Based on the model in \eqref{eq:statmodel}, the likelihood is given by
\begin{equation}
   \p\left(\yb\mid\tb\right) = (2 \pi)^{-d/2} |\Sigmav|^{-1/2} \exp\left(-\frac{1}{2} \left(\yb - \latent\left(\tb\right)\right)^\top \Sigmav^{-1} \left(\yb - \latent\left(\tb\right)\right)\right), \label{eq:truelike}
\end{equation}
where $\latent\left(\tb\right) = \left(\eta\left(\zb_1^o\right), \ldots, \eta\left(\zb_d^o\right)\right)^\top$ is the vector of expected model outputs at the field data design inputs, and $\Sigmav$ is a diagonal covariance matrix with diagonal entries equal to $\sigma^2$. \tcb{In this work, the intrinsic simulation variance $r(\cdot)$ and the field-data variance $\sigma^2$ are treated as distinct quantities, similar to the formulation in \cite{Yuan2013}. However, in some applications, portions of the variability observed in the field data may arise from the same stochastic mechanisms represented within the simulation model, in which case alternative formulations may be more appropriate.} In the proposed adaptive design procedure, the field-data covariance $\Sigmav$ is treated as a known input. \tcb{When $\Sigmav$ is unknown, one may instead specify or estimate it using replicated field observations, domain information, or an auxiliary statistical model, and then proceed conditionally on that estimate. The estimation of $\Sigmav$ itself is outside the scope of the present work and may require additional modeling assumptions, particularly in settings where replicated field observations are unavailable.}


Markov chain Monte Carlo (MCMC) methods are widely used in Bayesian calibration to sample from the posterior distribution \citep{Gilks1995}. 
Since the normalizing constant in \eqref{eq:posterior}, $\int\limits_{\Theta} p\left(\yb\mid\tb^\prime\right) p\left(\tb^\prime\right) {\rm d \tb^\prime}$, is independent of $\tb$ and often intractable, MCMC typically operates on the unnormalized posterior $\tilde{p}(\tb\mid\yb)$, which represents the posterior up to a constant. In this work, we adopt a similar perspective, focusing on learning the shape of the posterior without explicitly computing the normalizing constant. 
Following prior work \cite{Kandasamy2015}, \cite{Kandasamy2017}, \cite{Jarvenpa2019}, \cite{Jarvenpa2021}, \cite{Surer2023}, we treat the unnormalized posterior as the quantity of interest and quantify uncertainty in its estimation. Although one could instead define uncertainty over the normalized posterior, doing so would require discretizing the parameter space $\Theta$, which is computationally intensive. To remain tractable, our strategy targets regions with high unnormalized posterior values, which determine the overall shape of the posterior. For brevity, we refer to the unnormalized posterior simply as ``the posterior'' throughout the paper.

Overall, we summarize the key settings under which our sequential design is constructed:
(i) A design input is shared by both the simulation model and the field experiment, with field data observed at a finite set of such inputs.
(ii) The simulation model exhibits input-dependent noise, which is modeled using a hetGP surrogate (Section~\ref{sec:GP}; see also the discussion of homoscedastic noise above).
(iii) There exists a calibration parameter value that aligns the expected simulation model output with the field observations.
(iv) Replication \tcb{of simulation evaluations} at the same \tcb{input} locations is a modeling choice that helps distinguish signal from \tcb{intrinsic simulation} noise. To automatically balance exploration and replication, we further propose strategies (Section~\ref{sec:explorevsexploit}).
(v) New simulation runs are acquired sequentially, one at a time, as part of the adaptive data collection procedure (Section~\ref{sec:adaptive}).

\subsection{Adaptive design for efficient data collection}
\label{sec:adaptive}

Throughout this paper, we use subscript indices with $\siminput$ (e.g., $\siminput_i$) to denote input locations included in the proposed design. Let $\siminput_1, \ldots, \siminput_{n_t}$ represent the $n_t$ distinct input locations selected by stage $t$ of the design process. At each input $\siminput_i$, the simulation model is evaluated $a_i$ times, producing replicate outputs $\simout(\siminput_i)^\ell$ for $\ell = 1, \ldots, a_i$. The adaptive procedure begins with an initial design, denoted by $\mathcal{D}_1 = \{(\siminput_i, \simout(\siminput_i)^\ell) : \ell = 1, \ldots, a_i, i = 1, \ldots, n_1\}$, which includes $n_1$ unique input locations, each evaluated with $a_i$ replicates. The initial design can be generated using Latin hypercube sampling (LHS) \citep{Kleijnen2009}, allocating a fixed number of replicates to each point.
At each stage $t = 1, \ldots, T$, the simulation dataset $\mathcal{D}_t$ is used to train an emulator (Section~\ref{sec:GP}) and to guide input selection through the proposed acquisition function (Section~\ref{sec:acquisition}).
We derive two closed-form expressions to calculate this acquisition function: one for sampling from unexplored regions and one for allocating an additional replicate to previously evaluated input locations (Section~\ref{sec:ivar_criteria}). The proposed adaptive design balances exploration of the input space with exploitation of promising regions through an automated strategy to enable more efficient posterior learning (Section~\ref{sec:explorevsexploit}).
Once a new simulation is performed at the selected input, the dataset is updated to $\mathcal{D}_{t+1} = \{(\siminput_i, \simout(\siminput_i)^\ell): \ell = 1, \ldots, a_i, i = 1, \ldots, n_{t+1}\}$ to add the new data point. This adaptive process continues for the $T$ stages, iteratively refining the posterior inference by incorporating simulation data in the most informative regions for the calibration task. 

\subsection{Gaussian process regression with replication}
\label{sec:GP}

At stage~$t$, the simulation model has been evaluated at $n_t$ unique input locations, $\siminput_1, \ldots, \siminput_{n_t}$. The vector $\ab_t = (a_1, \ldots, a_{n_t})^\top$ records the number of replicates at each input. One way to model stochastic simulation outputs is to treat all individual evaluations $\simout(\siminput_i)^\ell$, for $\ell = 1, \ldots, a_i$ and $i = 1, \ldots, n_t$, as independent observations of a GP. This full-sample formulation places a GP prior on the mean function $\eta(\cdot)$ and directly models the entire set of $\sum\limits_{i=1}^{n_t} a_i$ simulation outputs. However, this approach can be computationally expensive, as the cost of GP inference scales cubically with the number of simulation data points.

To alleviate this, we adopt an aggregated formulation that models only the $n_t$ unique input locations, using the sample means and variances of replicates to inform both the mean function and the input-dependent noise. This approach significantly reduces computational complexity---from cubic in the total number of simulation outputs to cubic in the number of unique input locations---while preserving the predictive properties of the full model. Importantly, it has been shown that the predictive distributions under the full and aggregated formulations are equivalent under Gaussian assumptions \citep{Ankenman2009, Binois2018}.

At each location $\siminput_i$, we use the sample mean $\frac{\sum\limits_{\ell=1}^{a_i}\simout(\siminput_i)^\ell}{a_i}$ of the $a_i$ replicates as the observed response. 
The covariance structure between two input locations is defined by the kernel function $k_{t}(\cdot, \cdot) = \tau_{t} c_{t}(\cdot, \cdot)$, where $\tau_{t}$ is the scaling parameter and $c_{t}(\cdot, \cdot)$ represents the correlation function. Common choices for $c_{t}(\cdot, \cdot)$ include the Gaussian and Mat\'ern kernels \citep{Rasmussen2005, santner2018design}. 
In this study, we adopt the Gaussian kernel in a separable form, expressed as $c_{t}(\zb_i, \zb_{i^\prime}) = \prod\limits_{\iota=1}^{q+p} \exp\left(-\frac{(z_{i,\iota} - z_{i^\prime,\iota})^2}{2\rho_{t,\iota}}\right)$. This formulation allows for independent scaling across dimensions, governed by the lengthscales $\pmb{\rho}_{t} = (\rho_{t,1}, \ldots, \rho_{t,q+p})^\top$ in each dimension. The choice of kernel function does not affect the fundamental reasoning behind our proposed acquisition function, and our implementation supports different kernel options for users to choose from.

Let $\Kb_{t}$ be an $n_t \times n_t$ covariance matrix, where the $(i, i')$th entry is given by $k_{t}(\zb_i,\zb_{i'})$ for $1 \leq i, i' \leq n_t$. Under the GP prior, the vector of expected outputs, $\latent_{t} = \left(\eta(\siminput_1), \ldots, \eta(\siminput_{n_t})\right)^\top$, follows a multivariate normal (MVN) distribution with mean $\mathbf{0}$ and covariance matrix $\Kb_{t}$, i.e., $\latent_{t}\sim\mathcal{MVN}(\mathbf{0}, \Kb_{t})$. The cross-kernel evaluations between a new input $\zb$ and the $n_t$ previously evaluated inputs $\zb_{1}, \ldots, \zb_{n_t}$ are collected in the vector $\kb_{t}(\zb) = (k_{t}\left(\zb, \zb_1\right),$ $ \ldots, k_{t}\left(\zb, \zb_{n_t}\right))^\top$. The sample means of simulation outputs are stored in $\bar{\pmb{\simout}}_{t} = \left(\frac{\sum\limits_{\ell=1}^{a_1}\simout(\siminput_1)^\ell}{a_1}, \ldots, \frac{\sum\limits_{\ell=1}^{a_{n_t}}\simout(\siminput_{n_t})^\ell}{a_{n_t}}\right)^\top$. 
Using the conditional properties of the MVN distribution, the predictive equations at $\zb$ are characterized by the mean $m_{t}(\zb)$ and variance $\varsigma^2_{t}(\zb)$ such that
\begin{equation} 
    \begin{aligned} \label{eq:gp_prediction}
        & m_{t}(\zb) = \kb_{t}(\zb)^\top \Kb_{t}(\ab_t)^{-1} \bar{\pmb{\simout}}_{t} \quad {\rm and } \quad  \varsigma^2_{t}(\zb) = k_{t}(\zb, \zb) - \kb_{t}(\zb)^\top \Kb_{t}(\ab_t)^{-1} \kb_{t}(\zb), \text{ where } \\
        & \hspace{2cm}  \Kb_{t}(\ab_t) = \Kb_{t} + \Vb(\ab_t) \text{ and } \Vb(\ab_t) = {\rm diag}\left(\frac{r(\zb_1)}{a_1}, \ldots, \frac{r(\zb_{n_t})}{a_{n_t}}\right). 
    \end{aligned}
\end{equation}
We note that an emulator is constructed at each stage $t$ using the simulation dataset $\mathcal{D}_t$.
The kernel hyperparameters (the scaling parameter $\tau_{t}$ and the lengthscales $\pmb{\rho}_{t}$) are obtained by optimizing the likelihood at each stage using $\mathcal{D}_t$. 
Indexing the hyperparameters by $t$ (and hence the kernel function $k_{t}(\cdot, \cdot)$, the correlation function $c_{t}(\cdot, \cdot)$, and the covariance matrix $\Kb_{t}$) reflects their dependence on the dataset $\mathcal{D}_t$. 
While the equations above assume that the intrinsic noise variance function $r(\zb_i)$ is known, this is rarely the case in practice. To derive the two expressions of the acquisition function for exploration and replication in Section~\ref{sec:ivar_criteria}, we assume $r(\zb_i)$ is known; the inference procedure is discussed in Section~\ref{sec:inference}.
As a result, at each stage $t$, the total number of kernel hyperparameters does not increase with $t$ (e.g., a single scaling parameter $\tau_t$ and $q+p$ lengthscales in $\pmb{\rho}_t$, along with additional hyperparameters for the intrinsic noise variance function inferred following the procedure described in Section~\ref{sec:inference}).


\section{Acquisition Criterion: Integrated Variance}
\label{sec:acquisition}

Simulation evaluations are guided by an acquisition function, evaluated either for exploration of new inputs or for replication at existing input locations. At each stage $t$, the choice between exploration and replication is made to reduce overall uncertainty in the posterior estimate. The corresponding analytical expressions are detailed in Section~\ref{sec:ivar_criteria}.
To guide the acquisition process, we make use of both the expectation $\E[\tilde{p}(\tb\mid\yb)]$ and variance $\V[\tilde{p}(\tb\mid\yb)]$ of the unnormalized posterior density $\tilde{p}(\tb\mid\yb)$ at each parameter $\tb$. The expectation serves as a surrogate for the posterior density, while the variance quantifies the uncertainty in this estimate. \tcb{We build a hetGP emulator as in Section~\ref{sec:GP} to model the simulation output using the dataset $\mathcal{D}_t$. The quantities $\E[\tilde{p}(\tb\mid\yb)]$ and $\V[\tilde{p}(\tb\mid\yb)]$ are then obtained by propagating the predictive uncertainty of this emulator into the estimation of $\tilde{p}(\tb\mid\yb)$.} 
These two quantities form the basis of the acquisition function proposed in the following section.
The expressions are derived by extending Lemma~3.1 from \cite{Surer2024}. Specifically, we replace the deterministic simulation model output with the expected value of the output from the stochastic simulation model. The resulting expressions are:
    \begin{gather}
        \E\left[\tilde{p}\left(\tb\mid\yb\right)\right] = f_\mathcal{N}\left(\yb; \muv_t\left(\tb\right), \Sigmav + \Sv_t\left(\tb\right) \right) p\left(\tb\right), \text{ \rm and} \label{expectedpostfinal}\\
        \V\left[\tilde{p}\left(\tb\mid\yb\right)\right] = \left(\frac{1}{2^d\pi^{d/2}|\Sigmav|^{1/2}} f_\mathcal{N}\left(\yb; \muv_t\left(\tb\right), \frac{1}{2}\Sigmav + \Sv_t\left(\tb\right)\right) \right. \notag\\ \left.- \left(f_\mathcal{N}\left(\yb; \muv_t\left(\tb\right),\Sigmav + \Sv_t\left(\tb\right)\right)\right)^2\right)p\left(\tb\right)^2. \label{variancepostfinal}
    \end{gather}
Here, $\muv_t\left(\tb\right) = \left(m_{t}\left(\zb_1^o\right), \ldots, m_{t}\left(\zb_d^o\right)\right)^\top$ denotes the vector of predictive means at the observed field data inputs, where each $\zb_j^o = \left({\xf_j}^\top, \tb^\top\right)^\top$ combines the $j$th field input $\xf_j$ with the parameter value $\tb$. The predictive covariance matrix $\Sv_t\left(\tb\right)$ captures the emulator's uncertainty, where the $j$th diagonal entry is the variance $\varsigma^2_{t}(\zb_j^o)$, and the $(j, j')$th off-diagonal entry is the predictive covariance given by
${\rm cov}_t(\zb_j^o, \zb_{j^\prime}^o) = k_t(\zb_j^o, \zb_{j^\prime}^o) -  \kb_{t}(\zb_j^o)^\top \Kb_{t}(\ab_t)^{-1} \kb_{t}(\zb_{j^\prime}^o)$.
In the above expressions, $f_\mathcal{N}(\mathbf{a}; \mathbf{b}, \mathbf{C})$ denotes the probability density function of a multivariate normal distribution with mean $\mathbf{b}$ and covariance matrix $\mathbf{C}$, evaluated at the point $\mathbf{a}$.

\subsection{Expressions for Exploration and Replication}
\label{sec:ivar_criteria}

During stage $t$, the simulation dataset $\mathcal{D}_{t} = \{(\siminput_i, \simout(\siminput_i)^\ell): \ell = 1, \ldots, a_i, i = 1, \ldots, n_{t}\}$ is used to build an emulator, which is then used to define the acquisition function. Let $\zb^c$ denote a candidate input considered by the acquisition function. In the exploration case, the candidate corresponds to a previously unobserved input, denoted $\breve{\zb}$, such that $\breve{\zb} \notin \{\zb_1, \ldots, \zb_{n_t}\}$ and $\zb^c = \breve{\zb}$. In the replication case, the candidate input matches one of the existing inputs, i.e., $\zb^c = \zb_k$ for some $k \in \{1, \ldots, n_t\}$, where $\zb_k$ is the $k$th previously selected unique input.

We propose selecting an input that minimizes overall uncertainty in the posterior estimate to facilitate learning the posterior distribution. At each stage, we measure the value of evaluating the simulation model at a candidate input $\zb^c$ via the following integrated variance (IVAR) criterion 
\begin{align} \label{eq:IVAR_explore}
    \begin{split}
        {\rm IVAR}(\zb^c) &= \int\limits_{\tb \in \Theta} \mathbb{E}_{\zeta^c |\mathcal{D}_t}\left(\mathbb{V}_{\latent(\tb)|\mathcal{D}_t^c}\left[\tilde{p}(\tb\mid\yb) \right]\right) d\tb.
    \end{split}
\end{align}
The simulation dataset $\mathcal{D}_t^c$ includes the current simulation data $\mathcal{D}_t$ along with a candidate input and its unobserved output. For exploration (i.e., $\zb^c = \breve{\zb}$), this takes the form $\mathcal{D}_t^c = \mathcal{D}_t \cup {(\breve{\zb}, \breve{\simout})}$, where $\breve{\simout} \coloneqq \simout(\breve{\zb})$ denotes the unknown simulation output at a new input $\breve{\zb}$. For replication (i.e., $\zb^c = \zb_k$), the dataset is augmented as $\mathcal{D}_t^c = \mathcal{D}_t \cup {(\zb_k, \outputk)}$, where $\outputk \coloneqq \simout(\zb_k)$ is an additional unseen replicate at an existing input $\zb_k$. In both cases, the expectation is taken with respect to unknown output $\zeta^c$---that is, $\zeta^c = \breve{\simout}$ for exploration and $\zeta^c = \outputk$ for replication---which remains random given the simulation dataset $\mathcal{D}_t$. The IVAR criterion is designed to learn the overall shape of the posterior density of model parameters by minimizing the aggregated variance of the posterior over the parameter space. Regions of negligible posterior density contribute very little to the total variance, whereas regions near the calibration region of interest exhibit higher posterior variance. The acquisition selects inputs that most effectively reduce this total uncertainty, favoring those near the high posterior-variance regions. This guides the acquisition toward the calibration region while avoiding areas where the posterior is negligible, even if the emulator’s predictive variance is large there. Overall, this strategy ensures that the posterior is closely approximated and predictive uncertainty is reduced where it matters most.

We now present the computation of ${\rm IVAR}(\zb^c)$ for the two cases $\zb^c = \breve{\zb}$ and $\zb^c = \zb_k$.
Exploration is crucial for directing efforts toward the calibration region of interest by carefully selecting unique inputs $\zb_1, \ldots, \zb_{n_t}$. 
Without exploration, computational resources would be wasted by replicating model evaluations outside the region of interest. 
The next result is used to compute ${\rm IVAR}(\zb^c)$ for the case $\zb^c = \breve{\zb}$, with the derivation provided in Appendix~\ref{proof:lemma3.3}.
\begin{lemma}\label{lemma:IVAR_explore}
Suppose $\zb^c = \breve{\zb}$. The surrogate model is assumed to be a GP emulator with input-dependent noise variance, and its predictive mean and variance are given by the SK equations in \eqref{eq:gp_prediction}. Let $\breve{\PHI}_t(\tb)$ be the $d \times d$ matrix with $j$th diagonal element $\frac{\text{cov}_{t}(\zf_j, \breve{\zb})^2}{\varsigma^2_{t}(\breve{\zb}) + r(\breve{\zb})}$ and $(j,j^\prime)$th element $\frac{\text{cov}_{t}(\zf_j, \breve{\zb})\text{cov}_{t}(\zf_{j^\prime}, \breve{\zb})}{\varsigma^2_{t}(\breve{\zb}) + r(\breve{\zb})}$.
${\rm IVAR}(\breve{\zb})$ is computed via
    \begin{gather} 
        \int\limits_{\tb \in \Theta} p(\tb)^2\left(\frac{f_\mathcal{N}\left(\yb; \, \muv_t(\tb), \, \frac{1}{2}\Sigmav + \Sv_t(\tb)\right)}{2^d \pi^{d/2} |\Sigmav|^{1/2}} -  \frac{f_\mathcal{N}\left(\yb; \, \muv_t(\tb), \, \frac{1}{2}\left(\Sigmav + \Sv_t(\tb) + \breve{\PHI}_{t}(\tb)\right)\right)}{2^d \pi^{d/2} \left|\Sigmav + \Sv_t(\tb) - \breve{\PHI}_{t}(\tb)\right|^{1/2}}\right)d\tb. \label{eq:IVAR_explore_derive}
    \end{gather}
\end{lemma}

While exploration expands coverage of the calibration region of interest to capture overall model behavior, replication improves statistical efficiency by enhancing signal-to-noise distinction and providing computational benefits. The next lemma establishes the derivation of ${\rm IVAR}(\zb^c)$ for the case $\zb^c = \zb_k$, with the full derivation provided in Appendix~\ref{proof:ivar_rep}.
\begin{lemma}\label{lemma:IVAR_exploit}
Suppose $\zb^c = \zb_k$. The surrogate model is assumed to be a GP emulator with input-dependent noise variance, and its predictive mean and variance are given by the SK equations in \eqref{eq:gp_prediction}. Let $\PHI^k_t(\tb)$ be a $d \times d$ matrix with the $(j,j^\prime)$th element $\kb_{t}^\top(\zf_j) \mathbf{B}_k\kb_{t}(\zf_{j^\prime})$, where $\mathbf{B}_k = \frac{\left(\Kb_{t}(\ab_t)^{-1}\right)_{.,k}\left(\Kb_{t}(\ab_t)^{-1}\right)_{k,.}}{a_k(a_k + 1)/r(\zb_k) - \left(\Kb_{t}(\ab_t)^{-1}\right)_{k,k}}$. Define a size $d$ vector $\muv^k_{t}(\tb)$ with the $j$th element $\mathbb{E}_{\outputk|\mathcal{D}_t}\left[m_{t+1}(\zf_j)\right]$, for $j = 1, \ldots, d$. Define a $d \times d$ covariance matrix $\GAMMA^k_t(\tb)$ with the $j$th diagonal element $\mathbb{V}_{\outputk|\mathcal{D}_t}\left[m_{t+1}(\zf_j)\right]$ and $(j,j^\prime)$th element $\mathbb{C}_{\outputk|\mathcal{D}_t}\left[m_{t+1}(\zf_j), m_{t+1}(\zf_{j^\prime})\right]$.
${\rm IVAR}(\zb_k)$ is computed via
    \begin{gather} 
     \int\limits_{\tb \in \Theta} p(\tb)^2 \left(\frac{f_\mathcal{N}\left(\yb; \, \muv^k_{t}(\tb), \, \frac{1}{2}\Sigmav + \Sv_t(\tb) - \PHI^k_t(\tb) + \GAMMA^k_t(\tb) \right)}{2^{d}\pi^{d/2}|\Sigmav|^{1/2}} \right. \notag\\ \left.- \frac{f_\mathcal{N}\left(\yb; \, \muv^k_{t}(\tb), \, \frac{1}{2}\left(\Sigmav + \Sv_t(\tb) - \PHI^k_t(\tb)\right) + \GAMMA^k_t(\tb)\right)}{2^{d}\pi^{d/2}|\Sigmav + \Sv_t(\tb) - \PHI^k_t(\tb)|^{1/2}}\right)d\tb. \label{eq:IVAR_exploit_derive}
    \end{gather}
\end{lemma}

Computing the IVAR criterion for both exploration, as given in \eqref{eq:IVAR_explore_derive}, and exploitation, as shown in \eqref{eq:IVAR_exploit_derive}, involves integration over the parameter space $\Theta$. One approach to approximating high-dimensional integrals is to sum over uniformly distributed reference grids, as demonstrated in \cite{Surer2023} and \cite{Surer2024}. However, as the dimensionality $p$ increases, the reference grid size must also increase, resulting in a higher computational cost for acquiring each additional point. In this study, we use importance sampling (IS) to approximate the integral. Using the IS estimator, we approximate ${\rm IVAR}(\zb^c)$ via
\begin{align} \label{eq:approximate_IVAR}
    \begin{split}
        {\rm IVAR}(\zb^c) &= \int\limits_{\tb \in \Theta} p(\tb)^2 g_t\left(\tb, \zb^c\right)d\tb \approx \sum_{l=1}^s \omega^{l} p\left(\tb^{l}\right)^2 g_t\left(\tb^{l}, \zb^c\right),
    \end{split}
\end{align}
where $g_t(\tb, \zb^c)$ denotes the expression inside the large brackets in \eqref{eq:IVAR_explore_derive} when $\zb^c = \breve{\zb}$, and in \eqref{eq:IVAR_exploit_derive} when $\zb^c = \zb_k$. The importance weights $\omega^{l}$ are given by
\begin{align} \label{eq:importance_weight}
    \begin{split}
        \omega^{l} = \frac{1}{p\left(\tb^{l}\right)^2 \, \VA{l}\left[\tilde{p}\left(\tb^{l} \mid \yb\right)\right]}\mathrel{\scalebox{1.5}{$/$}} 
        \sum_{l'=1}^s \frac{1}{p\left(\tb^{l'}\right)^2 \, \VA{l'}\left[\tilde{p}\left(\tb^{l'} \mid \yb\right)\right]},
    \end{split}
\end{align}
where $\tb^{l} \sim q(\cdot)$ for $l = 1, \ldots, s$. The importance distribution $q(\cdot)$ is proportional to the square of the prior multiplied by the variance of the posterior, that is, $q(\cdot) \propto p\left(\tb^{l}\right)^2 \VA{l}\left[\tilde{p}\left(\tb^{l} \mid \yb\right)\right]$. Since a single evaluation is unlikely to alter the variance surface significantly, and the expected variance is expected to remain similar to the current variance surface, this choice of importance distribution is reasonable. If the prior is proper and bounded (i.e., $p(\tb) < \infty$ and $\int\limits_{\tb \in \Theta} p(\tb) d\tb = 1$), then $q(\cdot)$ defines a valid probability density function, aside from the normalization constant. Since the normalizing constant of $q(\cdot)$ is not available, we normalize the weights as shown in \eqref{eq:importance_weight}. To approximate the high-dimensional integrals, we obtain $s$ samples from $q(\cdot)$ using MCMC \tcb{with burn-in and thinning}.

We identify the input that minimizes the IVAR criterion for exploration and replication separately by solving $\min\limits_{\zb^c \in \mathcal{L}_t} {\rm IVAR}(\zb^c)$, where $\mathcal{L}_t$ is a discrete set of candidate inputs. In the replication case, $\mathcal{L}_t$ consists of previously evaluated simulation inputs, $\siminput_1, \ldots, \siminput_{n_t}$, and the candidate replicate is the one that minimizes the IVAR criterion. In the exploration case, we generate a discrete candidate set using LHS to mitigate the challenges of direct optimization and select the point that minimizes IVAR as the candidate new point. In the following section, we discuss strategies for choosing between the candidate replicate and the candidate new point (i.e., between exploitation and exploration) at a given stage $t$.

\begin{figure}[ht]
    \centering
    \begin{subfigure}{1\textwidth}
        \includegraphics[width=1\textwidth]{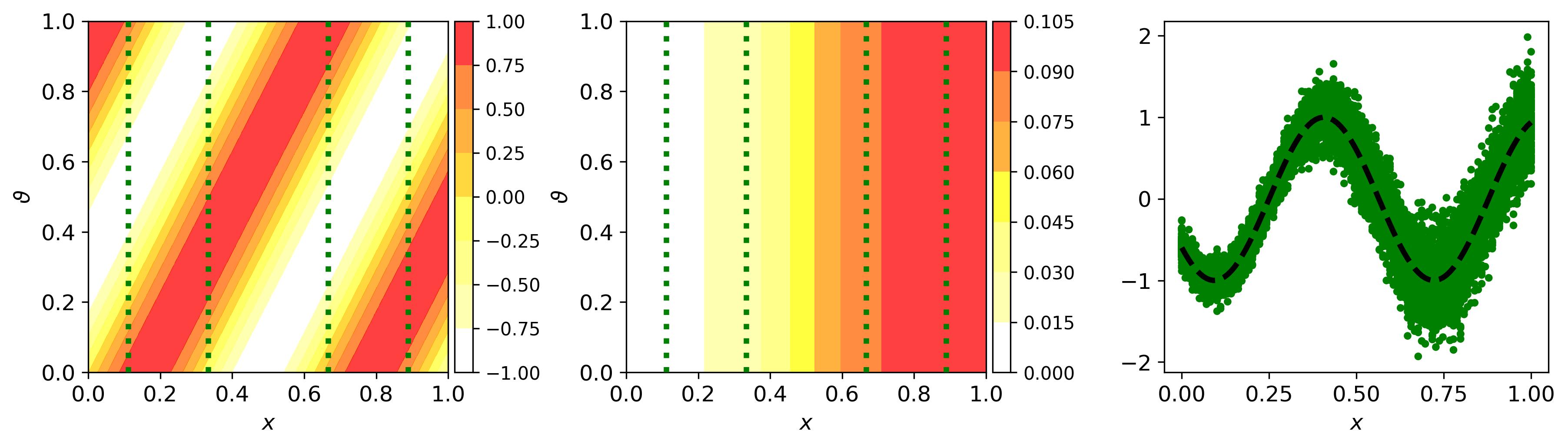}
    \end{subfigure}
    \caption{Illustrative example of a simulation model $\eta(\zb) = \sin(10x - 5\vartheta)$, where $x \in [0,1]$, $\vartheta \in [0, 1]$, and $\zb = (x, \vartheta)^\top$. The left panel visualizes $\eta(\zb)$ over the input space $(x, \vartheta)$, while the middle panel depicts the variance $r(\zb)$ of the intrinsic uncertainty. The field data is generated as $y(x^o) = \eta(\zb^o) + \epsilon$, where $\epsilon \sim {\rm N}(0, 0.1)$ and $\zb^o = (x^o, \theta = 0.5)^\top$. Green dotted lines indicate the locations of the field data design inputs. In the right panel, circle markers show \tcb{100} replicates of the simulation model across values of $x$ at $\theta=0.5$, and the dashed line indicates the corresponding expected value.} 
    \label{fig:figure1}
\end{figure}

\begin{figure}[t]
    \centering
    \begin{subfigure}{0.75\textwidth}
        \includegraphics[width=1\textwidth]{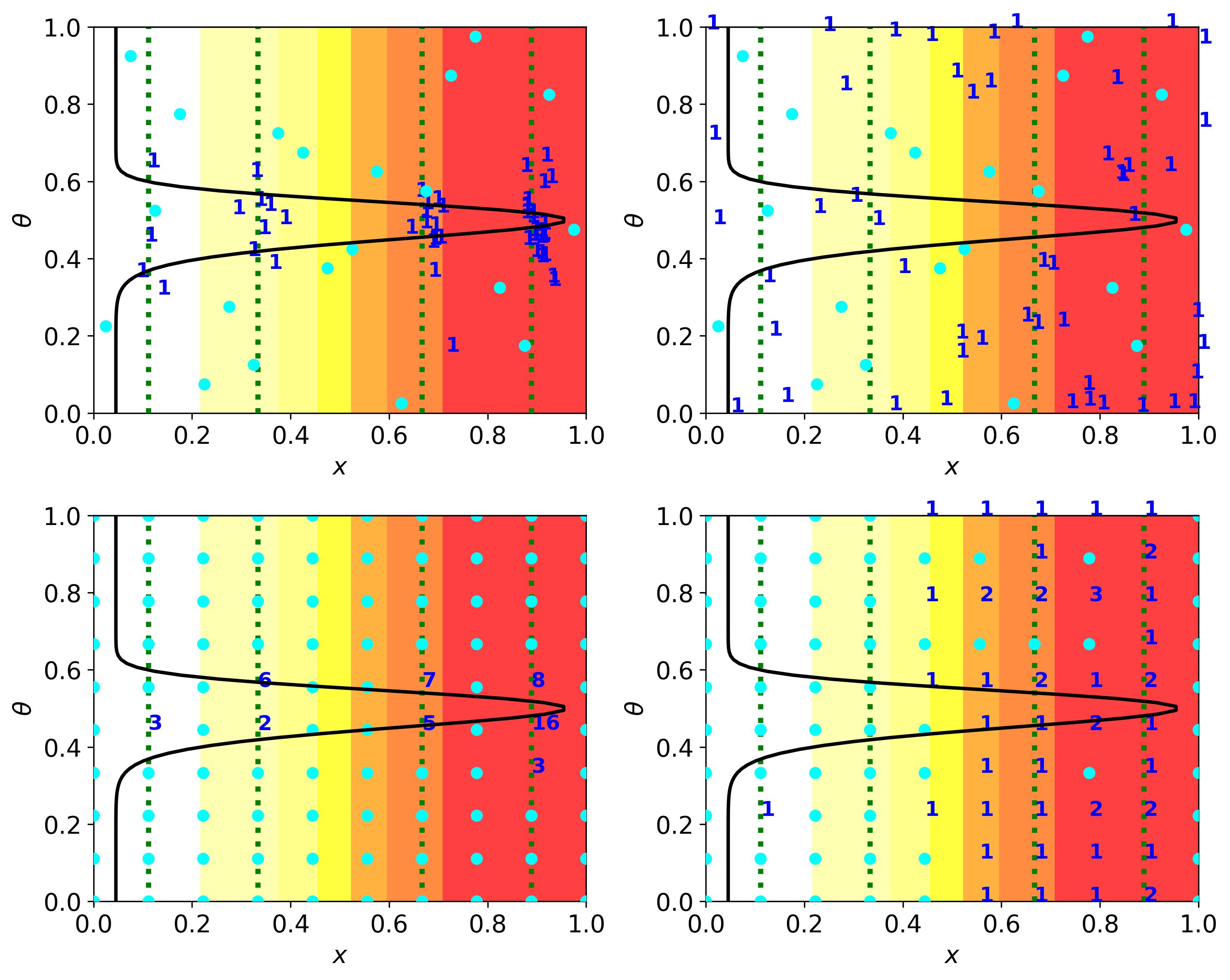}
    \end{subfigure}
    \caption{Illustration of points acquired using the IVAR (left) and IMSE (right) exploration (top) and replication (bottom) cases. Cyan markers indicate the initial sample points, while blue markers represent the acquired points, with numbers denoting replicates. 
    \tcb{The green dotted lines indicate the field data design inputs, $x_1^o=0.11$, $x_2^o=0.33$, $x_3^o=0.67$, and $x_4^o=0.89$, with the observations generated as $y(x_j^o) = \eta(\zb_j^o) + \epsilon$, where $\epsilon \sim {\rm N}(0, 0.1)$ and $\zb_j^o = (x_j^o, \theta = 0.5)^\top$. The black curve depicts the true likelihood function corresponding to the observed field-data vector $(-0.98, 0.74, -0.85, 0.11)^\top$, assuming full knowledge of the relationship between $x$, $\theta$, and $\mathbb{E}[y(x)]$.}}
    \label{fig:figure2}
\end{figure}

Before combining these decisions within the proposed procedure, we illustrate ${\rm IVAR}(\zb^c)$ separately for the two cases: $\zb^c = \breve{\zb}$ (exploration) and $\zb^c = \zb_k$ (replication), using the example in Figure~\ref{fig:figure1}. We similarly illustrate the corresponding exploration and replication forms for the IMSE acquisition function, enabling a direct comparison between the two criteria. In this example, field data is collected at four design inputs. The input region on the right exhibits higher noise than the region on the left. We begin with the exploration case. The procedure is initialized with 20 LHS samples from the input domain $[0,1]\times[0,1]$, each with five replicates ($n_1 = 20$ and $a_1 = \cdots = a_{n_1} = 5$). It then sequentially acquires 50 new points ($T=50$) using ${\rm IVAR}(\breve{\zb})$ to explore the input space. At each stage, a candidate set $\mathcal{L}_t$ of 1,000 LHS samples is generated from the same input domain, which is then evaluated to decide which point to acquire next. The top row of Figure~\ref{fig:figure2} visualizes the points acquired by IVAR and IMSE for exploration.
The black curve is added to Figure~\ref{fig:figure2} to show the true likelihood as a function of $\theta$, highlighting the parameter region of interest. Under the assumption of a uniform prior, the posterior is determined entirely by this likelihood.
IMSE distributes points across the entire input space to construct an accurate approximation of the simulation model. In contrast, IVAR concentrates sampling around the field data locations in the design input space, while in the parameter space, it favors regions with higher likelihood to refine posterior learning. Since uncertainty is greater on the right side, more points are acquired from that region. 

We next consider the replication case using an initial $10 \times 10$ grid in the input space, with each point having two replicates. We then sequentially acquire 50 additional replicates using ${\rm IVAR}(\zb_k)$ to perform replication. The bottom row of Figure~\ref{fig:figure2} visualizes the points acquired by IVAR and IMSE for replication. IMSE favors locations with higher intrinsic uncertainty to ensure broad coverage of uncertain regions. However, because it does not account for the calibration objective, it spreads samples across the entire parameter space and neglects the inputs on the left, as these do not correspond to regions of high uncertainty. In contrast, IVAR prioritizes the parameter region of interest and the field data design points and allocates more replicates to areas of higher uncertainty to improve statistical efficiency. We note that the top and bottom rows in Figure~\ref{fig:figure2} use different initial designs purely to illustrate the behavior of the exploration and replication cases. Specifically, the exploration panel uses a smaller, space-filling initial design to allow the acquisition function to cover the region of interest, while the replication panel employs a more structured grid to clearly illustrate how additional replicates are allocated once the input space is already well-explored. In practice, the proposed methods do not require different initial designs. However, exploration without replication may fail to adequately reduce uncertainty in key regions, while replication without exploration risks overlooking important areas. This motivates our sequential framework, which adaptively balances exploration and replication at each stage, as described in the following section.

\subsection{Choosing Between Exploration and Replication}
\label{sec:explorevsexploit}

Section~\ref{sec:ivar_criteria} introduces the IVAR acquisition function and derives two analytical expressions for evaluating it in the exploration and replication cases.
This section discusses how to effectively balance exploration and replication within the adaptive design framework. At stage $t$, the decision must be made whether to explore a new input or replicate an existing one. A straightforward approach is to independently identify the inputs that minimize ${\rm IVAR}(\breve{\zb})$ for exploration and ${\rm IVAR}(\zb_k)$ for replication, then select the one with the lowest IVAR value. However, this myopic selection strategy fails to account for the impact of future choices on current selections. To address this, we adopt a lookahead procedure inspired by \cite{Binois2019}, which incorporates future decision-making considerations.

The lookahead procedure determines the decision at the current stage by anticipating the choices made over the next $h$ stages, for which there are multiple possible strategies. In contrast to the myopic strategy discussed above, which tends to prioritize exploring new points over replication, our approach is inherently biased toward replication.
With a lookahead horizon of $h$, the procedure considers $h+1$ possible paths, each evaluating a sequence of decisions spanning the current stage and the subsequent $h$ stages.
In each path, the initial decision is either to replicate or explore. 
If the initial decision is exploration, the remaining $h$ stages are set to replication. Conversely, if the initial step is replication, the procedure determines at which future stage exploration should be introduced. Since there are $h$ distinct ways to introduce exploration, each of these $h$ remaining paths selects a new point at a different future stage.
Within a path, the decision on which point to select is guided by either ${\rm IVAR}(\breve{\zb})$ or ${\rm IVAR}(\zb_k)$. 
The integrated variance is computed at the end of the horizon $h$ for each path, and the path with the lowest uncertainty is chosen to determine the current decision. 
If the selected path begins with a new point, exploration is performed at stage $t$. Otherwise, replication is chosen. 
Since replication is the initial decision for $h$ out of the $h+1$ paths, a longer planning horizon inherently promotes replication. For $h=0$, the procedure chooses either exploration or replication, depending on which option most reduces uncertainty. For $h=-1$, it defaults to exploration.

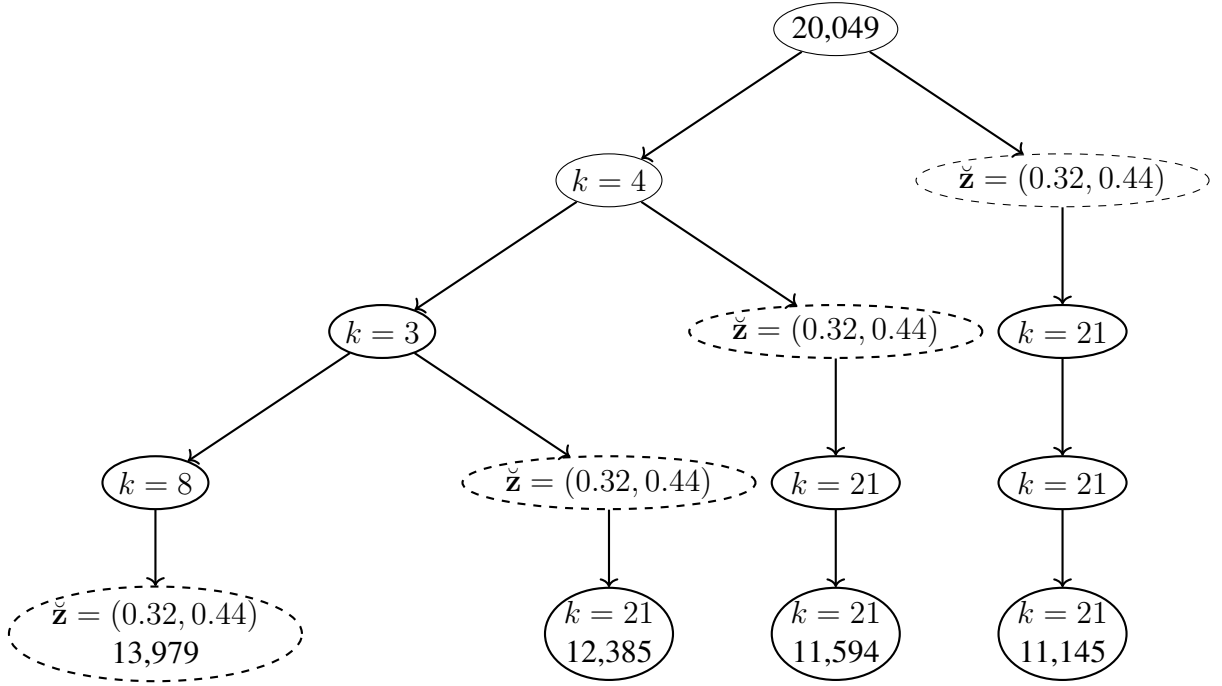
\begin{figure}[h!]
\centering
\begin{tikzpicture}[
  level distance=2cm,   
  sibling distance=6cm,    
  solid node/.style={
    ellipse, 
    draw, 
    align=center, 
    inner sep=0pt,   
    minimum size=0.7cm, 
  },
  dashed node/.style={
    ellipse, 
    draw, 
    dashed, 
    align=center, 
    inner sep=0pt,  
    minimum size=0.7cm, 
  },
  edge from parent/.style={draw, ->, thick}
]
  \node[solid node] {20,049}
    child {node[solid node] {$k=4$}
      child {node[solid node] {$k=3$}
        child {node[solid node] {$k=8$}
            child {node[dashed node] {$\breve{\zb}=(0.32, 0.44)$\\13,979}}}
        child {node[dashed node] {$\breve{\zb}=(0.32, 0.44)$}
            child {node[solid node] {$k=21$\\12,385}}}}
      child {node[dashed node] {$\breve{\zb}=(0.32, 0.44)$}
        child {node[solid node] {$k=21$}
            child {node[solid node] {$k=21$\\11,594}}}}}
    child {node[dashed node] {$\breve{\zb} = (0.32, 0.44)$}
      child {node[solid node] {$k=21$}
        child {node[solid node] {$k=21$}
            child {node[solid node] {$k=21$\\11,145}}}}};
\end{tikzpicture}
\caption{Illustration of the lookahead procedure with $h=3$. Each node represents a selected input, and the dashed node indicates the step where exploration occurs. For terminating nodes, the second line displays the associated IVAR value.}
\label{fig:figure3}
\end{figure}
Figure~\ref{fig:figure3} illustrates the lookahead procedure using the example from Figure~\ref{fig:figure1}. We begin with the same initial design used in the top-row example of Figure~\ref{fig:figure2}, consisting of $n_1=20$ unique inputs. The total posterior uncertainty at the root node of Figure~\ref{fig:figure3} is 20,049. 
At stage $t=1$, with a lookahead horizon of $h=3$, we evaluate four possible paths, each beginning with either replication or exploration.
The first path follows an explore–replicate–replicate–replicate decision sequence. If a new point is selected in the first step, the remaining three stages choose replication at that same location, indexed as $k=21$. 
Since this point is near the field data design input and within the parameter region of interest, initially exploring this region and then exploiting it improves posterior learning. 
The remaining three paths follow these decision sequences: (1) replicate-explore-replicate-replicate, (2) replicate-replicate-explore-replicate, and (3) replicate-replicate-replicate-explore. Among the four paths, the one that begins with exploration results in the lowest uncertainty (11,145). Consequently, the procedure selects exploration at stage $t=1$.

The horizon $h$ can either remain fixed throughout the procedure or be selected based on available computational resources, as larger values of $h$ incur higher costs. Alternatively, $h$ can be adjusted dynamically to align with evolving design goals. In this work, we introduce two approaches for setting $h$ at each stage, indexing it by stage $t$ to enable adaptive adjustments. The first approach aims to regulate the ratio of unique inputs to total collected outputs. Let $\rho$ denote the target ratio. We refer to this approach as the target-based scheme, where the horizon $h_t$ is updated dynamically based on the observed ratio. Specifically,
\begin{align} \label{eq:target_approach}
    \begin{split}
h_{t+1} =
\begin{cases}
    h_{t} + 1, & \text{if } \frac{n_t}{\sum\limits_{i=1}^{n_t} a_i} > \rho \text{ and a new point is chosen,} \\
    \max\{h_t - 1, -1\}, & \text{if } \frac{n_t}{\sum\limits_{i=1}^{n_t} a_i} < \rho \text{ and a replicate is chosen,} \\
    h_{t}, & \text{otherwise}.
\end{cases}
    \end{split}
\end{align}
This scheme ensures that when the ratio exceeds the target 
$\rho$, the horizon increases to encourage more replication, whereas if the ratio falls below $\rho$, the horizon decreases, promoting more exploration.

The second approach adjusts the horizon $h$ based on a replication criterion that minimizes the total posterior variance (TOTVAR) as a function of the replicates $\ab_t$, where
\begin{align} \notag 
    \begin{split}
        {\rm TOTVAR}(\ab_t) &= \int\limits_{\tb \in \Theta} \V\left[\tilde{p}\left(\tb|\yb\right)\right] d\tb = \int\limits_{\tb \in \Theta}\left(\frac{f\left(\tb\right)}{2^d\pi^{d/2}|\Sigmav|^{1/2}}  - g\left(\tb\right)^2\right)p\left(\tb\right)^2 d\tb, \\ 
         \text{with} \quad f(\tb) &= f_\mathcal{N}\left(\yb; \muv_{t}\left(\tb\right), \frac{1}{2}\Sigmav + \Sv_{t}(\tb)\right) \quad \text{and} \quad g(\tb) = f_\mathcal{N}\left(\yb; \muv_{t}\left(\tb\right),\Sigmav + \Sv_{t}(\tb)\right).
    \end{split}
\end{align}
The allocation criterion proposed by \cite{Surer2025+} was originally developed for the batch-sequential setting to determine an optimal allocation of replicates for high-dimensional stochastic simulation outputs. In this study, we adapt it to our setting, as further detailed in Appendix~\ref{sec:allocation_adaptive}. According to this criterion, the optimal budget allocation $a_i^*$ should be 
\begin{gather}
       a_i^*  \propto \sqrt{C_{i}(\ab_t)}, \quad i = 1, \ldots, n_t, \quad  \text{where} \label{eq:ivar_allocation} \\
        \begin{aligned}
       C_{i}(\ab_t) & = \int\limits_{\tb \in \Theta}  \Bigg( \frac{f(\tb)}{2^d\pi^{d/2}|\Sigmav|^{1/2}}  \left(-\frac{1}{2} {\rm Tr}\left(\dot{\mathbf{N}}(\tb)^{-1} \Mb_i(\tb) \right) + \frac{1}{2}\mathbf{h}(\tb)^\top \dot{\mathbf{N}}(\tb)^{-1} \Mb_i(\tb) \dot{\mathbf{N}}(\tb)^{-1} \mathbf{h}(\tb)\right) \\ &\quad- 2 g(\tb)^2 \left(-\frac{1}{2} {\rm Tr}\left(\mathbf{N}(\tb)^{-1}\Mb_i(\tb) \right) + \frac{1}{2}\mathbf{h}(\tb)^\top \mathbf{N}(\tb)^{-1} \Mb_i(\tb) \mathbf{N}(\tb)^{-1} \mathbf{h}(\tb)\right) \Bigg) p(\tb)^2 d\tb, 
    \end{aligned} \nonumber
\end{gather}
with $\dot{\mathbf{N}}(\tb) = 0.5\Sigmav + \Sv_{t}(\tb)$, $\mathbf{N}(\tb) = \Sigmav + \Sv_{t}(\tb)$, $\mathbf{h}(\tb) = \yb-\muv_{t}(\tb)$, and $\Mb_i(\tb)$ is a $d \times d$ matrix with $(j, j^\prime)$th element $-r(\zb_i)\kb^\top_{t}(\zb_j^o) \Kb_{t}(\ab_t)^{-1}  \mathbf{J}^{(i,i)} \Kb_{t}(\ab_t)^{-1} \kb_{t}(\zb_{j^\prime}^o)$. $\mathbf{J}^{(i,i)}$ is a $n_t \times n_t$ matrix with one in the $(i,i)$th entry and zeros elsewhere. At stage $t$, we determine the proposed allocations following \eqref{eq:ivar_allocation}, assuming that $\sum\limits_{i=1}^{n_t} a_i$ simulation evaluations are feasible. We then compare the optimal allocation $a_i^*$ with the actual $a_i$. If the number of replicates exceeds $a_i^*$, the horizon is reduced to promote exploration; otherwise, it is increased to encourage replication, following the rule below:
\begin{align} \label{eq:adaptive_approach}
    \begin{split}
        h_{t+1} \sim {\rm Uniform}\{a_1^\prime, \ldots, a_{n_t}^\prime\} \quad {\rm with} \quad a_i^\prime = \max\{0, a_i^* - a_i\}.
    \end{split}
\end{align}
We refer to this approach as the adaptive scheme.

\subsection{Inference}
\label{sec:inference}

Our derivations thus far assume that the intrinsic variance $r(\cdot)$ is known. In practice, however, $r(\cdot)$ is unknown and must be estimated for any input. Following \cite{Binois2018}, we employ hetGP to model the log variances, $\log \mathbf{\Lambda}_{t}$, as the mean output of a GP on latent variables, $\mathbf{\Delta}_{t} = (\delta_1, \ldots, \delta_{n_t})^\top$. The inverse covariance matrix is given by $\Kb_{t}(\ab_t)^{-1} = \tau_{t}^{-1}(\Cb_{t} + \mathbf{A}_t^{-1}\mathbf{\Lambda}_{t})^{-1}$, where the $(i, i')$th entry of $\Cb_{t}$ is defined as $c_{t}(\zb_i, \zb_{i'})$ for $1 \leq i, i' \leq n_t$, and $\mathbf{A}_t = {\rm diag}(a_1, \ldots, a_{n_t})$. The GP prior on the latent variables, $\mathbf{\Delta}_{t} \sim \mathcal{MVN}\left(\mathbf{0}, \tau_{t}^g\left(\Cb_{t}^g + g_{t} \mathbf{A}_t^{-1}\right)\right)$, implies that $\log \mathbf{\Lambda}_{t} = \Cb_{t}^g \left(\Cb_{t}^g + g_{t} \mathbf{A}_t^{-1}\right)^{-1} \mathbf{\Delta}_{t}$. Here, $\Cb_{t}^g$ is the correlation matrix based on a Gaussian kernel with lengthscale parameters $\pmb{\rho}^g_{t}$, while $g_{t} > 0$ and $\tau_{t}^g$ denote the nugget and scaling parameters, respectively. Inference requires estimating $\pmb{\rho}_{t}$, $\pmb{\rho}^g_{t}$, $\mathbf{\Delta}_{t}$, and $g_{t}$, with the scaling parameters $\tau_{t}$ and $\tau_{t}^g$ determined via plug-in maximum likelihood estimation. The joint log-likelihood is then optimized using its gradient with respect to these parameters, as described in \cite{Binois2018}. For our experiments, we implement our design procedure using the hetGPy \citep{OGara2025} Python package to construct emulators.

\section{Experiments}
\label{sec:experiment}
Section~\ref{sec:synthetic} assesses the performance on synthetic simulation models. Section~\ref{sec:epidemic} focuses on an application with an epidemiological simulation model. The proposed sequential approach is implemented in the Python package Parallel Uncertainty Quantification (PUQ), accessible at \hb{{https://github.com/parallelUQ/}}, along with example scripts for reproducibility.


\subsection{Benchmark with Synthetic Simulation Models}
\label{sec:synthetic}

We assess the performance of the proposed sequential method using the IVAR criterion across a range of synthetic simulation models. As a baseline, we consider the sequential design of \citet{Binois2019}, which relies on the IMSE criterion. In calibration settings, another common objective is to learn the field observations throughout the design space. 
Motivated by this goal, we include an additional benchmark, denoted $\text{IMSE}^y$, which selects inputs to minimize the aggregated predictive uncertainty when predicting field observations.
Appendix~\ref{sec:experiment_details} provides the definitions of IMSE and $\text{IMSE}^y$, along with the corresponding exploration and replication cases.

The first example, adapted from \citet{Ranjan2011}, features a two-dimensional design input $\xb = \left(x_1, x_2\right)^\top \in [0, 1]^2$ ($q = 2$) and a one-dimensional calibration parameter $\vartheta \in [0, 1]$ ($p = 1$). 
\tcb{The second example is based on the Park function \citep{park1991,synthlinks}, with a two-dimensional design input $\xb = \left(x_1, x_2\right)^\top \in [0, 1]^2$ and a two-dimensional calibration parameter $\tg = \left(\vartheta_1, \vartheta_2\right)^\top \in [0, 1]^2$ ($q = 2$, $p = 2$).} Full details on the data generation mechanisms are provided in Appendix~\ref{sec:experiment_details}. 
In addition, we compare performance using modified versions of three widely studied functions—unimodal, bimodal, and Branin—chosen for their varying number of modes in the parameter space: one, two, and three, respectively. These additional results are described in Appendix~\ref{sec:additional_experiments}.
\tcb{They serve as controlled benchmarks for evaluating the proposed strategy under posterior distributions with distinct geometries, including unimodal, multimodal, and non-identifiable structures.}
To assess performance, we impose distinct intrinsic variance structures for each example. 
In all examples, we assume a uniform prior over the input space. \tcb{Appendix~\ref{sec:homGP} provides further discussion of replication and its importance for heteroscedastic noise estimation.}

To initialize the sequential procedure, we draw an initial sample of size $n_1 = 30$ using LHS over the joint input space $[\mathcal{X}, \Theta]$, with each point replicated 5 times. 
The procedure then continues for a total of $T = 200$ acquisitions, starting from the 150 initial points, resulting in a total of 350 inputs. 
For exploration, at each stage, a candidate set $\mathcal{L}_t$ of size $|\mathcal{L}_t| = 300$ is constructed from two equal halves. The first half (150 points) is independently sampled via LHS over the joint space $[\mathcal{X}, \Theta]$ to encourage exploration of the entire input space. The second half (150 points) is designed to emphasize the exploitation of the observed field locations. 
To do this, 150 parameter values are generated via LHS in the parameter space $\Theta$, and 150 field data design inputs are created depending on the example.
\tcb{For the first two examples}, the 150 points are approximately evenly distributed across the four observed field locations in a randomized fashion (each location receives either 37 or 38 replicates, with the specific assignment varying across stages) and then randomly shuffled, yielding a $150 \times 2$ matrix in $\mathcal{X}$-space. \tcb{For the subsequent examples in Appendix~\ref{sec:additional_experiments}}, all 150 points are allocated to the single observed field location ($150 \times 1$ matrix). Each field design input is then paired with one of the sampled parameter values, resulting in 150 candidate inputs.
For replication, the candidate set is $\mathcal{L}_t = \{\zb_1, \ldots, \zb_{n_t}\}$.

We assess performance across 30 independent experimental replications. In each replication, all methods (IVAR, IMSE, and $\text{IMSE}^y$) share the same initial design to ensure a fair comparison. For each example, the observed field data are generated according to \eqref{eq:statmodel} using a data-generating calibration parameter $\tb = \tb^\ast$. This parameter is held constant across all 30 experimental replications within a given example to ensure that differences in performance are attributable to the acquisition strategy rather than variability in the data-generating mechanism (see Appendix~\ref{sec:experiment_details} for the specific values of $\tb^\ast$). To evaluate accuracy, we compute the mean absolute difference (MAD) between the estimated posterior $\hat{p}_t(\thetav\mid\yb)$ and true posterior $\tilde{p}(\thetav\mid\yb)$ over a reference set of calibration parameters. Specifically, we draw a reference set of calibration parameters, $\Theta_{\rm ref}$, of size 1000 from the true posterior distribution. At each stage $t$, we calculate the MAD as ${\rm MAD}_t = \frac{1}{|\Theta_{\rm ref}|}\sum\limits_{\thetav \in \Theta_{\rm ref}} \left|\tilde{p}(\thetav\mid\yb) - \hat{p}_t(\thetav\mid\yb)\right|$. Here, $\hat{p}_t(\thetav\mid\yb)$ is the unnormalized posterior estimate at stage $t$ obtained using the emulator built from the data $\mathcal{D}_t$, which is computed via $\E\left[\tilde{p}\left(\tb\mid\yb\right)\right]$ in \eqref{expectedpostfinal}. \tcb{We note that while posterior inference is invariant to multiplicative normalizing constants, the proposed criterion is designed to assess emulator fidelity to the target unnormalized surface, since inaccuracies in the emulator estimates of the simulation mean and intrinsic noise variance propagate directly to this quantity.}
To complement MAD, we compute a Kullback–Leibler (KL)-type measure ${\rm KL}_t = -\frac{1}{|\Theta_{\rm ref}|}\sum\limits_{\thetav \in \Theta_{\rm ref}} \log \hat{p}_t(\thetav\mid\yb)$ which approximates the KL divergence between the true posterior and its estimate up to an additive constant, and quantifies how well the estimated posterior assigns probability mass to points drawn from the true posterior. The KL results are reported in Appendix~\ref{sec:experiment_details}.
To characterize acquisition behavior, Table~\ref{tab:coverage_width_12} summarizes the width of the acquired inputs and the coverage of the data-generating calibration parameters for each acquisition strategy and example. For each experimental replicate, we compute the upper and lower $\alpha=0.10$ quantiles of each acquired input and report the average width across replicates. For the calibration parameters, we also assess whether the interval includes the data-generating value and report both marginal and joint coverage, averaged over 30 replicates.
\begin{table}[ht]
\centering
\scriptsize
\setlength{\tabcolsep}{4pt}
\caption{\tcb{Coverage and width of empirical $(1-\alpha)$ quantile intervals for the acquired calibration parameters and design inputs.}}
\label{tab:coverage_width_12}
\begin{tabular}{ll||cc|cc|cc||cc|cc|cc}
\toprule & & \multicolumn{6}{c||}{1st example ($q=2,p=1$)} & \multicolumn{6}{c}{2nd example ($q=2,p=2$)} \\
\cmidrule(r){3-8}
\cmidrule(l){9-14}
& &
\multicolumn{2}{c|}{IMSE} & \multicolumn{2}{c|}{IMSE$^y$} & \multicolumn{2}{c||}{IVAR} & \multicolumn{2}{c|}{IMSE} & \multicolumn{2}{c|}{IMSE$^y$} & \multicolumn{2}{c}{IVAR}
\\

\cmidrule(r){3-4}
\cmidrule(r){5-6}
\cmidrule(r){7-8}
\cmidrule(r){9-10}
\cmidrule(r){11-12}
\cmidrule(l){13-14}

$h$ & Parameter & Cov. & Width & Cov. & Width & Cov. & Width & Cov. & Width & Cov. & Width & Cov. & Width \\ \midrule
adapt  & $\vartheta_1$ & 1.00 & 0.99 & 1.00 & 0.42 & 1.00 & 0.13 & 1.00 & 0.99 & 1.00 & 0.37 & 1.00 & 0.89 \\
target & $\vartheta_1$ & 1.00 & 0.98 & 1.00 & 0.36 & 1.00 & 0.14 & 1.00 & 0.97 & 1.00 & 0.39 & 1.00 & 0.88 \\
adapt  & $\vartheta_2$ & \multicolumn{6}{c||}{--} & 1.00 & 0.98 & 1.00 & 0.33 & 1.00 & 0.51 \\
target & $\vartheta_2$ & \multicolumn{6}{c||}{--} & 1.00 & 0.97 & 1.00 & 0.35 & 1.00 & 0.53 \\
adapt  & Joint & \multicolumn{6}{c||}{--} & 1.00 & -- & 1.00 & -- & 1.00 & -- \\
target & Joint & \multicolumn{6}{c||}{--} & 1.00 & -- & 1.00 & -- & 1.00 & -- \\
\midrule
$h$ & Design input & \multicolumn{6}{c||}{Width} & \multicolumn{6}{c}{Width} \\
\midrule
adapt & $x_1$ & \multicolumn{2}{c|}{0.99} & \multicolumn{2}{c|}{0.99} & \multicolumn{2}{c||}{0.68} & \multicolumn{2}{c|}{0.98} & \multicolumn{2}{c|}{0.98} & \multicolumn{2}{c}{0.76}\\
target & $x_1$ & \multicolumn{2}{c|}{0.98} & \multicolumn{2}{c|}{0.98} & \multicolumn{2}{c||}{0.68} & \multicolumn{2}{c|}{0.97} & \multicolumn{2}{c|}{0.97} & \multicolumn{2}{c}{0.77} \\
adapt & $x_2$ & \multicolumn{2}{c|}{0.99} & \multicolumn{2}{c|}{0.98} & \multicolumn{2}{c||}{0.68} & \multicolumn{2}{c|}{0.99} & \multicolumn{2}{c|}{0.98} & \multicolumn{2}{c}{0.78} \\
target & $x_2$ & \multicolumn{2}{c|}{0.98} & \multicolumn{2}{c|}{0.98} & \multicolumn{2}{c||}{0.68} & \multicolumn{2}{c|}{0.97} & \multicolumn{2}{c|}{0.97} & \multicolumn{2}{c}{0.78}\\
\bottomrule
\end{tabular}
\end{table}

To approximate the integrals in \eqref{eq:approximate_IVAR}, we apply importance sampling using $s = 100$ samples from the parameter space. Specifically, we generate samples from the importance distribution using the emcee Python package \citep{emcee}. We discard the first 100 iterations as burn-in and thin the chains by retaining every 20th sample to reduce autocorrelation. We use a target-based scheme with a ratio $\rho = 0.2$ (i.e., an average of 5 replicates per unique input) and an adaptive scheme to determine the horizon $h_t$, as described in Section~\ref{sec:explorevsexploit}.

\begin{figure}[!ht]
\centering
    \includegraphics[width=0.8\textwidth]{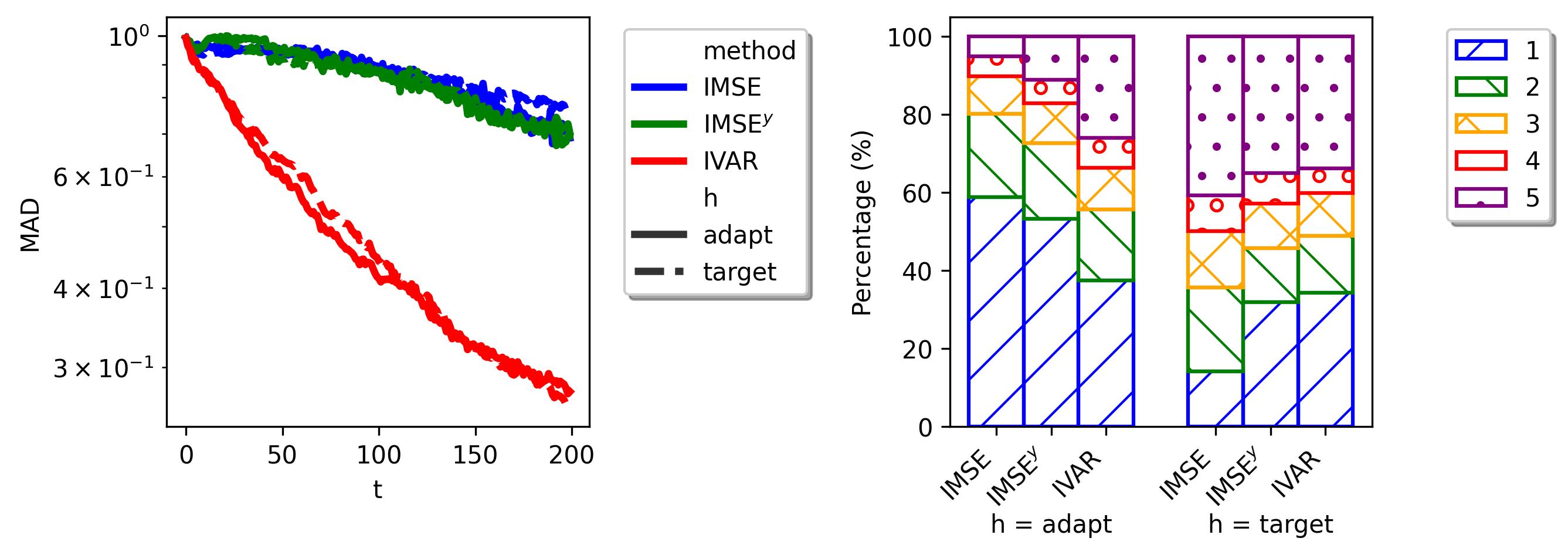}
    \caption{Comparison of acquisition functions for the first example.}
    \label{fig:figure4}
\end{figure}
\begin{figure}[!ht]
\centering
    \includegraphics[width=0.9\textwidth]{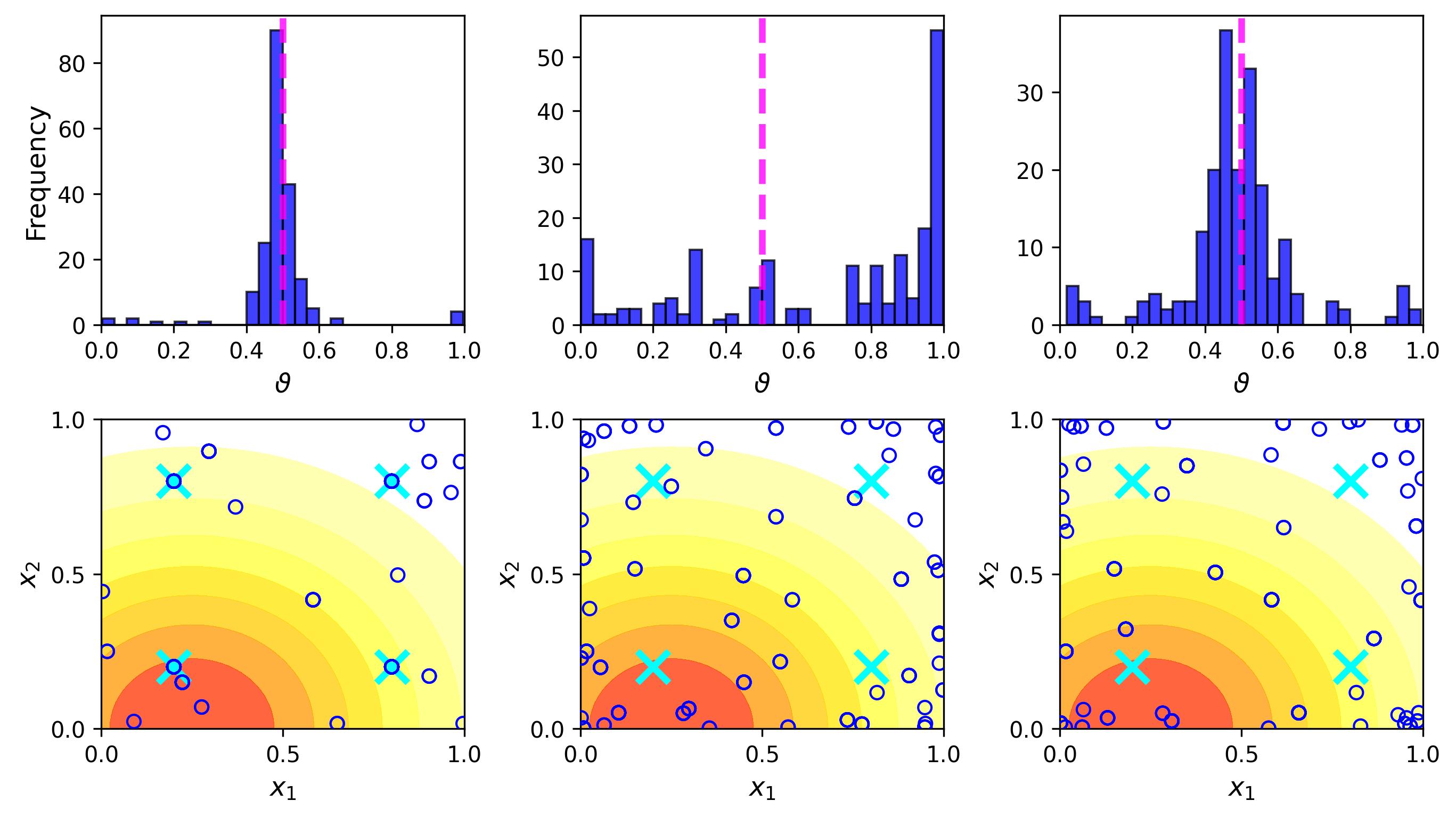}
    \caption{Illustration of acquired parameters (top row) and design inputs (bottom row) selected using IVAR (left), IMSE (middle), and $\text{IMSE}^y$ (right) for the first example. The magenta dashed line (top) indicates the data-generating parameter $\theta^\ast = 0.50$, while the background color (bottom) illustrates the intrinsic variance $r(\cdot)$ across the design input space at $\theta^\ast = 0.50$. Cyan cross markers (bottom) indicate the design input locations at which the field data are observed.}
    \label{fig:figure5}
\end{figure}
\begin{figure}[!ht]
\centering
    \includegraphics[width=1\textwidth]{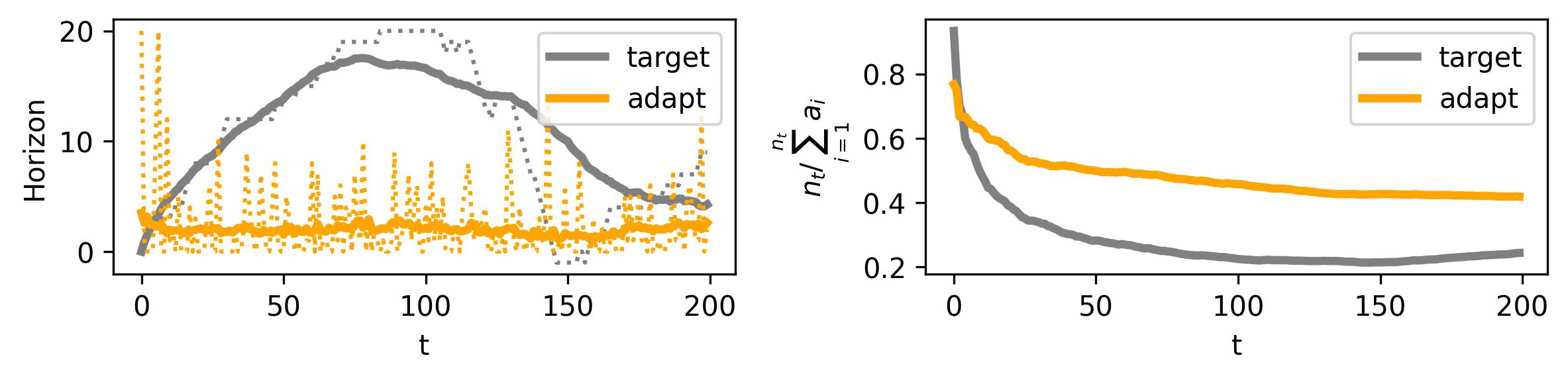}
    \caption{The left panel shows the evolution of the horizon over stages, where thin dotted lines represent an individual experimental replicate and thick lines indicate the median across 30 replicates for both the adaptive and target-based schemes. The right panel shows the ratio of unique input locations to the total number of acquired inputs, averaged over 30 replicates.}
    \label{fig:figure6}
\end{figure}

\begin{figure}[!ht]
\centering
    \includegraphics[width=0.8\textwidth]{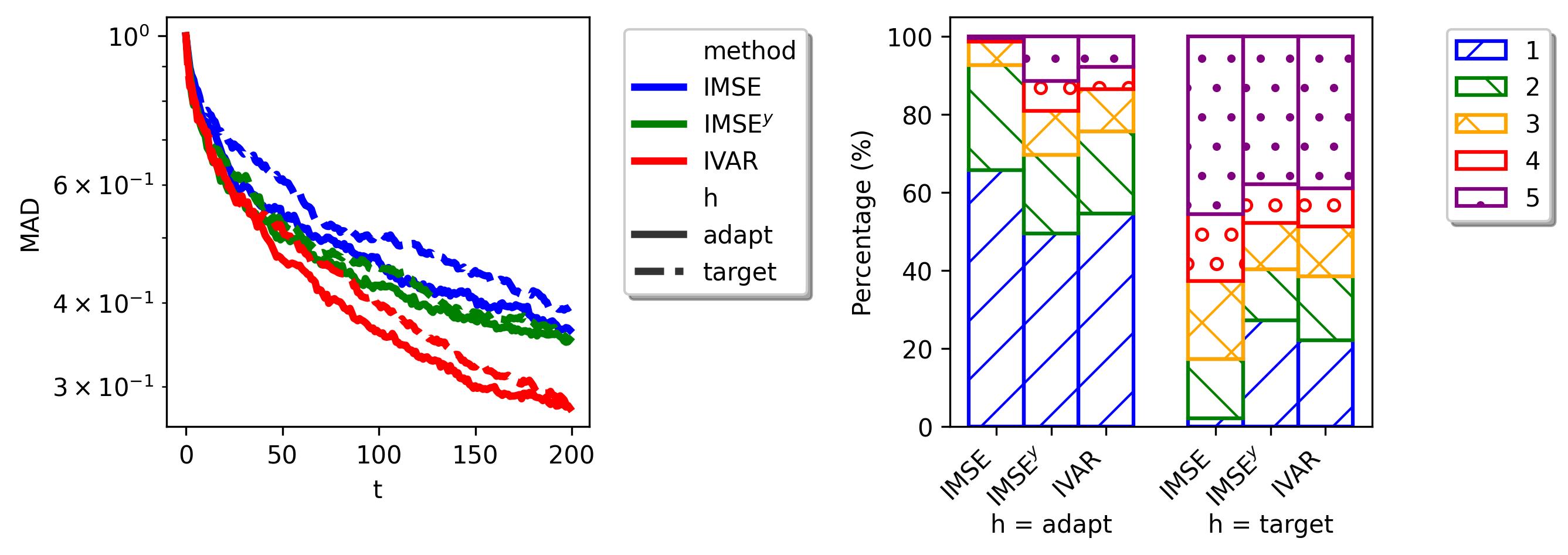}
    \caption{\tcb{Comparison of acquisition functions for the second example.}}
    \label{fig:figure7_R2}
\end{figure}

\begin{figure}[!ht]
\centering
    \includegraphics[width=0.9\textwidth]{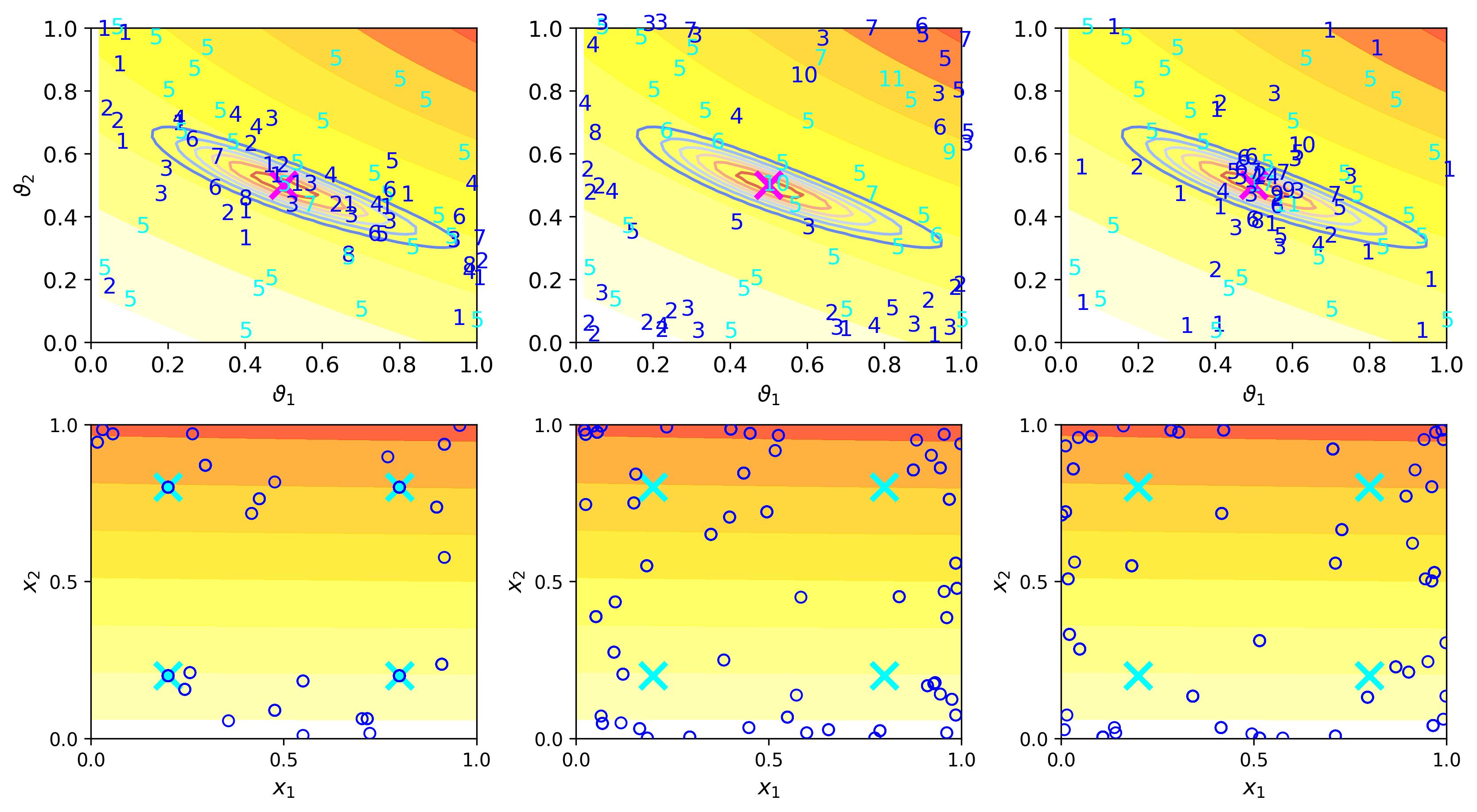}
    \caption{\tcb{Illustration of acquired parameters (top row) and design inputs (bottom row) selected using IVAR (left), IMSE (middle), and $\text{IMSE}^y$ (right) for the second example. Cyan markers denote the initial sample points, and blue markers denote the acquired points.  The magenta cross (top) indicates the data-generating parameter $\tb^\ast = (0.50, 0.50)^\top$, while the cyan cross (bottom) indicates the design input locations at which the field data are observed. Contours represent the true posterior density for reference. In the top row, the background color represents the aggregated intrinsic variance over the field data design inputs. In the bottom row, the background color represents the intrinsic variance $r(\cdot)$ across the design input space at $\tb^\ast$.}}
    \label{fig:figure8_R2}
\end{figure}

The left panels of Figures~\ref{fig:figure4} and \ref{fig:figure7_R2} show the MAD values, averaged over 30 experimental replicates, for different acquisition functions in the first and second examples, respectively. The right panels show the proportion of input locations that were repeated 1, 2, 3, 4, or at least 5 times over $T=200$ stages, also averaged over 30 replicates. Figures~\ref{fig:figure5} and \ref{fig:figure8_R2} illustrate the acquired calibration parameters and design inputs from a single replicate of the first and second examples, respectively, under the IVAR, IMSE, and $\text{IMSE}^y$ criteria with the target-based scheme. For the first example, the left panel of Figure~\ref{fig:figure6} shows how the horizon evolves across stages under the target-based and adaptive schemes using the IVAR criterion, while the right panel shows the ratio of unique inputs to the total design size.

Overall, the IVAR acquisition function consistently achieves superior posterior inference compared to IMSE and $\text{IMSE}^y$. The IMSE criterion, designed to construct globally accurate emulators, prioritizes coverage of the entire input space. Consequently, it often selects points near the domain boundaries and oversamples regions with high predictive uncertainty. However, this emphasis on emulator fidelity comes at the cost of insufficient coverage in the calibration region of interest. This behavior is further reflected in Table~\ref{tab:coverage_width_12}, which shows that IMSE produces the widest spread of acquired inputs, consistent with its global emphasis.
The $\text{IMSE}^y$ criterion selects inputs near the maximum likelihood estimate of the calibration parameters while simultaneously attempting to cover the entire design input space to accurately learn the field observations. As a result, it produces a wide spread across the design input space, but a narrow width among the acquired parameters, as shown in Table~\ref{tab:coverage_width_12}.
However, this localized focus around the point estimate can be limiting in examples with multimodal posteriors—such as the bimodal and Branin cases in Appendix~\ref{sec:additional_experiments}—where $\text{IMSE}^y$ often identifies a single mode and then predominantly exploits that region. As a result, other high-posterior regions remain underexplored.
In contrast, IVAR strategically targets regions around the field data design inputs and promotes a more balanced and focused exploration of the parameter space. By sampling not only in areas of peak posterior density but also in regions with moderate and lower density, IVAR enhances learning of the posterior distribution more effectively than strategies that either overexplore or focus too narrowly. In summary, the objective of IVAR is not point identification of a single true data-generating parameter, but rather to efficiently learn the shape of the posterior distribution that reflects uncertainty given the available field data. When field observations are weakly informative or when the simulation model structure or field configuration induces symmetry, the resulting posterior may be multimodal or non-identifiable—as in the unimodal, bimodal, and Branin examples. IVAR is designed to adaptively allocate simulation effort to these posterior-supported regions as much as the field data allow, rather than forcing a unique estimate when identifiability is not warranted. In applications where domain knowledge constrains the parameter space, such information can be incorporated through the prior distribution, which in turn steers IVAR away from scientifically implausible regions.

The target-based scheme tends to replicate input locations more frequently than the adaptive scheme. As shown in Figure~\ref{fig:figure6}, its median horizon gradually increases until around stage $t=100$. This pattern suggests that the horizon increases to encourage more replication to satisfy the target ratio. In contrast, the adaptive scheme—owing to its stochastic nature—exhibits more abrupt, stage-to-stage fluctuations in the horizon. Although its median horizon remains relatively stable and below 5 over 200 stages, individual replicates occasionally exceed $h = 10$, reflecting variability in decision-making. The more structured progression of the target-based scheme enables it to achieve the desired target ratio earlier. Meanwhile, the adaptive scheme maintains a higher ratio of unique parameter values relative to the total design size, indicating greater exploratory behavior. 

\subsection{Application to an Epidemiological Simulation Model}
\label{sec:epidemic}

We demonstrate the proposed sequential procedure using a discrete compartmental model commonly employed in epidemiology to simulate the spread of infectious diseases. These models, particularly during the COVID-19 pandemic, have offered critical insights into transmission dynamics and the effects of interventions such as social distancing, mask usage, and vaccination \citep{Yang2020}. The Susceptible-Infected-Recovered (SIR) model serves as a foundational example. In this model, the infection rate controls the transition from the susceptible (S) to the infected (I) compartment, while the removal rate governs the shift from the infected (I) to recovered (R). Given specific values for these rates, the SIR model tracks the evolution of susceptible, infected, and recovered populations over time.

In the SIR model, the calibration parameter is two-dimensional, $\tg = (\vartheta_1, \vartheta_2)^\top$ with $p = 2$, where $\vartheta_1 \in [0.1, 0.3]$ represents the infection rate and $\vartheta_2 \in [0.05, 0.15]$ represents the removal rate. Each parameter is rescaled to the $[0, 1]$ interval and assigned a uniform prior.
The design input vector is defined as $\xb = (x_1, x_2)^\top  \in [0,1]^2$, where $x_1$ and $x_2$ are normalized design inputs corresponding to the initial numbers of susceptible ($S_0$) and infected individuals ($I_0$), respectively; thus, $\xb$ has dimension $q=2$. The design input $\xb$ is linearly mapped to integer-valued initial conditions of the stochastic SIR model, with $S_0 \in [200, 250] \cap \mathbb{N}$ and $I_0 \in [0,100] \cap \mathbb{N}$. The total population size ($M$) is fixed at $M=350$, and the remainder of the population is assigned to the recovered compartment (i.e., $R_0 = M - S_0 - I_0$). For a given simulation input $\zb = \left(\xb^\top, \tg^\top\right)^\top$, the model returns the average number of infected individuals over a simulation horizon of 75 time units. Data are generated according to the model in \eqref{eq:statmodel}, with the true calibration parameter set to $\tb = \tb^*$. We use $\tb^*$ as the midpoint of prior ranges. For field data design inputs, we use the normalized design locations $\xf_1 = (0.04, 0.98)^\top$, $\xf_2 = (0.24, 0.24)^\top$, $\xf_3 = (0.74, 0.74)^\top$, and $\xf_4 = (0.98, 0.04)^\top$. Since the expected simulation output $\eta(\cdot)$ is unknown, we perform 1000 independent replications of the model and use the sample mean as an estimate of the expected value to generate the observed field data.

\begin{figure}[!ht]
\centering
    \includegraphics[width=1\textwidth]{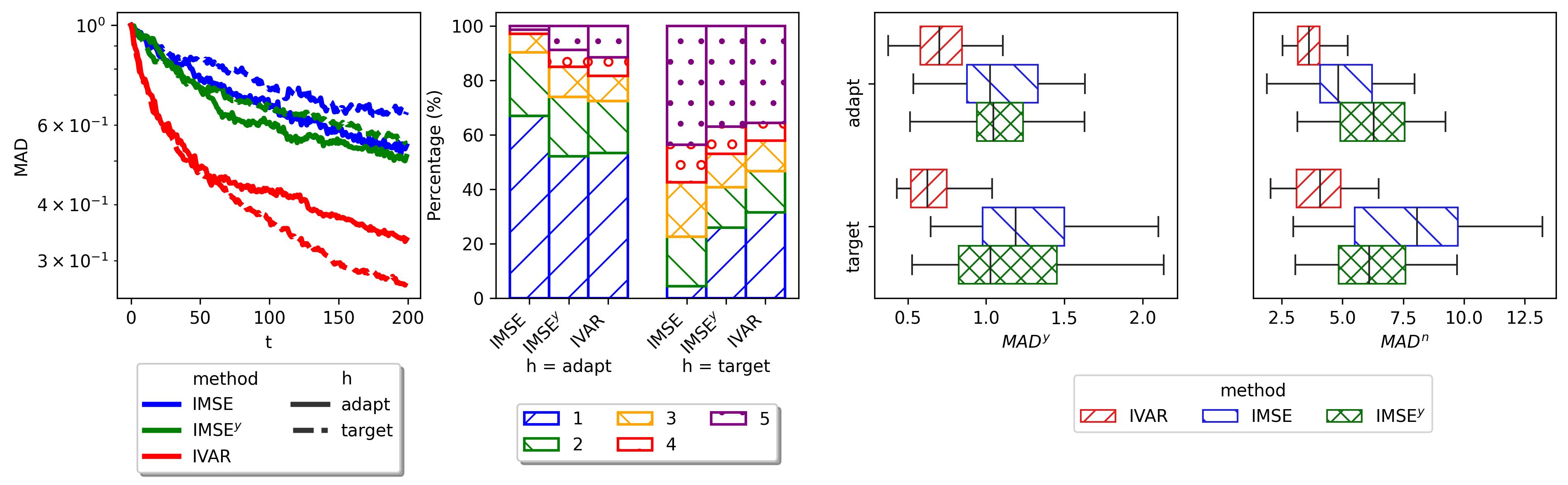}
    \caption{Comparison of different acquisition functions for the SIR model.}
    \label{fig:figure9}
\end{figure}
\begin{figure}[!ht]
\centering
    \includegraphics[width=1\textwidth]{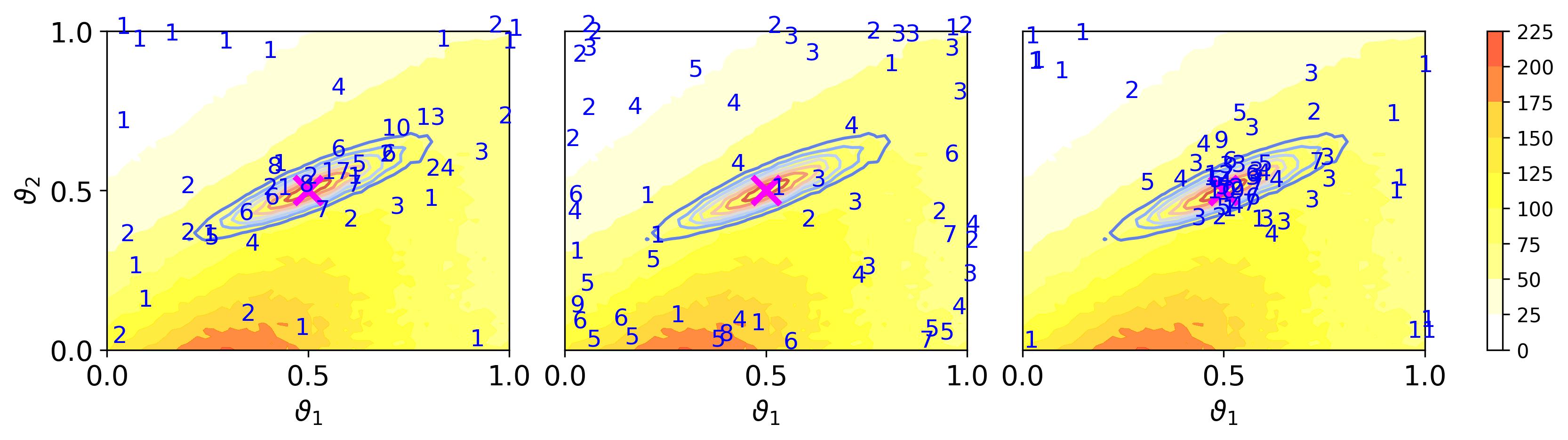}
    \caption{Illustration of the acquired parameters selected using IVAR (left), IMSE (center), and $\text{IMSE}^y$ (right). Contours represent the true posterior density $\tilde{p}(\thetav\mid\yb)$ as a reference, while the background color indicates the intrinsic noise aggregated over the field data design inputs. The magenta marker indicates the data-generating parameter $\tb^\ast$.}
    \label{fig:figure10}
\end{figure}
We generate an initial sample of size $n_1 = 40$ using LHS, with 5 replications per point, and acquire $T=200$ inputs. A candidate set $\mathcal{L}_t$ of size 400 is constructed following the same procedure as in Section~\ref{sec:synthetic}. We also set $s=100$ and $\rho = 0.2$, consistent with Section~\ref{sec:synthetic}. The IVAR, IMSE, and $\text{IMSE}^y$ criteria are each repeated 30 times, with average MAD values reported in Figure~\ref{fig:figure9}. In addition to evaluating overall posterior prediction accuracy with MAD, we also consider ${\rm MAD}^y$ and ${\rm MAD}^n$, defined as, ${\rm MAD}_t^y = \frac{1}{d|\Theta_{\rm ref}|} \sum\limits_{j=1}^d \sum\limits_{\thetav \in \Theta_{\rm ref}} \left|\eta(\zb_j^o) - m_t(\zb_j^o)\right|$ and ${\rm MAD}_t^n = \frac{1}{d|\Theta_{\rm ref}|}\sum\limits_{j=1}^d \sum\limits_{\thetav \in \Theta_{\rm ref}} \left|r(\zb_j^o) - \hat{r}_t(\zb_j^o)\right|$, where $\zb_j^o = \left({\xf_j}^\top, \tb^\top\right)^\top$, and $m_t(\cdot)$ and $\hat{r}_t(\cdot)$ represent the mean and noise estimates at stage $t$ predicted by the associated emulator. The two plots on the right-hand side of Figure~\ref{fig:figure9} display ${\rm MAD}^y$ and ${\rm MAD}^n$ at the final stage of the procedure, obtained across 30 replicates. Figure~\ref{fig:figure10} visualizes the parameters acquired using the target-based scheme for a single experimental replicate. Table~\ref{tab:coverage_width_sir} summarizes the width of the empirical $(1 - \alpha)$ quantile interval of the acquired inputs and reports the proportion of experimental replicates in which this interval contains the data-generating parameter.

By collecting inputs near the field data design points and within the parameter region of interest, IVAR more effectively predicts both the average simulation output and the associated noise in this critical region, outperforming alternative approaches. As a result, it enables more efficient and accurate posterior learning. While IVAR encourages sampling around the most likely regions of the calibration parameters, it also goes beyond this objective to learn the full shape of the posterior distribution. By avoiding over-localization around the parameter estimate, it exhibits a balanced behavior—reflected in the intermediate interval widths of the acquired parameters in Table~\ref{tab:coverage_width_sir} and the spread of points in Figure~\ref{fig:figure10}—that enables exploration beyond a single mode through sampling of moderate- and lower-density posterior regions. Consistent with the findings in Section~\ref{sec:synthetic}, the target-based horizon scheme increases the number of replications across all methods. 

\begin{table}[ht]
\centering
\scriptsize
\setlength{\tabcolsep}{4pt}
\caption{Width of empirical $(1 - \alpha)$ quantile intervals of acquired inputs and marginal and joint coverage of the data-generating calibration parameters for the SIR model.}
\label{tab:coverage_width_sir}
\begin{tabular}{l l || cc cc cc}
\cmidrule(lr){3-8}
 &  & \multicolumn{2}{c}{IMSE}
    & \multicolumn{2}{c}{$\text{IMSE}^y$}
    & \multicolumn{2}{c}{IVAR} \\
\cmidrule(lr){3-4}\cmidrule(lr){5-6}\cmidrule(lr){7-8}
Metric & Input
 & adapt & target
 & adapt & target
 & adapt & target \\
\midrule
Width    & $\vartheta_1$ & 0.99 & 0.97 & 0.51 & 0.50 & 0.84 & 0.81 \\
         & $\vartheta_2$ & 0.98 & 0.97 & 0.33 & 0.35 & 0.46 & 0.44 \\
         & $x_1$         & 0.98 & 0.97 & 0.98 & 0.98 & 0.95 & 0.94 \\
         & $x_2$         & 0.98 & 0.97 & 0.98 & 0.97 & 0.96 & 0.95 \\
\midrule         
Coverage & $\vartheta_1$ & 1.00 & 1.00 & 0.97 & 0.93 & 1.00 & 1.00 \\
         & $\vartheta_2$ & 1.00 & 1.00 & 1.00 & 1.00 & 1.00 & 1.00 \\
         & Joint         & 1.00 & 1.00 & 0.97 & 0.93 & 1.00 & 1.00 \\
\bottomrule
\end{tabular}
\end{table}

\section{Conclusion}
\label{sec:conc}

This work highlights the advantages of IVAR-based sequential design for efficient posterior learning in simulation-based calibration tasks. By strategically allocating samples near the field data inputs and within the parameter region of interest, IVAR improves the accuracy of predictions for both simulation means and noise levels in this critical region. As a result, it outperforms alternative approaches in posterior learning. Additionally, we propose two lookahead strategies to balance exploration and replication. The target-based strategy can be considered when greater replication is desirable, although the most suitable choice may vary depending on the application. 

This work opens several avenues for future research. 
One important direction is to examine the role of the initial design. While our framework is initialized with a space-filling design to ensure broad coverage of the input space and avoid imposing strong assumptions about where information is most valuable, the proposed acquisition strategy adaptively prioritizes field data design input locations. The preference for these locations could then be incorporated into the initial or subsequent designs. Assessing the impact of different initial designs on performance remains an important area for future investigation.
The proposed IVAR criterion relies on marginalizing the Gaussian likelihood over the Gaussian predictive distribution of the emulator. As a potential extension, one could consider deriving the uncertainty on the log-posterior scale and then transforming back via exponentiation. Developing the IVAR criterion under this log-posterior formulation would be an interesting direction for future work, allowing a systematic comparison between density-scale and log-density-scale acquisition strategies.
A further promising direction is to extend the proposed IVAR-based design criterion to settings with non-Gaussian noise. While the definition of the IVAR criterion in \eqref{eq:IVAR_explore} does not rely on Gaussian assumptions, the analytical derivations and closed-form expressions currently rely on Gaussianity of both simulation and field observations; adapting these derivations to alternative noise models is another line of future development.
An additional avenue is the integration of the acquisition function into joint designs for field and simulation experiments, where careful coordination is crucial to maximize the use of limited experimental resources. 
Extending the proposed IVAR-based acquisition strategy to select multiple inputs per stage (batch acquisitions) is another important direction, particularly for expensive simulation models, as it enables parallel evaluations and can reduce overall wall-clock time. Investigating how to construct effective batches that balance exploration and replication, while managing their computational cost, remains an important avenue for future research.


\subsection*{Acknowledgement}

\hb{We gratefully acknowledge the computing resources provided on Bebop, a high-performance computing cluster operated by the Laboratory Computing Resource Center at Argonne National Laboratory.}

\subsection*{Disclosure of Interest}

\hb{No potential competing interest was reported by the author.}

\subsection*{Data Availability Statement}

The data that support the findings of this study are openly available in the Python software package \hb{Parallel Uncertainty Quantification (PUQ) at {https://github.com/parallelUQ/}}.

\subsection*{Funding}

\hb{This work was supported by the National Science Foundation (NSF) under Grants OAC 2004601 and PHY 2402275.}

\bibliographystyle{apalike}
\bibliography{refs}

@phdthesis{park1991,
  author = {Park, Joon Seong},
  title  = {Tuning Complex Computer Codes to Data and Optimal Designs},
  school = {University of Illinois at Urbana-Champaign},
  year   = {1991}
}

@article{Yuan2013,
    title = {A sequential approach for stochastic computer model calibration and prediction},
    journal = {Reliability Engineering \& System Safety},
    volume = {111},
    pages = {273-286},
    year = {2013},
    issn = {0951-8320},
    doi = {https://doi.org/10.1016/j.ress.2012.11.004},
    url = {https://www.sciencedirect.com/science/article/pii/S0951832012002335},
author = {Jun Yuan and Szu Hui Ng}
}

@article{emcee,
       author = {{Foreman-Mackey}, Daniel and {Hogg}, David W. and {Lang}, Dustin and {Goodman}, Jonathan},
        title = "{emcee: The MCMC Hammer}",
      journal = {Publications of the Astronomical Society of the Pacific},
         year = 2013,
        month = mar,
       volume = {125},
       number = {925},
        pages = {306},
          doi = {10.1086/670067},
archivePrefix = {arXiv},
       eprint = {1202.3665},
 primaryClass = {astro-ph.IM},
       adsurl = {https://ui.adsabs.harvard.edu/abs/2013PASP..125..306F}
}

@article{Lartaud2025,
author = {Paul Lartaud and Philippe Humbert and Josselin Garnier},
title = {Solving {Bayesian} Inverse Problems Using {Gaussian} Process Regression with Goal-Oriented Active Learning},
journal = {Technometrics},
volume = {0},
number = {0},
pages = {1--14},
year = {2025},
publisher = {Taylor \& Francis},
doi = {10.1080/00401706.2025.2561745},
URL = {https://doi.org/10.1080/00401706.2025.2561745},
eprint = {https://doi.org/10.1080/00401706.2025.2561745}
}

@misc{synthlinks,
    author = {Sonia Surjanovic and Derek Bingham},
    title = {Virtual Library of Simulation Experiments: Test Functions and Data Sets},
    howpublished = "\url{https://www.sfu.ca/~ssurjano/index.html}",
    year = {2013}, 
    note = "[Online; accessed 06-June-2024]"
}

@article{OGara2025, 
doi = {10.21105/joss.07518}, 
url = {https://doi.org/10.21105/joss.07518}, 
year = {2025}, 
publisher = {The Open Journal}, 
volume = {10}, 
number = {106}, 
pages = {7518}, 
author = {David O'Gara and Mickaël Binois and Roman Garnett and Ross A. Hammond},
title = {{hetGPy}: Heteroskedastic {Gaussian} Process Modeling in {Python}}, journal = {Journal of Open Source Software} 
}

@ARTICLE{Lam2008,
  author={Chen Quin Lam and William I. Notz},
  journal={Statistics and Applications}, 
  title={Sequential Adaptive Designs in Computer Experiments for Response Surface Model Fit}, 
  year={2008},
  volume={6},
  number={1},
  pages={207--233}}

@article{Kleijnen2009,
title = {Kriging metamodeling in simulation: A review},
journal = {European Journal of Operational Research},
volume = {192},
number = {3},
pages = {707-716},
year = {2009},
issn = {0377-2217},
doi = {https://doi.org/10.1016/j.ejor.2007.10.013},
url = {https://www.sciencedirect.com/science/article/pii/S0377221707010090},
author = {Jack P.C. Kleijnen},
}

@ARTICLE{Ranjan2011,
  author = {Pritam Ranjan and Wilson Lu and Derek Bingham and Shane Reese and Brian J. Williams and Chuan-Chih Chou and Forrest Doss and Michael Grosskopf and James Paul Holloway},
  title = {Follow-Up Experimental Designs for Computer Models and Physical Processes},
  journal = {Journal of Statistical Theory and Practice},
  year = {2011},
  volume = {5},
  number = {1},
  pages = {119--136},
  doi = {10.1080/15598608.2011.10412055}
}

@article{Surer2025+,
    author = {\"Ozge S\"urer},
    title = {Batch Sequential Experimental Design for Calibration of Stochastic Simulation Models},
    journal = {Technometrics},
    volume = {68},
    number = {1},
    pages = {1--13},
    year = {2026},
    publisher = {ASA Website},
    doi = {10.1080/00401706.2025.2520860},
    URL = {https://doi.org/10.1080/00401706.2025.2520860},
    eprint = {https://doi.org/10.1080/00401706.2025.2520860}
}

@inproceedings{Kersting2007,
author = {Kersting, Kristian and Plagemann, Christian and Pfaff, Patrick and Burgard, Wolfram},
title = {Most likely heteroscedastic {Gaussian} process regression},
year = {2007},
isbn = {9781595937933},
publisher = {Association for Computing Machinery},
address = {New York, NY, USA},
booktitle = {Proceedings of the 24th International Conference on Machine Learning},
pages = {393–400},
numpages = {8},
location = {Corvalis, Oregon, USA},
series = {ICML '07}
}

@inproceedings{Gredilla2011,
author = {L\'{a}zaro-Gredilla, Miguel and Titsias, Michalis K.},
title = {Variational heteroscedastic {Gaussian} process regression},
year = {2011},
isbn = {9781450306195},
publisher = {Omnipress},
address = {Madison, WI, USA},
booktitle = {Proceedings of the 28th International Conference on International Conference on Machine Learning},
pages = {841–848},
numpages = {8},
location = {Bellevue, Washington, USA},
series = {ICML'11}
}

@inproceedings{ChenZhou2015,
    title = {Sequential experimental designs for stochastic kriging},
    author = {Xi Chen and Qiang Zhou},
    note = {Publisher Copyright: {\textcopyright} 2014 IEEE.; 2014 Winter Simulation Conference, WSC 2014 ; Conference date: 07-12-2014 Through 10-12-2014},
    year = {2015},
    month = jan,
    day = {23},
    language = {English (US)},
    series = {Proceedings - Winter Simulation Conference},
    publisher = {Institute of Electrical and Electronics Engineers Inc.},
    pages = {3821--3832},
    editor = {Andreas Tolk and Levent Yilmaz and Diallo, {Saikou Y.} and Ryzhov, {Ilya O.}},
    booktitle = {Proceedings of the 2014 Winter Simulation Conference, WSC 2014},
}

@article{ChenZhou2017,
    title = {{Sequential design strategies for mean response surface metamodeling via stochastic kriging with adaptive exploration and exploitation}},
    author = {Xi Chen and Qiang Zhou},
    journal = {European Journal of Operational Research},
    volume = {262},
    issue = {2},
    pages = {575--585},
    year = {2017}
}

@article{Sung2024,
  title={A review on computer model calibration},
  author={Sung, Chih-Li and Tuo, Rui},
  journal={WIREs Computational Statistics},
  volume={16},
  number={1},
  pages={e1645},
  year={2024}
}

@ARTICLE{Yang2020,
  author = {Yang, Haoxiang and S\"{u}rer, \"{O}zge and Duque, Daniel and Morton, David P. and Singh, Bismark and Fox, Spencer J. and Pasco, Remy and Pierce, Kelly and Rathouz, Paul and Valencia, Victoria and Du, Zhanwei and Pignone, Michael and Escott, Mark E. and Adler, Stephen I. and Johnston, S. Claiborne and Meyers, Lauren Ancel},
  title = {Design of {COVID-19} Staged Alert Systems to Ensure Healthcare Capacity with Minimal Closures},
  journal = {Nature Communications},
  year = {2021},
  volume = {12},
  number = {1},
  pages = {3767}
}

@article{Surer2023,
  title = {Sequential {Bayesian} experimental design for calibration of expensive simulation models},
  author = {\"Ozge S\"urer and Matthew Plumlee and Stefan M. Wild},
  journal = {Technometrics},
  volume = {66},
  number = {2},
  pages = {157--171},
  year = {2024}
}

@article{Baker2022,
  title = {Analyzing Stochastic Computer Models: A Review with Opportunities},
  author = {Evan Baker and Pierre Barbillon and Arindam Fadikar and Robert B. Gramacy and Radu Herbei and David Higdon and Jiangeng Huang and Leah R. Johnson and Pulong Ma and Anirban Mondal and Bianica Pires and Jerome Sacks and Vadim Sokolov},
  journal = {Statistical Science},
  volume = {37},
  issue = {1},
  pages = {64 -- 89},
  year = {2022},
}

@article{Ankenman2009,
  title = {Stochastic Kriging for Simulation Metamodeling},
  author = {Ankenman, Bruce and Nelson, Barry L. and Staum, Jeremy},
  journal = {Operations Research},
  volume = {58},
  issue = {2},
  pages = {371--382},
  year = {2009}
}

@book{gramacy2020surrogates,
	address = {New York, NY},
	title = {Surrogates: {Gaussian} Process Modeling, Design, and Optimization for the Applied Sciences},
	isbn = {9780367415426},
	shorttitle = {Surrogates},
	publisher = {CRC Press ; Taylor \& Francis Group},
	author = {Gramacy, Robert B.},
	year = {2020},
}

@book{santner2018design,
	address = {New York, NY},
	edition = {2nd ed. 2018},
	series = {Springer {Series} in {Statistics}},
	title = {The {Design} and {Analysis} of {Computer} {Experiments}},
	isbn = {9781493988471},
	publisher = {Springer New York : Imprint: Springer},
	author = {Santner, Thomas J. and Williams, Brian J. and Notz, William I.},
	year = {2018},
}

@article{Jalali2017,
    title = {Comparison of Kriging-based algorithms for simulation optimization with heterogeneous noise},
    journal = {European Journal of Operational Research},
    volume = {261},
    number = {1},
    pages = {279-301},
    year = {2017},
    issn = {0377-2217},
    doi = {https://doi.org/10.1016/j.ejor.2017.01.035},
    url = {https://www.sciencedirect.com/science/article/pii/S037722171730070X},
    author = {Hamed Jalali and Inneke {Van Nieuwenhuyse} and Victor Picheny},
}

@inbook{Frazier2018,
author = {Peter I. Frazier},
year = {2018},
title = {Bayesian Optimization},
booktitle = {Recent Advances in Optimization and Modeling of Contemporary Problems},
chapter = {11},
pages = {255-278},
doi = {10.1287/educ.2018.0188},
URL = {https://pubsonline.informs.org/doi/abs/10.1287/educ.2018.0188},
}

@ARTICLE{Surer2024,
  author={\"O. S\"urer},
  title = {Simulation Experiment Design for Calibration via Active Learning},
  journal = {Journal of Quality Technology},
  volume = {57},
  number = {1},
  pages = {16--34},
  year = {2025},
  publisher = {Taylor \& Francis},
  doi = {10.1080/00224065.2024.2391780},
}

@ARTICLE{Plumlee2017,
  author={Matthew Plumlee},
  journal={Journal of the American Statistical Association}, 
  title={Bayesian Calibration of Inexact Computer Models}, 
  year={2017},
  volume={112},
  number={519},
  pages={1274--1285}}

@ARTICLE{Bayarri2007,
  author={Maria J. Bayarri and James O Berger and Rui Paulo and Jerry Sacks and John A Cafeo and James Cavendish and Chin-Hsu Lin and Jian Tu},
  journal={Technometrics}, 
  title={A Framework for Validation of Computer Models}, 
  year={2007},
  volume={49},
  number={2},
  pages={138--154}}

@ARTICLE{Bayarri2009,
  author={Maria J. Bayarri and James O Berger and F. Liu},
  journal={Bayesian Analysis}, 
  title={{Modularization in Bayesian analysis, with emphasis on analysis of computer models}}, 
  year={2009},
  volume={4},
  number={1},
  pages={119--150}}

@ARTICLE{Ohagan2001,
  author={Kennedy, M. C. and O’Hagan, A},
  journal={Journal of the Royal Statistical Society. Series B (Statistical Methodology)}, 
  title={Bayesian Calibration of Computer Models}, 
  year={2001},
  volume={63},
  number={3},
  pages={425--464}}

@article{Binois2019,
   title={Replication or Exploration? {Sequential} Design for Stochastic Simulation Experiments},
   volume={61},
   number={1},
   pages={7--23},
   journal={Technometrics},
   author={Micka\"el Binois and Jiangeng Huang and Robert B. Gramacy and Mike Ludkovski},
   year={2019}}

@article{Binois2018,
   title={Practical Heteroscedastic {Gaussian} Process Modeling for Large Simulation Experiments},
   volume={27},
   number={4},
   pages={808--821},
   journal={Journal of Computational and Graphical Statistics},
   author={Micka\"el Binois and Robert B. Gramacy and Mike Ludkovski},
   year={2018}}

@article{Jarvenpa2019,
   title={Efficient Acquisition Rules for Model-Based Approximate {Bayesian} Computation},
   volume={14},
   ISSN={1936-0975},
   number={2},
   journal={Bayesian Analysis},
   publisher={Institute of Mathematical Statistics},
   author={J\"arvenp\"a\"a, Marko and Gutmann, Michael U. and Pleska, Arijus and Vehtari, Aki and Marttinen, Pekka},
   year={2019},
   month={Jun} }

@article{Jarvenpa2021,
   title={Parallel {Gaussian} Process Surrogate {Bayesian} Inference with Noisy Likelihood Evaluations},
   volume={16},
   number={1},
   journal={Bayesian Analysis},
   publisher={International Society for Bayesian Analysis},
   author={Marko J\"arvenp\"a\"a and Michael U. Gutmann and Aki Vehtari and Pekka Marttinen},
   year={2021},
   pages = {147 -- 178}
   }

@inproceedings{Kandasamy2015, 
    author = {Kandasamy, Kirthevasan and Schneider, Jeff and P\'{o}czos, Barnab\'{a}s}, 
    title = {Bayesian Active Learning for Posterior Estimation}, 
    year = {2015}, 
    isbn = {9781577357384}, 
    publisher = {AAAI Press}, booktitle = {Proceedings of the 24th International Conference on Artificial Intelligence}, 
    pages = {3605–3611}, 
    numpages = {7}, 
    location = {Buenos Aires, Argentina}, series = {IJCAI'15} }

@article{Kandasamy2017,
title = {Query efficient posterior estimation in scientific experiments via Bayesian active learning},
journal = {Artificial Intelligence},
volume = {243},
pages = {45-56},
year = {2017},
issn = {0004-3702},
doi = {https://doi.org/10.1016/j.artint.2016.11.002},
url = {https://www.sciencedirect.com/science/article/pii/S0004370216301394},
author = {Kirthevasan Kandasamy and Jeff Schneider and Barnabás Póczos}
}

@book{Rasmussen2005, 
    author = {Rasmussen, Carl Edward and Williams, Christopher K. I.}, title = {Gaussian Processes for Machine Learning (Adaptive Computation and Machine Learning)}, year = {2005}, isbn = {026218253X}, publisher = {The MIT Press} }

@book{Gilks1995, 
    author = {Walter R. Gilks and Sylvia Richardson and David J. Spiegelhalter}, 
    title = {{Markov Chain Monte Carlo} in Practice (1st ed.)}, 
    year = {1995}, 
    publisher = {Chapman and Hall/CRC} 
}

@article{Jenny2014,
 author = {Brynjarsd\'ottir, Jenny and O'Hagan, Anthony},
 journal = {Inverse Problems},
 number = {11},
 pages = {114007},
 publisher = {IOP Publishing},
 title = {Learning about physical parameters: the importance of model discrepancy},
 volume = {30},
 year = {2014}
}

@article{Tuo2015,
    title = {Efficient calibration for imperfect computer models},
    author = {Rui Tuo and C. F. Jeff Wu},
    journal = {The Annals of Statistics},
    number = 6,
    volume = 43,
    year = {2015},
    pages = {2331 -- 2352}
}

\appendix

\section{Appendix}

\label{app:code}
\subsection{Proof of Lemma~3.1}
\label{proof:lemma3.3}

To compute the criterion, we first establish general results for GPs. Recall that $m_{t}(\zb)$ and $\varsigma^2_{t}(\zb)$ refer to the emulator mean and variance for any $\zb$ at stage $t$. Let $\varsigma_{t+1}^2(\zb)$ denote the variance after observing a (hypothetical) simulation data point $(\breve{\zb}, \breve{\zeta})$.
Given this hypothetical observation, we can express the covariance matrix as
\begin{align}\label{updatedK_explore}
    \begin{split}
        \Kb_{t+1}(\ab_{t+1}) = \left[ {\begin{array}{cc} \Kb_{t}(\ab_t) & \kb_{t}(\breve{\zb}) \\
                        \kb_{t}(\breve{\zb})^\top & k_{t}(\breve{\zb}, \breve{\zb}) + r\left(\breve{\zb}\right) \\
                        \end{array} } \right],
    \end{split}      
\end{align}
where $\ab_{t+1} = (a_1, \ldots, a_{n_t}, 1)^\top$. The inverse of $\Kb_{t+1}(\ab_{t+1})$ is obtained using the partitioned inverse formula as
\begin{align}\notag
    \begin{split}
         \Kb_{t+1}(\ab_{t+1})^{-1} = \breve{l} \left[{\begin{array}{cc} \Kb_{t}(\ab_t)^{-1} \breve{l}^{-1} + \Kb_{t}(\ab_t)^{-1} \kb_{t}(\breve{\zb}) \kb_{t}(\breve{\zb})^\top \Kb_{t}(\ab_t)^{-1} & -\Kb_{t}(\ab_t)^{-1} \kb_{t}(\breve{\zb}) \\
            -\kb_{t}(\breve{\zb})^\top \Kb_{t}(\ab_t)^{-1} & 1\\
            \end{array} } \right]
    \end{split}      
\end{align}
where $\breve{l} = \left(\varsigma^2_t(\breve{\zb}) + r(\breve{\zb})\right)^{-1} $. Substituting $\Kb_{t+1}(\ab_{t+1})^{-1}$ into the variance expression gives
    \begin{align}\label{var_future}
        \begin{split}
            \varsigma^2_{t+1}(\zb) &= k_{t}(\zb, \zb) - \left[\kb_{t}(\zb)^\top, k_{t}(\zb, \breve{\zb})\right] \Kb_{t+1}(\ab_{t+1})^{-1} \left[ {\begin{array}{cc} \kb_{t}(\zb)\\
            k_{t}(\zb, \breve{\zb}) \\
                \end{array} } \right] \\
            &= \varsigma^2_{t}(\zb) - \frac{{\rm cov}_{t}(\zb, \breve{\zb})^2}{\varsigma^2_t(\breve{\zb}) + r(\breve{\zb})}.
        \end{split} 
    \end{align}
    Similarly, for the covariance function, we have
    \begin{align}\label{cov_future}
        \begin{split}
        {\rm cov}_{t+1}(\zb, \zb^\prime) = {\rm cov}_{t}(\zb, \zb^\prime) - \frac{{\rm cov}_{t}(\zb, \breve{\zb}){\rm cov}_{t}(\zb^\prime, \breve{\zb})}{\varsigma^2_t(\breve{\zb}) + r(\breve{\zb})}.
        \end{split} 
    \end{align}
By following an analogous argument for the mean function, we arrive at
\begin{align}\label{mean_future_explore}
    \begin{split}
        m_{t+1}(\zb) &= \left[\kb_{t}(\zb)^\top, k_{t}(\zb, \breve{\zb})\right] \Kb_{t+1}(\ab_{t+1})^{-1}  \left[ {\begin{array}{cc} \bar{\pmb{\simout}}_{t} \\
        \breve{\zeta} \\
            \end{array} } \right] \\
        &= m_{t}(\zb) + \frac{\text{cov}_{t}(\zb, \breve{\zb})}{\varsigma_{t}^2(\breve{\zb}) + r(\breve{\zb})}\left(\breve{\zeta} - m_{t}(\breve{\zb})\right).
    \end{split} 
\end{align}

The expectation, variance, and covariance of \eqref{mean_future_explore} are given by
    \begin{align}
        \begin{split}
        \mathbb{E}_{\outputnew |\mathcal{D}_t}\left[m_{t+1}(\zb)\right] = m_{t}(\zb), \quad 
        \mathbb{V}_{\outputnew |\mathcal{D}_t}\left[m_{t+1}(\zb)\right] = \frac{\text{cov}_{t}(\zb, \breve{\zb})^2}{\varsigma^2_{t}(\breve{\zb}) + r(\breve{\zb})}, {\rm and} \\
        \mathbb{C}_{\outputnew |\mathcal{D}_t}[m_{t+1}(\zb), m_{t+1}(\zb')] = \frac{{\rm cov}_{t}(\zb, \breve{\zb}){\rm cov}_{t}(\zb', \breve{\zb})}{\varsigma^2_t(\breve{\zb}) + r(\breve{\zb})}.
    \end{split}
\end{align}
Using the distribution $\breve{\zeta} \mid \mathcal{D}_t \sim \mathcal{N}\left(m_{t}(\breve{\zb}), \, \varsigma^2_{t}(\breve{\zb}) + r(\breve{\zb})\right)$ and the transformation in \eqref{mean_future_explore}, we obtain
    \begin{align}\label{mean_distr1}
        \begin{split}
            m_{t+1}(\zb) \mid \mathcal{D}_t \sim \mathcal{N}\left(m_{t}(\zb), \, \frac{\text{cov}_{t}(\zb, \breve{\zb})^2}{\varsigma^2_{t}(\breve{\zb}) + r(\breve{\zb})}\right).
       \end{split} 
    \end{align}
Extending this result to the multivariate case, \eqref{mean_distr1} implies
    \begin{align}\label{mean_distribution_explore}
        \begin{split}
            \muv_{t+1}\left(\tb\right) \mid \mathcal{D}_t \sim \mathcal{MVN}\left(\muv_{t}(\tb), \breve{\PHI}_t(\tb)\right),
       \end{split} 
    \end{align}  
where $\muv_t\left(\tb\right) = \left(m_{t}\left(\zb_1^o\right), \ldots, m_{t}\left(\zb_d^o\right)\right)^\top$ denotes the vector of predictive means at the field data design inputs and $\breve{\PHI}_t(\tb)$ is a $d \times d$ covariance matrix. The $j$th diagonal entry of $\breve{\PHI}_t(\tb)$ is $\frac{\text{cov}_{t}(\zf_j, \breve{\zb})^2}{\varsigma^2_{t}(\breve{\zb}) + r(\breve{\zb})}$ and the off-diagonal $(j,j^\prime)$th entry is $\frac{\text{cov}_{t}(\zf_j, \breve{\zb})\text{cov}_{t}(\zf_{j^\prime}, \breve{\zb})}{\varsigma^2_{t}(\breve{\zb}) + r(\breve{\zb})}$. 

To derive ${\rm IVAR}(\breve{\zb})$ in \eqref{eq:IVAR_explore}, we first compute the expectation $\mathbb{E}_{\outputnew |\mathcal{D}_t}\left(\mathbb{V}_{\latent(\tb)|\DatCan}\left[\tilde{p}(\tb\mid\yb)\right]\right)$, where $\DatCan = \mathcal{D}_t $ $\cup (\breve{\zb}, \breve{\zeta})$.        
Using the identity $\mathbb{V}_{\latent(\tb)|\DatCan}\left[\tilde{p}(\tb\mid\yb)\right] = \mathbb{V}_{\latent(\tb)|\DatCan}\left[\tilde{p}(\yb\mid\tb)\right]p(\tb)^2$ and the derivation from \eqref{variancepostfinal}, we obtain  $\mathbb{V}_{\latent(\tb)|\DatCan}\left[\tilde{p}(\yb\mid\tb)\right]$ as
    \begin{align} \notag 
        \begin{split}
              & \frac{1}{2^{d}\pi^{d/2}|\Sigmav|^{1/2}}f_\mathcal{N}\left(\yb; \, \muv_{t+1}(\tb), \, \frac{1}{2}\Sigmav + \Sv_{t+1}(\tb)\right)  - \left(f_\mathcal{N}\left(\yb; \, \muv_{t+1}(\tb), \, \Sigmav + \Sv_{t+1}(\tb)\right)\right)^2.
        \end{split}
    \end{align}

We have $\Sv_{t+1}(\tb) = \Sv_{t}(\tb) - \breve{\PHI}_t(\tb)$ as implied by \eqref{var_future} and \eqref{cov_future}, and note that $\mathbf{S}_{t+1}(\tb)$ does not depend on $\breve{\zeta}$. Substituting $\mathbf{S}_{t+1}(\tb)$ with $\mathbf{S}_{t}(\tb) - \breve{\PHI}_t(\tb)$ and applying \eqref{mean_distribution_explore}, we obtain $\mathbb{E}_{\outputnew |\mathcal{D}_t}\left(\mathbb{V}_{\latent(\tb)|\DatCan}\left[\tilde{p}(\yb\mid\tb)\right]\right)$ 
    \begin{align} \label{eq:IVAR_explore_inner}
        \begin{split}
             & \int \frac{1}{2^{d}\pi^{d/2}|\Sigmav|^{1/2}}f_\mathcal{N}\left(\yb; \, \muv_{t+1}(\tb), \, \frac{1}{2}\Sigmav + \mathbf{S}_{t}(\tb) - \breve{\PHI}_t(\tb)\right) f_\mathcal{N}\left(\muv_{t+1}(\tb); \, \muv_{t}(\tb), \, \breve{\PHI}_t(\tb)\right) d\muv_{t+1}(\tb) \\& - \int \left(f_\mathcal{N}\left(\yb; \, \muv_{t+1}(\tb), \, \Sigmav + \mathbf{S}_{t}(\tb) - \breve{\PHI}_t(\tb)\right)\right)^2 f_\mathcal{N}\left(\muv_{t+1}(\tb); \, \muv_{t}(\tb), \, \breve{\PHI}_t(\tb)\right) d\muv_{t+1}(\tb).
        \end{split}
    \end{align}

The remainder of the proof follows the approach in \cite{Surer2023}, which we include here for completeness. To simplify notation, we omit the dependence on $\tb$ in $\muv_{t}(\tb)$ and $\muv_{t+1}(\tb)$. By defining $\mathbf{L} \coloneqq \frac{1}{2}\Sigmav + \Sv_t(\tb) - \breve{\PHI}_t(\tb)$, $\mathbf{M} \coloneqq  \Sigmav + \Sv_t(\tb) - \breve{\PHI}_t(\tb)$, and setting $a_1 \coloneqq \frac{2^{-d}\pi^{-d/2}|\Sigmav|^{-1/2}}{(2\pi)^d|\mathbf{L}\breve{\PHI}_t(\tb)|^{1/2}}$, $a_2 \coloneqq \frac{(2\pi)^{-3d/2}}{|\mathbf{M}\breve{\PHI}_t(\tb)\mathbf{M}|^{1/2}}$, and assuming that $\mathbf{L}$ and $\mathbf{M}$ are invertible, \eqref{eq:IVAR_explore_inner} is equivalently expressed as 
    \begin{align} 
        \begin{split} \label{eq:gnew}
        & a_1 \int \exp\left\{-\frac{1}{2} \left(\left(\yb - \muv_{t+1}\right)^\top \mathbf{L}^{-1} \left(\yb - \muv_{t+1}\right) + \left(\muv_{t+1} - \muv_{t}\right)^\top \breve{\PHI}_t(\tb)^{-1} \left(\muv_{t+1} - \muv_{t}\right) \right)\right\}d\muv_{t+1} \\
        & - a_2 \int \exp\left\{-\frac{1}{2} \left(2\left(\yb - \muv_{t+1}\right)^\top \mathbf{M}^{-1} \left(\yb - \muv_{t+1}\right) + \left(\muv_{t+1} - \muv_{t}\right)^\top \breve{\PHI}_t(\tb)^{-1} \left(\muv_{t+1} - \muv_{t}\right) \right)\right\}d\muv_{t+1}.
        \end{split}
    \end{align}
    Defining $\mathbf{v} \coloneqq \muv_t - \muv_{t+1}$ and $\mathbf{z} \coloneqq \yb - \muv_t$, we can write \eqref{eq:gnew} in matrix notation as
    \begin{align}\label{jointlikematrix} 
            \begin{split}
                =& \frac{1}{2^{d}\pi^{d/2}|\Sigmav|^{1/2}} \int f_\mathcal{N}\left(\left[{\begin{array}{c} \mathbf{v} \\
                \mathbf{z} \\\end{array}} \right]; \,  \mathbf{0}, \, \left[ {\begin{array}{cc} \breve{\PHI}_t(\tb) & -\breve{\PHI}_t(\tb)\\
                -\breve{\PHI}_t(\tb) & \mathbf{L} + \breve{\PHI}_t(\tb)\\
                \end{array} } \right]  \right)d \mathbf{v} \\
                &-\frac{1}{2^{d}\pi^{d/2}|\mathbf{M}|^{1/2}} \int f_\mathcal{N}\left(\left[{\begin{array}{c} \mathbf{v} \\
                \mathbf{z} \\\end{array}} \right]; \,  \mathbf{0}, \, \left[ {\begin{array}{cc} \breve{\PHI}_t(\tb) & -\breve{\PHI}_t(\tb)\\
                -\breve{\PHI}_t(\tb) & \frac{1}{2}\mathbf{M} + \breve{\PHI}_t(\tb)\\
                \end{array} } \right]  \right)d \mathbf{v}.
            \end{split}
        \end{align}
    Marginalizing over $\mathbf{v}$ yields $\mathbb{E}_{\outputnew |\mathcal{D}_t}\left(\mathbb{V}_{\latent(\tb)|\DatCan}\left[\tilde{p}(\yb\mid\tb)\right]\right)$ as
    \begin{align} \label{IVARpart2}
        \begin{split}
            \frac{f_\mathcal{N}\left(\yb; \, \muv_{t}(\tb), \, \frac{1}{2}\Sigmav + \Sv_t(\tb) \right)}{2^{d}\pi^{d/2}|\Sigmav|^{1/2}} - \frac{f_\mathcal{N}\left(\yb; \, \muv_{t}(\tb), \, \frac{1}{2}\left(\Sigmav + \Sv_t(\tb) + \breve{\PHI}_t(\tb)\right)\right)}{2^{d}\pi^{d/2}|\Sigmav + \Sv_t(\tb) - \breve{\PHI}_t(\tb)|^{1/2}} .
        \end{split}
    \end{align}

Substituting \eqref{IVARpart2} into the definition of the IVAR criterion, $\displaystyle\int\limits_{\tb \in \Theta} \mathbb{E}_{\outputnew |\mathcal{D}_t}\left(\mathbb{V}_{\latent(\tb)|\DatCan}\left[\tilde{p}(\tb\mid\yb)\right]\right) d\tb$, yields \eqref{eq:IVAR_explore_derive}.
    
\subsection{Proof of Lemma~3.2}
\label{proof:ivar_rep}

We begin with the derivation of the variance $\varsigma^2_{t+1}(\zb)$ after observing the hypothetical replicate $\left(\zb_k, \zeta_k\right)$. Let $\ab_{t+1}$ be the $n_t \times 1$ vector of replicates after adding this data point into the simulation dataset such that $\ab_{t+1} = \ab_{t} + \mathbf{e}_k$. Here, $\mathbf{e}_k$ is an $n_t \times 1$ vector with all entries zero except for the $k$th element, which is one. We can write
\begin{align}\label{updatedK_exploit}
    \begin{split}
        \Kb_{t+1}(\ab_{t+1}) &= \Kb_{t}(\ab_t) - {\rm diag}\left(0, \ldots, 0, \frac{r(\zb_k)}{a_k (a_k + 1)}, 0, \ldots, 0\right) \\
        &= \Kb_{t}(\ab_t) + \mathbf{u}_1 \mathbf{u}_2^\top, \text{with} \quad \mathbf{u}_1 = -\frac{r(\zb_k)}{a_k}\mathbf{e}_k \quad \text{and} \quad \mathbf{u}_2 = \frac{1}{a_k + 1}\mathbf{e}_k.
    \end{split}      
\end{align}
The Sherman-Morrison formula gives the matrix inverse $\Kb_{t+1}(\ab_{t+1})^{-1}$ as
\begin{align}\label{Kinverse}
    \begin{split}
        \Kb_{t+1}(\ab_{t+1})^{-1} = \left(\Kb_{t}(\ab_t) + \mathbf{u}_1 \mathbf{u}_2^\top\right)^{-1} = \Kb_{t}(\ab_t)^{-1} + \mathbf{B}_k
    \end{split}      
\end{align}
where $\mathbf{B}_k = \frac{\left(\Kb_{t}(\ab_t)^{-1}\right)_{.,k}\left(\Kb_{t}(\ab_t)^{-1}\right)_{k,.}}{a_k(a_k + 1)/r(\zb_k) - \left(\Kb_{t}(\ab_t)^{-1}\right)_{k,k}}$.
Plugging \eqref{Kinverse} into the variance definition gives 
    \begin{align}\label{var_future_exploit}
        \begin{split}
            \varsigma^2_{t+1}(\zb) &= k_{t}(\zb, \zb) - \kb_{t}(\zb)^\top \Kb_{t+1}(\ab_{t+1})^{-1} \kb_{t}(\zb) \\
            &= \varsigma^2_{t}(\zb) - \kb_{t}(\zb)^\top \mathbf{B}_k\kb_{t}(\zb).
        \end{split} 
    \end{align}
Following the same logic, we derive ${\rm cov}_{t+1}(\zb, \zb')$ as 
    \begin{align}\label{cov_future_exploit}
        \begin{split}
            {\rm cov}_{t+1}(\zb, \zb') &= k_{t}(\zb, \zb') - \kb_{t}(\zb)^\top \Kb_{t+1}(\ab_{t+1})^{-1} \kb_{t}(\zb') \\
            &= {\rm cov}_{t}(\zb, \zb') - \kb_{t}(\zb)^\top \mathbf{B}_k\kb_{t}(\zb').
        \end{split} 
    \end{align}
Likewise, the updated emulator mean $m_{t+1}(\zb)$ is
\begin{align}\label{mean_future_exploit}
    \begin{split}
        m_{t+1}(\zb) &= \kb_{t}(\zb)^\top\Kb_{t+1}(\ab_{t+1})^{-1}  \left(\bar{\pmb{\simout}}_{t} + \mathbf{e}_k  \frac{\outputk - \bar{\zeta}_k}{a_k + 1}\right)\\
        &= m_t(\zb) + \kb_{t}(\zb)^\top\left(\left(\Kb_{t}(\ab_t)^{-1} + \mathbf{B}_k\right)\mathbf{e}_k  \frac{\outputk - \bar{\zeta}_k}{a_k + 1} + \mathbf{B}_k \bar{\pmb{\simout}}_{t} \right).\\
    \end{split} 
\end{align}
Here, $\bar{\zeta}_k$ denotes $k$th entry of $\bar{\pmb{\simout}}_{t}$.

Using the predictive distribution $\outputk \mid \mathcal{D}_t \sim \mathcal{N}\left(m_{t}(\zb_k), \, \varsigma^2_{t}(\zb_k) + r(\zb_k)\right)$ and the transformation in \eqref{mean_future_exploit}, we have $m_{t+1}(\zb) \mid \mathcal{D}_t \sim \mathcal{N}\left(\mathbb{E}_{\outputk|\mathcal{D}_t}\left[m_{t+1}(\zb)\right], \mathbb{V}_{\outputk|\mathcal{D}_t}\left[m_{t+1}(\zb)\right]\right)$ where
\begin{align} \notag
    \begin{split}
        \mathbb{E}_{\outputk|\mathcal{D}_t}\left[m_{t+1}(\zb)\right] &= m_t(\zb) + \left[\kb_{t}(\zb)^\top\left(\Kb_{t}(\ab_t)^{-1} + \mathbf{B}_k\right)\right]_k  \frac{m_t(\zb_k) - \bar{\zeta}_k}{a_k + 1} + \kb_{t}(\zb)^\top \mathbf{B}_k \bar{\pmb{\simout}}_{t} , \\
        \mathbb{V}_{\outputk|\mathcal{D}_t}\left[m_{t+1}(\zb)\right] &= \frac{\varsigma^2_t(\zb_k) + r(\zb_k)}{(a_k + 1)^2} \left[\kb_{t}(\zb)^\top\left(\Kb_{t}(\ab_t)^{-1} + \mathbf{B}_k\right) \right]_k^2. \\
    \end{split}
\end{align}
Let $\muv_{t+1}\left(\tb\right) = \left(m_{t+1}(\zf_1), \ldots, m_{t+1}(\zf_d)\right)^\top$ denote the vector of predictive means at the field data design inputs after observing the hypothetical data point $(\zb_k, \outputk)$. Let $\muv^k_{t}(\tb)$ be a vector with entries $\mathbb{E}_{\outputk|\mathcal{D}_t}\left[m_{t+1}(\zf_j)\right]$, for $j = 1, \ldots, d$. Define a $d \times d$ covariance matrix $\GAMMA^k_t(\tb)$, where the $j$th diagonal element is $\mathbb{V}_{\outputk|\mathcal{D}_t}\left[m_{t+1}(\zf_j)\right]$, and the $(j,j^\prime)$th element---denoted the covariance $\mathbb{C}_{\outputk|\mathcal{D}_t}\left[m_{t+1}(\zf_j), m_{t+1}(\zf_{j^\prime})\right]$---is given by
\begin{equation} \notag
         \frac{\varsigma^2_t(\zb_k) + r(\zb_k)}{(a_k + 1)^2} \left[\kb_{t}(\zf_j)^\top\left(\Kb_{t}(\ab_t)^{-1} + \mathbf{B}_k\right) \right]_k \left[\left(\Kb_{t}(\ab_t)^{-1} + \mathbf{B}_k\right)^\top \kb_{t}(\zf_{j^\prime})\right]_k.
\end{equation}
In the multivariate setting, we then obtain
\begin{equation}\label{mean_distribution_exploit}
        \muv_{t+1}\left(\tb\right) | \mathcal{D}_t \sim \mathcal{MVN}\left(\muv^k_{t}(\tb), \GAMMA^k_t(\tb)\right).
\end{equation}

To derive ${\rm IVAR}(\zb_k)$, we begin by computing the expectation $\mathbb{E}_{\outputk|\mathcal{D}_t}\left(\mathbb{V}_{\latent(\tb)|\DatCan}\left[\tilde{p}(\tb\mid\yb) \right]\right)$, where $\DatCan = \mathcal{D}_t $ $\cup \left(\zb_k, \zeta_k\right)$.        
We compute the variance $\mathbb{V}_{\latent(\tb)|\DatCan}\left[\tilde{p}(\tb\mid\yb)\right] = \mathbb{V}_{\latent(\tb)|\DatCan}\left[\tilde{p}(\yb\mid\tb)\right]p(\tb)^2,$
where, according to \eqref{variancepostfinal}, the term $\mathbb{V}_{\latent(\tb)|\DatCan}\left[\tilde{p}(\yb\mid\tb)\right]$ is given by
    \begin{align} \notag 
        \begin{split}
             & \frac{1}{2^{d}\pi^{d/2}|\Sigmav|^{1/2}}f_\mathcal{N}\left(\yb; \, \muv_{t+1}(\tb), \, \frac{1}{2}\Sigmav + \Sv_{t+1}(\tb)\right)  - \left(f_\mathcal{N}\left(\yb; \, \muv_{t+1}(\tb), \, \Sigmav + \Sv_{t+1}(\tb)\right)\right)^2.
        \end{split}
    \end{align}
Let $\PHI^k_t(\tb)$ be a $d \times d$ matrix with the $(j,j^\prime)$th element $\kb_{t}^\top(\zf_j) \mathbf{B}_k\kb_{t}(\zf_{j^\prime})$. Then, from \eqref{var_future_exploit} and \eqref{cov_future_exploit}, we have $\Sv_{t+1}(\tb) = \Sv_{t}(\tb) - \PHI^k_t(\tb)$.
Substituting $\mathbf{S}_{t+1}(\tb)$ with $\mathbf{S}_{t}(\tb) - \PHI^k_t(\tb)$ and using \eqref{mean_distribution_exploit}, we arrive at the following expression for $\mathbb{E}_{\outputk|\mathcal{D}_t}\left(\mathbb{V}_{\latent(\tb)|\DatCan}\left[\tilde{p}(\yb\mid\tb)\right]\right)$:
    \begin{align} \label{eq:IVAR_exploit_inner}
        \begin{split}
             & \int \frac{1}{2^{d}\pi^{d/2}|\Sigmav|^{1/2}}f_\mathcal{N}\left(\yb; \, \muv_{t+1}(\tb), \, \frac{1}{2}\Sigmav + \mathbf{S}_{t}(\tb) - \PHI^k_t(\tb)\right) f_\mathcal{N}\left(\muv_{t+1}(\tb); \, \muv^k_{t}(\tb), \, \GAMMA^k_t(\tb)\right) d\muv_{t+1}(\tb) \\& - \int \left(f_\mathcal{N}\left(\yb; \, \muv_{t+1}(\tb), \, \Sigmav + \mathbf{S}_{t}(\tb) - \PHI^k_t(\tb)\right)\right)^2 f_\mathcal{N}\left(\muv_{t+1}(\tb); \, \muv^k_{t}(\tb), \, \GAMMA^k_t(\tb)\right) d\muv_{t+1}(\tb).
        \end{split}
    \end{align}
For brevity, we omit the dependence on $\tb$ in $\muv^k_{t}(\tb)$ and $\muv_{t+1}(\tb)$. Let $\mathbf{L} \coloneqq \frac{1}{2}\Sigmav + \Sv_t(\tb) - \PHI^k_t(\tb)$, $\mathbf{M} \coloneqq  \Sigmav + \Sv_t(\tb) - \PHI^k_t(\tb)$, and define $a_1 \coloneqq \frac{2^{-d}\pi^{-d/2}|\Sigmav|^{-1/2}}{(2\pi)^d|\mathbf{L}\GAMMA^k_t(\tb)|^{1/2}}$, $a_2 \coloneqq \frac{(2\pi)^{-3d/2}}{|\mathbf{M}\GAMMA^k_t(\tb)\mathbf{M}|^{1/2}}$. Assuming $\mathbf{L}$ and $\mathbf{M}$ are invertible, we rewrite \eqref{eq:IVAR_exploit_inner} as 
    \begin{align} 
        \begin{split} \label{eq:gnew_exploit}
        & a_1 \int \exp\left\{-\frac{1}{2} \left(\left(\yb - \muv_{t+1}\right)^\top \mathbf{L}^{-1} \left(\yb - \muv_{t+1}\right) + \left(\muv_{t+1} - \muv^k_{t}\right)^\top \GAMMA^k_t(\tb)^{-1} \left(\muv_{t+1} - \muv^k_{t}\right) \right)\right\}d\muv_{t+1} \\
        & - a_2 \int \exp\left\{-\frac{1}{2} \left(2\left(\yb - \muv_{t+1}\right)^\top \mathbf{M}^{-1} \left(\yb - \muv_{t+1}\right) + \left(\muv_{t+1} - \muv^k_{t}\right)^\top \GAMMA^k_t(\tb)^{-1} \left(\muv_{t+1} - \muv^k_{t}\right) \right)\right\}d\muv_{t+1}.
        \end{split}
    \end{align}
    Setting $\mathbf{v} \coloneqq \muv^k_{t} - \muv_{t+1}$ and $\mathbf{z} \coloneqq \yb - \muv^k_{t}$, we express  \eqref{eq:gnew_exploit} in matrix form as
    \begin{align} \notag 
            \begin{split}
                =& \frac{2^{-d}\pi^{-d/2}|\Sigmav|^{-1/2}}{(2\pi)^d|\mathbf{L}\GAMMA^k_t(\tb)^{-1}|^{1/2}} \int \exp\left\{-\frac{1}{2}\left[{\begin{array}{c} \mathbf{v} \\
                \mathbf{z} \\\end{array}} \right]^\mathsf{T}\left[{\begin{array}{cc} \mathbf{L}^{-1} + \GAMMA^k_t(\tb)^{-1} & \mathbf{L}^{-1} \\
                \mathbf{L}^{-1} & \mathbf{L}^{-1} \\
                \end{array} } \right] \left[{\begin{array}{c} \mathbf{v} \\
                \mathbf{z} \\ \end{array} } \right]\right\} d \mathbf{v} \\
                &- \frac{(2\pi)^{-3d/2}}{2^{d/2}|\mathbf{M}|^{1/2}\left|\frac{1}{2}\mathbf{M}\GAMMA^k_t(\tb)\right|^{1/2}} \int \exp\left\{-\frac{1}{2}\left[{\begin{array}{c} \mathbf{v} \\
                \mathbf{z} \\\end{array}} \right]^\mathsf{T}\left[{\begin{array}{cc} 2\mathbf{M}^{-1} + \GAMMA^k_t(\tb)^{-1} & 2\mathbf{M}^{-1} \\
                2\mathbf{M}^{-1} & 2\mathbf{M}^{-1} \\
                \end{array} } \right] \left[{\begin{array}{c} \mathbf{v} \\
                \mathbf{z} \\ \end{array} } \right]\right\} d \mathbf{v} \\
                =& \frac{1}{2^{d}\pi^{d/2}|\Sigmav|^{1/2}} \int f_\mathcal{N}\left(\left[{\begin{array}{c} \mathbf{v} \\
                \mathbf{z} \\\end{array}} \right]; \,  \mathbf{0}, \, \left[ {\begin{array}{cc} \GAMMA^k_t(\tb) & -\GAMMA^k_t(\tb)\\
                -\GAMMA^k_t(\tb) & \mathbf{L} + \GAMMA^k_t(\tb)\\
                \end{array} } \right]  \right)d \mathbf{v} \\
                &-\frac{1}{2^{d}\pi^{d/2}|\mathbf{M}|^{1/2}} \int f_\mathcal{N}\left(\left[{\begin{array}{c} \mathbf{v} \\
                \mathbf{z} \\\end{array}} \right]; \,  \mathbf{0}, \, \left[ {\begin{array}{cc} \GAMMA^k_t(\tb) & -\GAMMA^k_t(\tb)\\
                -\GAMMA^k_t(\tb) & \frac{1}{2}\mathbf{M} + \GAMMA^k_t(\tb)\\
                \end{array} } \right]  \right)d \mathbf{v}.
            \end{split}
        \end{align}
    Marginalizing over $\mathbf{v}$ yields $\mathbb{E}_{\zeta^k|\mathcal{D}_t}\left(\mathbb{V}_{\latent(\tb)|\DatCan}\left[\tilde{p}(\yb\mid\tb)\right]\right)$ as
    \begin{equation}  \label{IVARexploitpart2}
            \frac{f_\mathcal{N}\left(\yb; \, \muv^k_{t}(\tb), \, \frac{1}{2}\Sigmav + \Sv_t(\tb) - \PHI^k_t(\tb) + \GAMMA^k_t(\tb) \right)}{2^{d}\pi^{d/2}|\Sigmav|^{1/2}} - \frac{f_\mathcal{N}\left(\yb; \, \muv^k_{t}(\tb), \, \frac{1}{2}\left(\Sigmav + \Sv_t(\tb) - \PHI^k_t(\tb)\right) + \GAMMA^k_t(\tb)\right)}{2^{d}\pi^{d/2}|\Sigmav + \Sv_t(\tb) - \PHI^k_t(\tb)|^{1/2}} .
    \end{equation}
Combining \eqref{IVARexploitpart2} with the definition of ${\rm IVAR}(\zb_k)$, $\displaystyle\int\limits_{\tb \in \Theta}  p(\tb)^2\mathbb{E}_{\zeta^k|\mathcal{D}_t}\left(\mathbb{V}_{\latent(\tb)|\DatCan}\left[\tilde{p}(\yb\mid\tb)\right]\right) d\tb$, yields the expression in \eqref{eq:IVAR_exploit_derive}.

\subsection{Allocation Rule for Adaptive Scheme}
\label{sec:allocation_adaptive}

The goal is to solve the following mathematical program optimally:
    \begin{subequations} \label{prob:repopt}
        \begin{align}
            \min_{\ab_t} \quad & {\rm TOTVAR}(\ab_t) && \label{eqn:objective_function} 
            \\
            \quad & \sum_{i=1}^{n_t} a_i \leq N, \label{eqn:cons1}  \\
            & a_i \in \mathbb{Z}^+, \forall i = 1, \ldots, n_t. \label{eqn:cons2} 
        \end{align}
    \end{subequations}
By relaxing the integrality constraint in \eqref{eqn:cons2}, the Lagrangian $L(\ab_t)$ is formulated as:
\begin{align} \notag 
    \begin{split}
        L(\ab_t) = {\rm TOTVAR}(\ab_t) + \lambda \left(N - \sum_{i=1}^{n_t} a_i\right).
    \end{split}
\end{align}  
The first-order optimality conditions are as follows
\begin{align} \notag 
    \begin{split}
        \frac{\partial L(\ab_t)}{\partial a_i} = \frac{\partial  {\rm TOTVAR}(\ab_t)}{\partial a_i} -\lambda = 0, \quad i = 1, \ldots, n_t.
    \end{split}
\end{align}     
We have
\begin{align} \notag 
    \begin{split}
        \frac{\partial {\rm TOTVAR}(\ab_t)}{\partial a_i} = \frac{1}{2^d\pi^{d/2}|\Sigmav|^{1/2}} \int\limits_{\tb \in \Theta} \frac{\partial f(\tb)}{\partial a_i} p(\tb)^2 d\tb - 2 \int\limits_{\tb \in \Theta} g(\tb) \frac{\partial g(\tb)}{\partial a_i} p(\tb)^2 d\tb,
    \end{split}
\end{align}
where 
\[
    \frac{\partial f(\tb)}{\partial a_i} = f(\tb) \frac{\partial \log f(\tb)}{\partial a_i} \quad {\rm and} \quad \frac{\partial g(\tb)}{\partial a_i} = g(\tb) \frac{\partial \log g(\tb)}{\partial a_i}.
\]
Let $\dot{\mathbf{N}}(\tb) = 0.5\Sigmav + \Sv_{t}(\tb)$, $\mathbf{N}(\tb) = \Sigmav + \Sv_{t}(\tb)$, and $\mathbf{h}(\tb) = \yb-\muv_{t}(\tb)$. By computing the partial derivatives, we obtain:
\begin{align} \label{partialfandg}
    \begin{split}
    &\frac{\partial \log f(\tb)}{\partial a_i} = -\frac{1}{2} \frac{\partial \log |\dot{\mathbf{N}}(\tb)|}{\partial a_i} -\frac{1}{2}\mathbf{h}(\tb)^\top\frac{\partial\dot{\mathbf{N}}(\tb)^{-1}}{\partial a_i}\mathbf{h}(\tb), \\
    &\frac{\partial \log g(\tb)}{\partial a_i} = -\frac{1}{2} \frac{\partial \log |\mathbf{N}(\tb)|}{\partial a_i} -\frac{1}{2}\mathbf{h}(\tb)^\top\frac{\partial \mathbf{N}(\tb)^{-1}}{\partial a_i}\mathbf{h}(\tb).
    \end{split}
\end{align}
Using $\frac{\partial \log |\mathbf{Y}|}{\partial x} = {\rm Tr}\left(\mathbf{Y}^{-1} \frac{\partial \mathbf{Y}}{\partial x}\right)$ and $\frac{\partial \mathbf{Y}^{-1}}{\partial x} = -\mathbf{Y}^{-1} \frac{\partial \mathbf{Y}}{\partial x} \mathbf{Y}^{-1}$, we rewrite \eqref{partialfandg} as
\begin{align} \label{eq:logderivative}
    \begin{split}
         \frac{\partial \log f(\tb)}{\partial a_i} &= -\frac{1}{2} {\rm Tr}\left(\dot{\mathbf{N}}(\tb)^{-1}\frac{\partial \dot{\mathbf{N}}(\tb)}{\partial a_i} \right) + \frac{1}{2}\mathbf{h}(\tb)^\top \dot{\mathbf{N}}(\tb)^{-1} \frac{\partial \dot{\mathbf{N}}(\tb)}{\partial a_i}\dot{\mathbf{N}}(\tb)^{-1} \mathbf{h}(\tb), \\
         \frac{\partial \log g(\tb)}{\partial a_i} &= -\frac{1}{2} {\rm Tr}\left(\mathbf{N}(\tb)^{-1}\frac{\partial \mathbf{N}(\tb)}{\partial a_i} \right) + \frac{1}{2}\mathbf{h}(\tb)^\top \mathbf{N}(\tb)^{-1} \frac{\partial \mathbf{N}(\tb)}{\partial a_i}\mathbf{N}(\tb)^{-1} \mathbf{h}(\tb),
    \end{split}
\end{align}
where $\frac{\partial \dot{\mathbf{N}}(\tb)}{\partial a_i} = \frac{\partial \mathbf{N}(\tb)}{\partial a_i} = \frac{\partial \Sv_{t}(\tb)}{\partial a_i}$.

Recall that the $j$th diagonal element of $\Sv_t(\tb)$ is $\varsigma^2_t(\zb_j^o)$, and the $(j,j^\prime)$th off-diagonal element is given by ${\rm cov}_t(\zb_j^o, \zb_{j^\prime}^o)$, where $\zb_j^o = \left({\xf_j}^\top, \tb^\top\right)^\top$ for $j = 1, \ldots, d$. We begin by deriving the expression for the derivative of the variance
\begin{align} 
    \begin{split} \notag 
    &\hspace{1cm} \frac{\partial \varsigma^2_{t}(\zb_j^o)}{\partial a_i} = - \kb^\top_{t}(\zb_j^o) \frac{\partial \Kb_{t}(\ab_t)^{-1}}{\partial a_i} \kb_{t}(\zb_j^o) \quad \rm{where} \\
    \frac{\partial \Kb_{t}(\ab_t)^{-1}}{\partial a_i} &= -\Kb_{t}(\ab_t)^{-1}  \frac{\partial \Kb_{t}(\ab_t)}{\partial a_i} \Kb_{t}(\ab_t)^{-1} \quad {\rm and} \quad
    \frac{\partial}{\partial a_i} \Kb_{t}(\ab_t) = -\frac{r(\zb_i)}{a_i^2}\mathbf{J}^{(i,i)},
    \end{split}
\end{align}
which results in
\begin{equation} \notag 
        \frac{\partial \varsigma^2_{t}(\zb_j^o)}{\partial a_i} = -\frac{r(\zb_i)}{a_i^2}\kb^\top_{t}(\zb_j^o) \Kb_{t}(\ab_t)^{-1}  \mathbf{J}^{(i,i)} \Kb_{t}(\ab_t)^{-1} \kb_{t}(\zb_j^o).
\end{equation}
Similarly, the derivative of the covariance function yields
\begin{equation}
     \frac{\partial {\rm cov}_{t}(\zb_j^o, \zb_{j^\prime}^o)}{\partial a_i} = -\frac{r(\zb_i)}{a_i^2}\kb^\top_{t}(\zb_j^o) \Kb_{t}(\ab_t)^{-1}  \mathbf{J}^{(i,i)} \Kb_{t}(\ab_t)^{-1} \kb_{t}(\zb_{j^\prime}^o).
\end{equation}
Let $\Mb_i(\tb)$ be a $d \times d$ matrix, where the $(j, j^\prime)$th element is given by $-r(\zb_i)\kb^\top_{t}(\zb_j^o) \Kb_{t}(\ab_t)^{-1}  \mathbf{J}^{(i,i)} \Kb_{t}(\ab_t)^{-1} \kb_{t}(\zb_{j^\prime}^o)$. 
Plugging $\Mb_i(\tb)$ into \eqref{eq:logderivative} and then solving first-order optimality conditions completes the proof.

\subsection{Details of Experiments}
\label{sec:experiment_details}

\begin{figure}[ht]
\centering
    \includegraphics[width=0.8\textwidth]{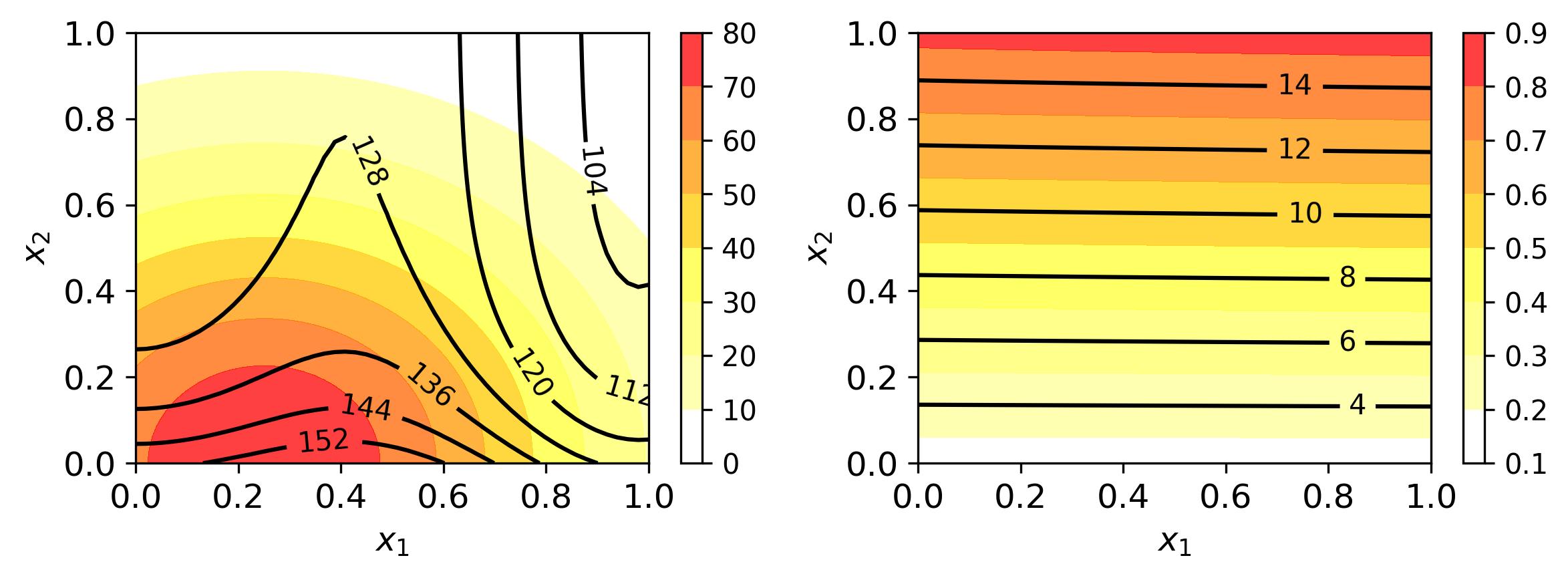}
    \caption{\tcb{Illustration of the noise variance $r(\cdot)$ (color shading) and the expected value of the simulation output $\eta(\cdot)$ (contour lines) across design input values. The left panel corresponds to the first example evaluated at $\theta^\ast=0.50$, while the right panel corresponds to the second example evaluated at $\tb^\ast=(0.50,0.50)^\top$.}}
    \label{fig:figure11}
\end{figure}

This section details the experimental setups for the synthetic functions described in Section~\ref{sec:synthetic}. 
\tcb{The first and second examples are adapted from \citet{Ranjan2011} and \citet{park1991}, respectively, to illustrate calibration settings in which the field inputs and calibration parameters interact.} Figure~\ref{fig:figure11} provides a visual reference for the noise variance $r(\cdot)$ and the expected value of the simulation output $\eta(\cdot)$ for both examples.

For the first example with $p=1$ and $q=2$, the expected simulation output is given by $\eta(\zb) = (30 + 5 \times x_1 \times \sin(5  x_1)) (6 \times \vartheta + 1 + \exp(-5 x_2))$ and the noise variance is $r(\zb) = 200 \vartheta \times f_\mathcal{N}(\xb; \mathbf{b}, \mathbf{C})$ with $\mathbf{b} = (0.25, 0)^\top$ and $\mathbf{C} = 0.2 \mathbf{I}_2$. The observed field data point is generated as $y(\xb_j^o) = \eta(\zb_j^o) + \epsilon$, for $j=1,2,3,4$, where $\xb_1^o = (0.2, 0.2)^\top$, $\xb_2^o = (0.2, 0.8)^\top$, $\xb_3^o = (0.8, 0.2)^\top$, $\xb_4^o = (0.8, 0.8)^\top$ and $\theta^\ast = 0.50$, and $\epsilon \sim {\rm N}(0, 10)$. 

\tcb{For the second example with $p=2$ and $q=2$, the expected simulation output is given by $\eta(\zb) = \frac{\vartheta_1}{2}\left(\sqrt{1+\frac{(x_1 + \vartheta_2^2) x_2}{\vartheta_1^2}} - 1\right) + (\vartheta_1 + 3x_2)\exp\left(1+\sin(\vartheta_2)\right)$ and the noise variance is $r(\zb) = 0.05 + 0.05 \eta(\zb)$. The observed field data point is generated as $y(\xb_j^o) = \eta(\zb_j^o) + \epsilon$, for $j=1,2,3,4$, where $\xb_1^o = (0.2, 0.2)^\top$, $\xb_2^o = (0.2, 0.8)^\top$, $\xb_3^o = (0.8, 0.2)^\top$, $\xb_4^o = (0.8, 0.8)^\top$ and $\tb^\ast = (0.50, 0.50)^\top$, and $\epsilon \sim {\rm N}(0, 0.5)$.}

We now provide the details of the acquisition functions included in our benchmark study in Section~\ref{sec:experiment}. To compute the IMSE at a given candidate input, we construct an emulator of the simulation model at each stage following the procedure described in Section~\ref{sec:GP}. The next input is selected to minimize the total uncertainty of the emulator. Specifically, at stage $t$, we quantify the value of evaluating the simulation model at a candidate input $\zb^c$ using the following IMSE criterion:
    \begin{align}\label{imse_general}
        \begin{split}
            \text{IMSE}(\zb^c) = \int_{\zb \in \mathcal{Z}} \varsigma^2_{t+1}(\zb) d\zb.
        \end{split} 
    \end{align}
Analogous to the proposed IVAR criterion, we evaluate the IMSE in two distinct settings corresponding to exploration and replication. For exploration, when the candidate input corresponds to a new location, $\zb^c = \breve{\zb}$, the IMSE criterion is given by (see \eqref{var_future} for the derivation of $\varsigma^2_{t+1}(\zb)$)
    \begin{align}\label{imse_explore}
        \begin{split}
            \text{IMSE}(\breve{\zb}) = \int_{\zb \in \mathcal{Z}} \left(\varsigma^2_{t}(\zb) - \frac{{\rm cov}_{t}(\zb, \breve{\zb})^2}{\varsigma^2_t(\breve{\zb}) + r(\breve{\zb})}\right) d\zb.
        \end{split} 
    \end{align}
For replication, when the candidate input coincides with an existing design point, $\zb^c = \zb_k$, the IMSE criterion becomes (see \eqref{var_future_exploit} for the derivation of $\varsigma^2_{t+1}(\zb)$)
        \begin{align}\label{imse_exploit}
        \begin{split}
            \text{IMSE}(\zb_k) = \int_{\zb \in \mathcal{Z}} \left(\varsigma^2_{t}(\zb) - \kb_{t}(\zb)^\top \mathbf{B}_k\kb_{t}(\zb)\right) d\zb.
        \end{split} 
    \end{align}
We use the IMSE implementation provided in the Python package hetGPy \citep{OGara2025}. While IMSE targets global uncertainty reduction of the emulator over the joint design–parameter space, $\text{IMSE}^y$ focuses on reducing predictive uncertainty of the field observations across the design space at the current parameter estimate. At each stage $t$, we first obtain the maximum likelihood estimate $\hat{\tb}_t$ of the parameter, and then evaluate the uncertainty conditional on this estimate. In practice, this is implemented by replacing $\mathcal{Z}$ in \eqref{imse_general}--\eqref{imse_exploit} with $\tilde{\mathcal{Z}}_t$, defined as $\tilde{\mathcal{Z}}_t = \left\{\zb = \left(\xb^\top, \tg^\top\right)^\top \in \mathcal{Z} : \tg=\hat{\tb}_t\right\}$.

\begin{figure}[ht]
\centering
    \includegraphics[width=0.9\textwidth]{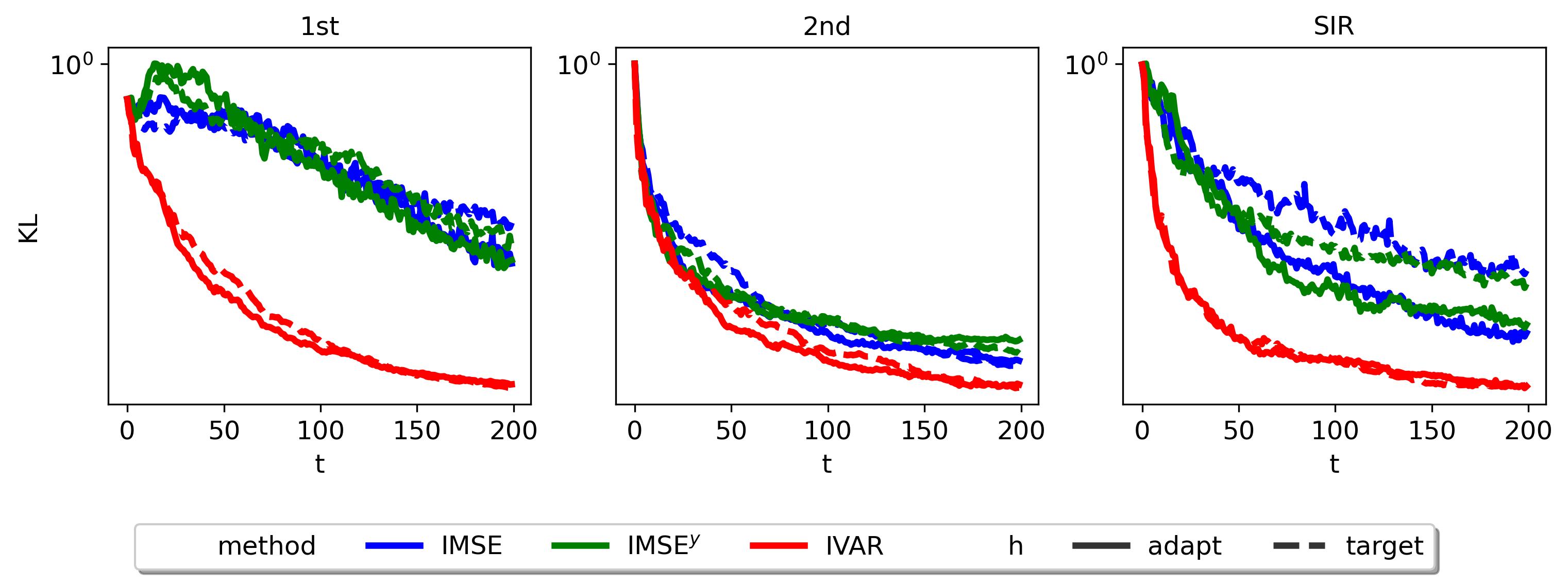}
    \caption{\tcb{KL-based comparison of different acquisition functions for three test problems: the first example, the second example, and the SIR function.}}
    \label{fig:figure12_R2}
\end{figure}

In addition to the MAD results presented in Sections~\ref{sec:synthetic}--\ref{sec:epidemic}, Figure~\ref{fig:figure12_R2} summarizes performance using a Kullback–Leibler (KL)-type measure computed for all examples and acquisition functions. Smaller values of this metric indicate that the estimated posterior assigns higher density to points drawn from the true posterior, reflecting a closer match and more accurate learning of the posterior distribution.

\subsection{\tcb{Additional Experiments}}
\label{sec:additional_experiments}

\begin{figure}[ht]
\centering
    \includegraphics[width=1\textwidth]{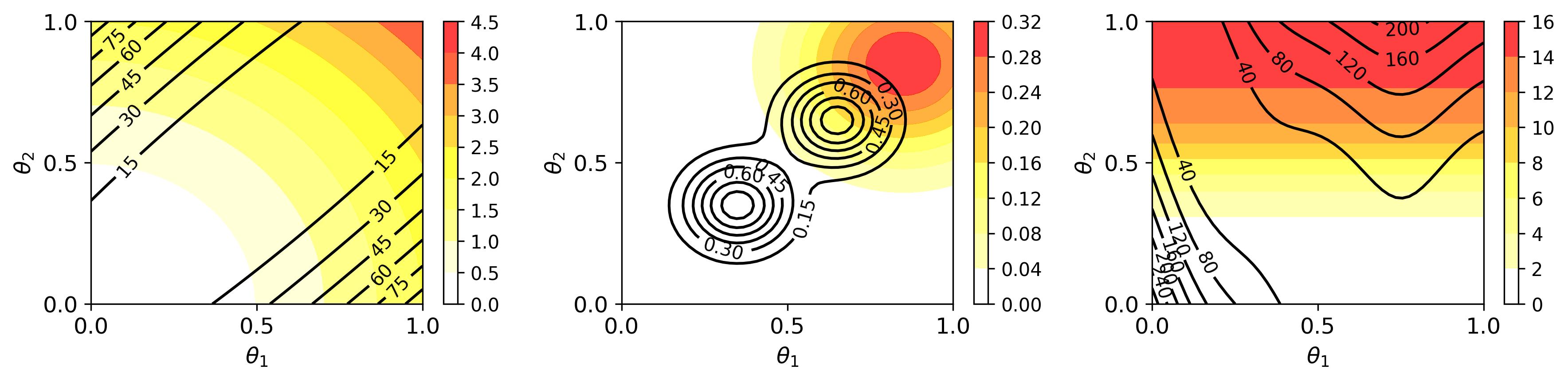}
    \caption{Illustration of the noise variance $r(\cdot)$ (color shading) and the expected value of the simulation output $\eta(\cdot)$ (contour lines) across parameter values at the field data design input $x^o=0.50$ for the unimodal (left), bimodal (middle), and Branin (right) functions.}
    \label{fig:figure12}
\end{figure}

We also evaluate performance using modified versions of three widely studied benchmark functions—unimodal, bimodal (\cite{Jarvenpa2019, Surer2023, Lartaud2025}), and Branin \citep{synthlinks}—to examine how our approach handles different posterior shapes. These three functions originally depend on two calibration parameters. To introduce a field dimension while preserving the posterior structure induced by the calibration parameters, we add a linear term in the design input $x$. In all examples, observed field data are generated according to \eqref{eq:statmodel}, with the calibration parameters set to the data-generating value $\tb = \tb^\ast$, as specified below. Figure~\ref{fig:figure12} illustrates the noise variance $r(\cdot)$ and the expected simulation output $\eta(\cdot)$ for the unimodal, bimodal, and Branin test functions.

\begin{table}[t]
\centering
\scriptsize
\setlength{\tabcolsep}{3pt}
\caption{Width of empirical $(1-\alpha)$ quantile intervals of acquired inputs and marginal and joint coverage of the data-generating calibration parameters.}
\label{tab:coverage_width_synth}
\begin{tabular}{l l || l c c c || c c c c}
 & & & \multicolumn{3}{c|}{Coverage} & \multicolumn{3}{c}{Width} \\
\toprule
 & $h$ &  & Bimodal & Branin & Unimodal &  & Bimodal & Branin & Unimodal \\
\midrule
\multirow{5}{*}{\rotatebox{90}{IMSE}} 
 & adapt  & $\vartheta_1$ & 1.00 & 1.00 & 1.00 & $\vartheta_1$  & 0.95 & 0.98 & 0.99 \\
 & target & $\vartheta_1$ & 1.00 & 0.83 & 1.00 & $\vartheta_1$ & 0.89 & 0.94 & 0.97 \\
 & adapt  & $\vartheta_2$ & 1.00 & 1.00 & 1.00 & $\vartheta_2$ & 0.95 & 0.98 & 0.99 \\
 & target & $\vartheta_2$ & 1.00 & 1.00 & 1.00 & $\vartheta_2$ & 0.89 & 0.91 & 0.97 \\
 & adapt  & Joint         & 1.00 & 1.00 & 1.00 & $x$ & 0.98 & 0.98 & 0.99 \\
 & target & Joint         & 1.00 & 0.83 & 1.00 & $x$ & 0.95 & 0.94 & 0.97 \\
\midrule
\multirow{5}{*}{\rotatebox{90}{$\text{IMSE}^y$}} 
 & adapt  & $\vartheta_1$ & 0.87 & 0.63 & 1.00 & $\vartheta_1$ & 0.21 & 0.57 & 0.40 \\
 & target & $\vartheta_1$ & 0.80 & 0.77 & 1.00 & $\vartheta_1$ & 0.20 & 0.68 & 0.40 \\
 & adapt  & $\vartheta_2$ & 0.87 & 0.87 & 1.00 & $\vartheta_2$ & 0.21 & 0.57 & 0.41 \\
 & target & $\vartheta_2$ & 0.83 & 0.87 & 1.00 & $\vartheta_2$ & 0.19 & 0.62 & 0.39 \\
 & adapt  & Joint         & 0.87 & 0.63 & 1.00 & $x$ & 0.98 & 0.97 & 0.98 \\
 & target & Joint         & 0.80 & 0.73 & 1.00 & $x$ & 0.97 & 0.96 & 0.97 \\
\midrule
\multirow{5}{*}{\rotatebox{90}{IVAR}}  
 & adapt  & $\vartheta_1$ & 1.00 & 1.00 & 1.00 & $\vartheta_1$ & 0.69 & 0.93 & 0.52 \\
 & target & $\vartheta_1$ & 1.00 & 1.00 & 1.00 & $\vartheta_1$ & 0.66 & 0.92 & 0.51 \\
 & adapt  & $\vartheta_2$ & 1.00 & 1.00 & 1.00 & $\vartheta_2$ & 0.68 & 0.97 & 0.53 \\
 & target & $\vartheta_2$ & 1.00 & 1.00 & 1.00 & $\vartheta_2$ & 0.65 & 0.95 & 0.50 \\
 & adapt  & Joint         & 1.00 & 1.00 & 1.00 & $x$  & 0.17 & 0.06 & 0.07 \\
 & target & Joint         & 1.00 & 1.00 & 1.00 & $x$  & 0.21 & 0.06 & 0.09 \\
\bottomrule
\end{tabular}
\end{table}
For the unimodal function with $p=2$ and $q=1$, the expected simulation output is given by $\eta(\zb) = 0.26 \times \left(\left(-10 + \vartheta_1 \times 20\right)^2 + \left(-10 + \vartheta_2 \times 20\right)^2\right) - 0.48 \times \left(-10 + \vartheta_1 \times 20\right) \times \left(-10 + \vartheta_2 \times 20\right) + (2 \times x - 1) $ and the noise variance is $r(\zb) = 0.01 + (\vartheta_1^2 + \vartheta_2^2) \times 2$. The observed field data point is generated as $y(x^o) = \eta(\zb^o) + \epsilon$, where $x^o = 0.5$, and $\tb^\ast = (0.5, 0.5)^\top$, and $\epsilon \sim {\rm N}(0, 0.1)$.

For the bimodal function with $p=2$ and $q=1$, the expected simulation output is given by $\eta(\zb) = \exp\left(-\frac{(\vartheta_1 - 0.35)^2 + (\vartheta_2 - 0.35)^2 }{0.15^2}\right) + \exp\left(-\frac{(\vartheta_1 - 0.65)^2 + (\vartheta_2 - 0.65)^2 }{0.15^2}\right) + (2 \times x - 1)$ and the noise variance is $r(\zb) = 0.1 \times f_\mathcal{N}(\tg; \mathbf{b}, \mathbf{C})$ with $\mathbf{b} = (0.85, 0.85)^\top$ and $\mathbf{C} = 0.05 \mathbf{I}_2$. The observed field data point is generated as $y(x^o) = \eta(\zb^o) + \epsilon$, where $x^o = 0.5$, and $\tb^\ast = (0.35, 0.35)^\top$, and $\epsilon \sim {\rm N}(0, 0.05)$.

For the Branin function with $p=2$ and $q=1$, the expected simulation output is given by $\eta(\zb) = \left( 15 \times \vartheta_2 - \frac{5.1}{4 \pi^2} (-5 + 15 \times \vartheta_1)^2 + \frac{5}{\pi} (-5 + 15 \times \vartheta_1) - 6 \right)^2  + 10 \left( 1 - \frac{1}{8 \pi} \right) \cos(-5 + 15 \times \vartheta_1) + 10 + (2\times x - 1)$ and the noise variance is $r(\zb) = 0.1 + \left(15 - 0.1\right) \times \frac{1}{1 + \exp(-10 \times (\vartheta_2 - 0.5))}$. The observed field data point is generated as $y(x^o) = \eta(\zb^o) + \epsilon$, where $x^o = 0.5$, and $\tb^\ast = (0.96, 0.16)^\top$, and $\epsilon \sim {\rm N}(0, 5)$.

\begin{figure}[!ht]
\centering
    \includegraphics[width=1\textwidth]{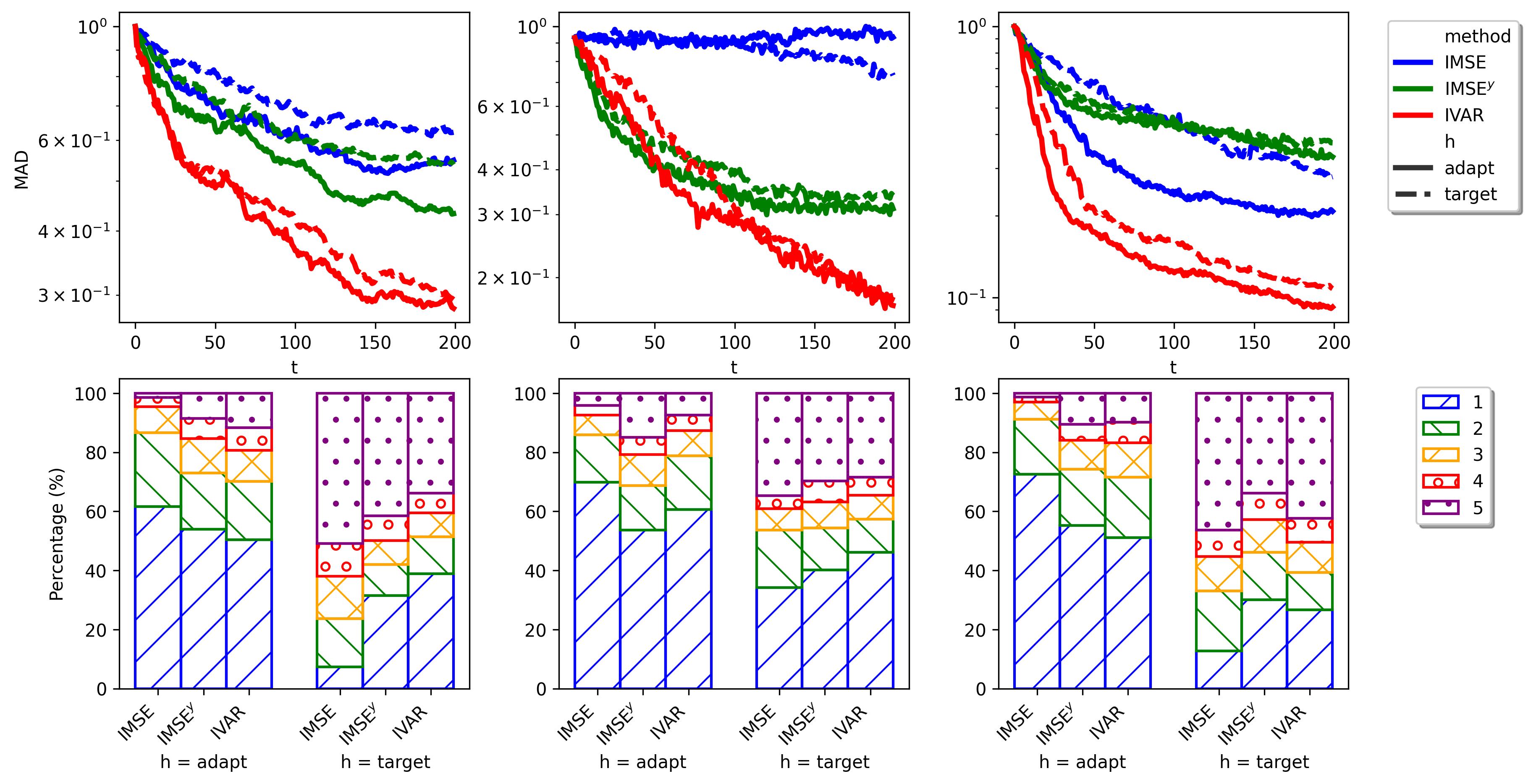}
    \caption{Comparison of different acquisition functions for the unimodal (left), bimodal (middle), and Branin (right) functions.}
    \label{fig:figure7}
\end{figure}

\begin{figure}[ht]
\centering
    \includegraphics[width=0.9\textwidth]{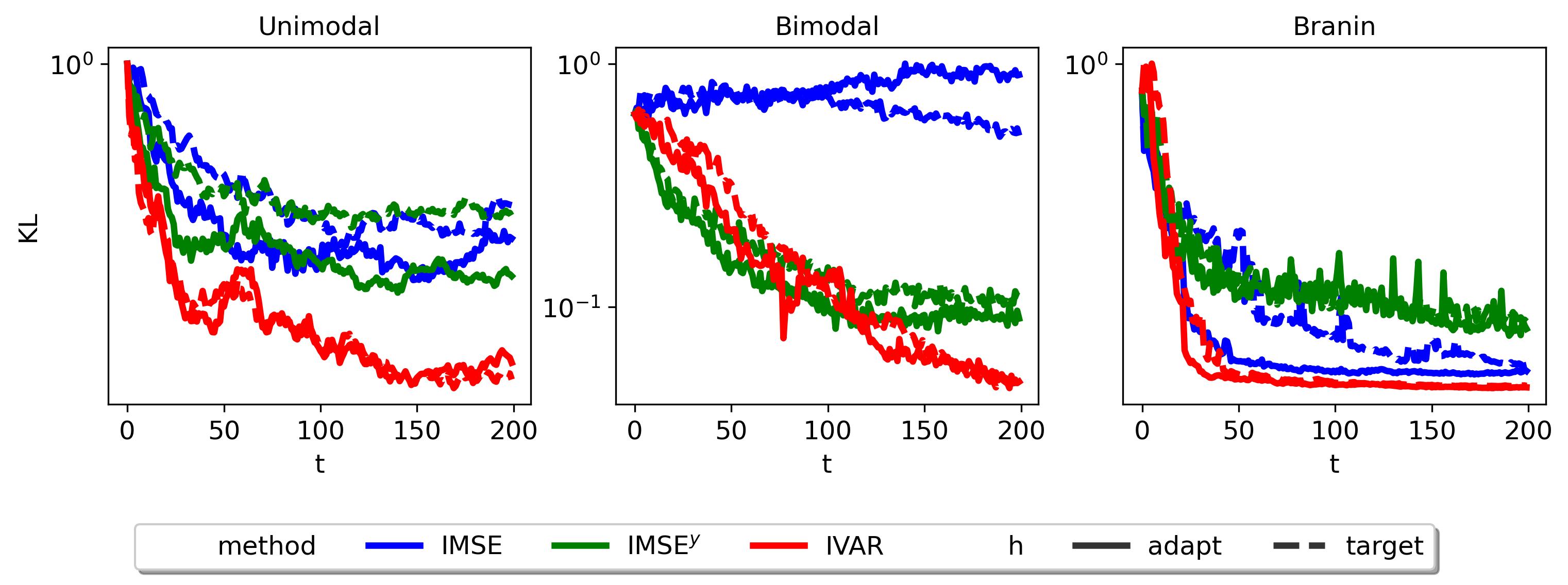}
    \caption{KL-based comparison of different acquisition functions for three test problems: unimodal, bimodal, and Branin functions.}
    \label{fig:figure13}
\end{figure}

\begin{figure}[!ht]
\centering
    \includegraphics[width=1\textwidth]{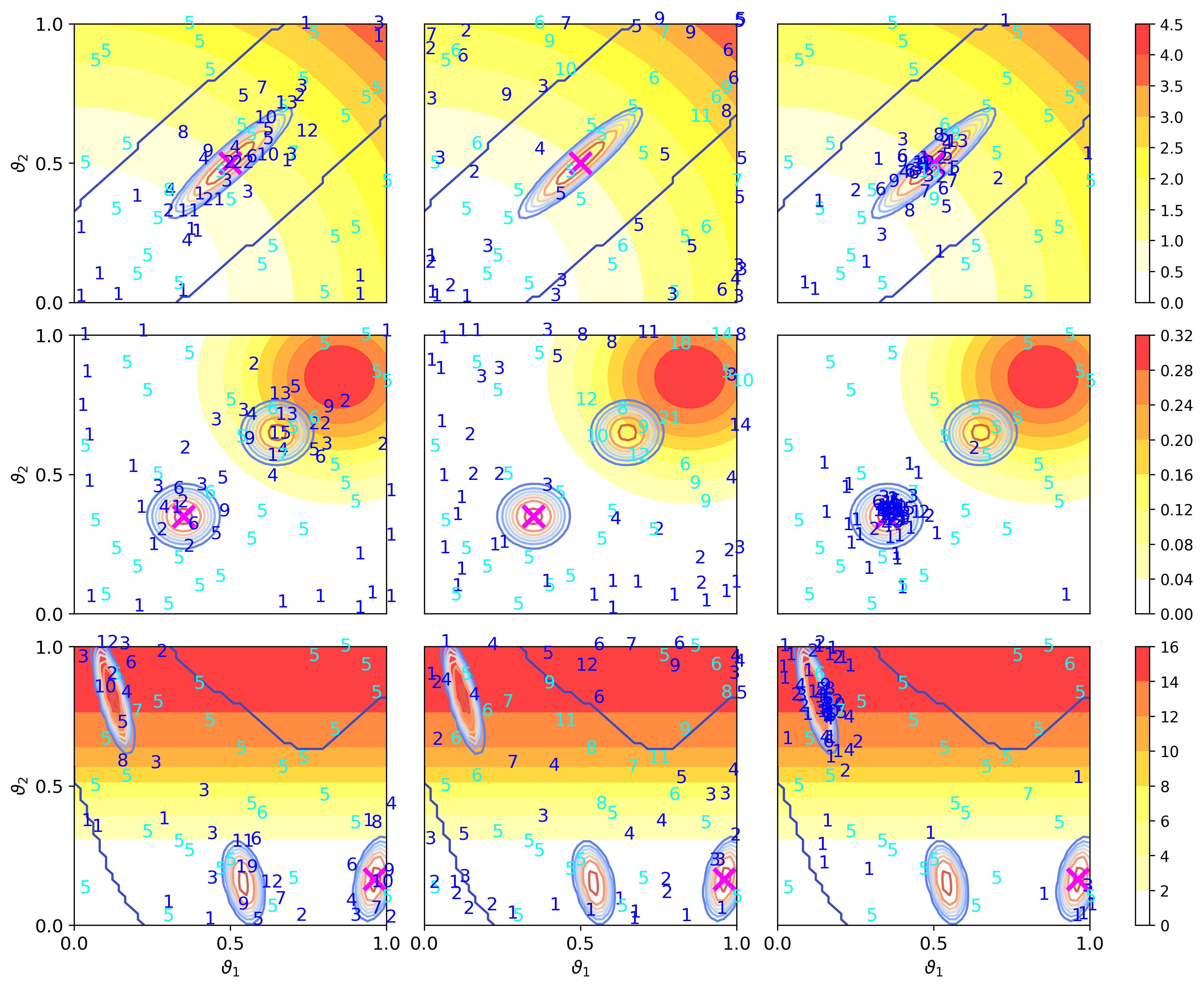}
    \caption{Illustration of the acquired parameters selected using IVAR (left), IMSE (center), and $\text{IMSE}^y$ (right) for the unimodal (top), bimodal (middle), and Branin (bottom) functions. Cyan markers indicate the initial sample points, and blue markers indicate the acquired parameters. Contours represent the true posterior density $\tilde{p}(\thetav\mid\yb)$ as a reference, while the background color indicates the intrinsic noise at the field data design input. The magenta marker indicates the data-generating parameter $\tb^\ast$.}
    \label{fig:figure8}
\end{figure}

Figure~\ref{fig:figure7} summarizes performance across the unimodal, bimodal, and Branin functions, reporting both the MAD values and the corresponding distribution of replication counts. Table~\ref{tab:coverage_width_synth} summarizes the widths of the empirical $(1-\alpha)$ quantile intervals for the acquired inputs and their associated coverage rates. To complement the MAD results, Figure~\ref{fig:figure13} reports a KL-type measure computed across all examples and acquisition functions. Finally, Figure~\ref{fig:figure8} displays the parameters acquired by each criterion for a single replicate of these examples.

\clearpage
\subsection{Impact of Replication and Noise Modeling}
\label{sec:homGP}

This section investigates the importance of replication and heteroscedastic noise estimation from both inferential and computational perspectives. The analysis uses two synthetic examples: the first example with a one-dimensional parameter space in Section~\ref{sec:synthetic} and the second example based on the Branin function in Appendix~\ref{sec:additional_experiments}. We compare the proposed approach with two homoscedastic GP (homGP) benchmarks, namely those with replication and without replication. In the former, we replace the hetGP in the proposed framework with a homGP, assuming an intrinsic noise variance that is independent of the input (i.e., $r(\cdot)$ in \eqref{eq:gp_prediction} is no longer input-dependent and is replaced by a single scalar value for all $\zb_i$). In the latter, only the exploration component of IVAR is used with homGP, and no replication is performed. For the methods with replications, we use the target-based scheme with a ratio $\rho=0.20$.

\begin{figure}[!ht]
\centering
    \includegraphics[width=1\textwidth]{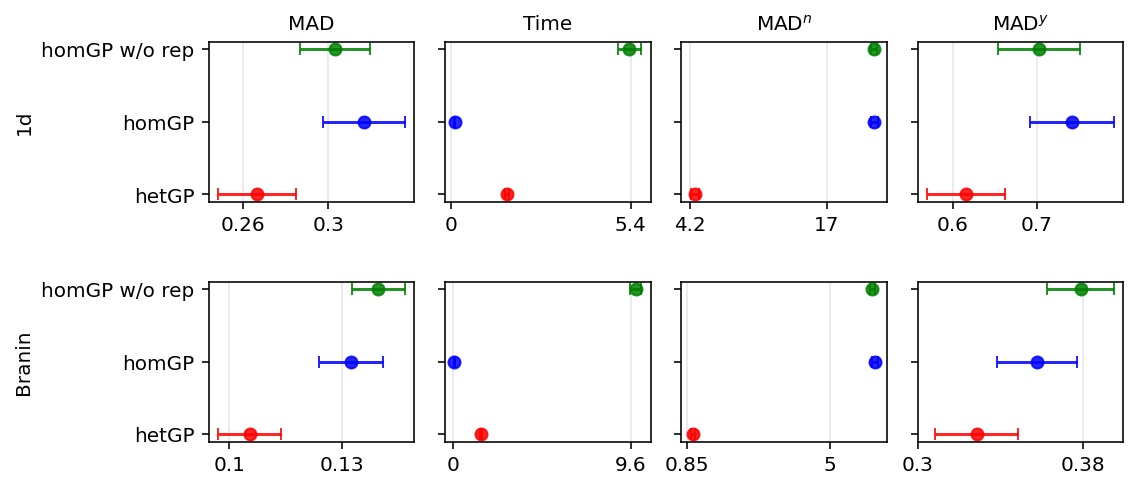}
    \caption{Comparison of the proposed procedure (bottom line) with homGP with replication (middle line) and homGP without replication (top line), shown for the one-dimensional example (top row) and the Branin function (bottom row). Circle markers show the mean across replications, with error lines indicating $\pm$ one standard error.}
    \label{fig:fighom}
\end{figure}

Performance is assessed across 30 independent experimental replications for a total of $T=200$ acquisitions, similar to the setting presented in Section~\ref{sec:synthetic}. In addition to MAD, we evaluate how well each criterion estimates the expected simulation output and intrinsic noise at the field data input locations via ${\rm MAD}^y$ and ${\rm MAD}^n$, as defined in Section~\ref{sec:epidemic}. In addition, we record the time required to build the emulators at each stage. Figure~\ref{fig:fighom} shows the final-stage values of MAD, computation time (seconds), ${\rm MAD}^n$, and ${\rm MAD}^y$.

The proposed procedure achieves the lowest MAD values and the most accurate estimates of both the mean response and intrinsic noise. The hetGP emulator is slightly more time-consuming than homGP with replication due to the additional estimation of the noise structure; however, this cost is negligible compared to the substantially higher build time in the no-replication scenario, where the absence of replication results in a much larger number of unique input locations and, consequently, a more expensive emulator construction. Overall, modeling heteroscedastic noise enables accurate characterization of input-dependent uncertainty, while replication improves inference and substantially reduces emulator training time.

\end{document}